\documentclass[%
 reprint,
nofootinbib,
 amsmath,amssymb,
 aps,
]{revtex4-2}

\usepackage{graphicx}
\usepackage{dcolumn}
\usepackage{bm}

\usepackage{slashed}
\usepackage{float}

\begin{document}


\title{Complete one-loop self-energies of the linear sigma model coupled to quarks at finite temperature and in a magnetic field}

\author{Adolfo Flores-Aguilar}
\author{Luis A. Hernández}
\author{J. Carlos Márquez}
\affiliation{Departamento de F\'isica, Universidad Aut\'onoma Metropolitana-Iztapalapa, Avenida San Rafael Atlixco 186, Ciudad de México 09340, Mexico.}
\author{R. Zamora}
\affiliation{Centro de Investigaci\'on y Desarrollo en Ciencias Aeroespaciales (CIDCA), Academia Politécnica Aeronáutica, Fuerza A\'erea de Chile, Casilla 8020744, Santiago, Chile.}
\affiliation{Facultad de Ingenier\'ia, Universidad San Sebasti\'an, Bellavista 7, Recoleta, Santiago, Chile.}

\date{\today}

\begin{abstract}

We present a complete calculation of the one-loop self-energies for all fields in the linear sigma model coupled to quarks at finite temperature and in the presence of a uniform magnetic field. The analysis consistently incorporates thermal and magnetic effects for both neutral and charged degrees of freedom, providing a unified framework valid for arbitrary values of the temperature and the field strength. The computation is performed using the Matsubara formalism to account for finite temperature effects and the Schwinger proper-time representation for charged propagators in a magnetic background. Special attention is given to loop contributions involving particles with different electric charges, for which the associated Schwinger phases do not cancel. We show that these terms can be systematically evaluated in coordinate space using the Ritus formalism, which provides the appropriate framework for treating external charged states in the presence of a magnetic background, and consistently expressed in momentum space. The resulting expressions exhibit a nontrivial interplay between thermal fluctuations and magnetic effects and allow for a clear separation between vacuum and matter contributions, providing a well-defined structure for the identification of ultraviolet divergences. Our results establish a consistent and systematic framework for the computation of thermomagnetic one-loop corrections in effective models of QCD, capturing the full interplay between thermal and magnetic effects for all dynamical degrees of freedom.

\end{abstract}

\maketitle


\section{Introduction \label{sec1}}

The strong magnetic fields generated in relativistic heavy-ion collisions~\cite{Skokov:2009qp,Brandenburg:2021lnj,STAR:2023jdd}, observed in compact stars~\cite{Ho:2011gy,Gusakov:2017uam,Igoshev:2021ewx}, and conjectured to have existed in the early universe~\cite{Grasso:2000wj,Subramanian:2009fu} have renewed interest in understanding the properties of strongly interacting matter under extreme conditions~\cite{Adhikari:2024bfa,Hattori:2023egw}. Numerous studies have explored the impact of magnetic fields on the QCD phase transition using lattice QCD simulations~\cite{DElia:2010abb,Bali:2011qj,Bali:2012zg,Bali:2014kia,Endrodi:2015oba,DElia:2021yvk,Endrodi:2024cqn}, effective models~\cite{Mizher:2010zb,Fraga:2012ev,Andersen:2014xxa,Ayala:2014gwa,Ayala:2015lta,Pagura:2016pwr,Bandyopadhyay:2020zte,Lo:2020ptj,Ayala:2021nhx,Moreira:2021ety,Farias:2021fci,Cao:2021rwx,Backes:2021mdt,Pelicer:2022hjp,Fernandez:2025fgp}, and other perturbative and non-perturbative approaches to the strong interaction~\cite{Zayakin:2008cy,Filev:2009xp,Gutierrez:2013sta,Rodrigues:2017cha,Ballon-Bayona:2020xtf,Gursoy:2021efc,Ahmad:2023ecw,Cai:2024eqa}. In addition, several works have investigated the influence of magnetic fields on hadronic matter~\cite{Andreichikov:2013zba,Machado:2013yaa,Mueller:2014tea,Liu:2014uwa,Loewe:2022aaw,Dominguez:2023bjb,Taya:2014nha,Simonov:2015xta,Gubler:2015qok,Hattori:2015aki,Yoshida:2016xgm,Zhang:2016qrl,Ayala:2016bbi,Ghosh:2016evc,Bali:2017ian,Ghosh:2017rjo,Coppola:2018vkw,Liu:2018zag,Aguirre:2018fbo,Ayala:2018zat,Avancini:2018svs,Das:2019ehv,Ayala:2020dxs,Carlomagno:2022inu,Das:2022mic,Ayala:2023llp,Hernandez:2025inu,Dominguez:2025nar}, as well as their effects on electromagnetic probes~\cite{Ayala:2017vex,Ayala:2019jey,Jia:2022awu,Jaber-Urquiza:2023swn,Castano-Yepes:2024vlj,Jia:2024ctx,Ayala:2025jpr} and on the early stages of relativistic heavy-ion collisions~\cite{Ayala:2024jvc}.

These developments highlight the relevance of studying the dynamics of strongly interacting systems in the presence of magnetic fields. Nevertheless, despite the considerable progress achieved over the past decades, several theoretical quantities remain uncalculated, either because of the analytical complexity involved or due to the computational effort required for their numerical evaluation. In this work, we contribute to this effort by analyzing the dynamics of an effective description of QCD, namely the linear sigma model coupled to quarks (LSMq).

Since the physical systems mentioned above are characterized by finite temperature, the study of magnetic effects in the dynamics of strongly interacting matter must also incorporate thermodynamic effects. We compute the one-loop corrections to the propagation of all the degrees of freedom in the LSMq, namely the neutral pion, the charged pions, the sigma meson, and the two lightest quark flavors. The corresponding self-energies encode the effects of a constant and uniform magnetic field together with the presence of a thermal bath. The calculation is performed without imposing restrictions on the magnetic-field strength and is valid for arbitrary temperatures. It is carried out using the Schwinger proper-time representation for propagators in a magnetic background combined with the imaginary-time formalism of finite-temperature field theory.

To the best of our knowledge, this is the first comprehensive computation of the complete one-loop self-energy structure of the LSMq including both thermal and magnetic effects. Our results therefore provide the most general one-loop thermomagnetic corrections to the propagators of the sigma meson, the neutral and charged pions, and the two lightest quark flavors within this model. These results are relevant for several applications. On the one hand, the self-energies constitute a key ingredient for studying the QCD phase transition beyond the mean-field approximation. On the other hand, they are essential for describing properties of the hadronic medium beyond the quantum ideal-gas approximation.

The paper is organized as follows. In Sec.~\ref{sec1} we introduce the linear sigma model coupled to quarks and discuss the features relevant for the present study. We present the Lagrangian density, describe the degrees of freedom of the model, and summarize the corresponding Feynman rules. In Sec.~\ref{sec3} we compute the one-loop self-energies for all the fields. We first present the Feynman diagrams contributing to each self-energy and then evaluate the corresponding contributions in the presence of a constant magnetic field and finite temperature. To the best of our knowledge, this is the first time that the self-energies of all the particles within the LSMq are computed without restrictions on the values of $T$ and $|eB|$. For charged particles, the effect of the magnetic field is incorporated for external states, which are expressed in the Ritus basis instead of plane waves. Finally, in Sec.~\ref{sec4} we summarize our results and present our conclusions.

Finally, we emphasize that this work has been written with the aim of making the derivation of all expressions as transparent and reproducible as possible. For this reason, we present the intermediate steps in detail, even when this leads to an extended exposition. We consider this level of detail to be essential, as part of the value of this work lies not only in reporting expressions that, to the best of our knowledge, have not been previously obtained in the literature, but also in providing a clear and systematic methodology to derive them. This approach is intended to facilitate future applications and extensions of these results in the study of strongly interacting matter under extreme conditions.

\section{Linear Sigma Model coupled to quarks \label{sec2}}

The LSMq is one of the most widely used effective models for describing the low-energy regime of QCD. A key feature of this renormalizable model is its ability to exhibit spontaneous chiral symmetry breaking. Its degrees of freedom consist of scalar and pseudoscalar mesons together with the two lightest quark flavors. The Lagrangian density is given by
\begin{align}
\mathcal{L} &= \frac{1}{2}(\partial_{\mu}\sigma)^2+\frac{1}{2}(\partial_{\mu} \pi_{0})^2+D_{\mu} \pi_- D^{\mu} \pi_{+}\nonumber \\
&+\frac{a^2}{2}(\sigma^2+\pi_0^2+2 \pi_{-}\pi_{+})-\frac{\lambda}{4} \left( \sigma^2+\pi_0^2+2 \pi_{-}\pi_{+}\right) \nonumber \\
&+i\bar{\psi} \slashed{\partial}\psi - g\bar{\psi} ( \sigma+i \gamma^{5} \vec{\tau}\cdot\vec{\pi} ) \psi,
\label{lagrangian}
\end{align}
where $\psi$ is an $SU(2)$ isospin doublet of quarks (up and down), $\sigma$ is an isospin singlet scalar field, and $\{\pi_+,\pi_-,\pi_0\}$ is an isospin triplet corresponding to the three pions. In Eq.~(\ref{lagrangian}), $\tau_i$ (with $i=+,-,3$)\footnote{The matrices $\tau_\pm$ are defined as $\tau_\pm=\frac{1}{\sqrt{2}}(\tau_1\pm\tau_2)$} denote the Pauli matrices acting in isospin space. The parameters $\lambda$ and $g$ correspond to the boson self-coupling and the Yukawa coupling between fermions and mesons, respectively, and $a^2$ is the squared mass parameter. Throughout this work we consider $a^2,\lambda,g>0$. The presence of an external magnetic field is taken into account in Eq.~(\ref{lagrangian}) through a minimal coupling for the charged fields via the replacement
\begin{equation}
\partial_\mu \rightarrow D_\mu=\partial_\mu+iqA_\mu,
\label{covariantder}
\end{equation}
where $A_\mu$ is the vector potential corresponding to a constant and uniform magnetic field oriented along the $\hat{z}$ direction and $q$ is the electric charge of the corresponding field.

In order to allow for spontaneous symmetry breaking, the $\sigma$ field develops a vacuum expectation value $v$, namely
\begin{equation}
\sigma \rightarrow \sigma + v.
\label{shift}
\end{equation}
After performing this shift, the Lagrangian can be written as
\begin{align}
\mathcal{L}&=\frac{1}{2}\partial_{\mu}\sigma \partial^{\mu}\sigma+\frac{1}{2}\partial_{\mu}\pi_{0}\partial^{\mu}\pi_{0}+D_{\mu}\pi_{-}D^{\mu}\pi_{+}\nonumber\\
&-\frac{1}{2}m_{\sigma}^{2}\sigma^{2}-\frac{1}{2}m_{\pi}^{2}\pi_{0}^{2}-m_{\pi}^{2}\pi_{-}\pi_{+}+i\bar{\psi}\slashed{\partial}\psi\nonumber\\
&-m_{f}\bar{\psi}\psi+\frac{a^2}{2}v^2-\frac{\lambda}{4}v^4+\mathcal{L}_{int}.
\label{linearsigmamodelSSB}
\end{align}
From Eq.~(\ref{linearsigmamodelSSB}) we identify the masses of the fields as
\begin{equation}
m_{\sigma}^{2}=3\lambda v^2-a^2, \qquad
m_{\pi}^{2}=\lambda v^2-a^2, \qquad
m_{f}=gv.
\label{masses}
\end{equation}
\begin{figure}[t]
    \centering
    \includegraphics[scale=0.61]{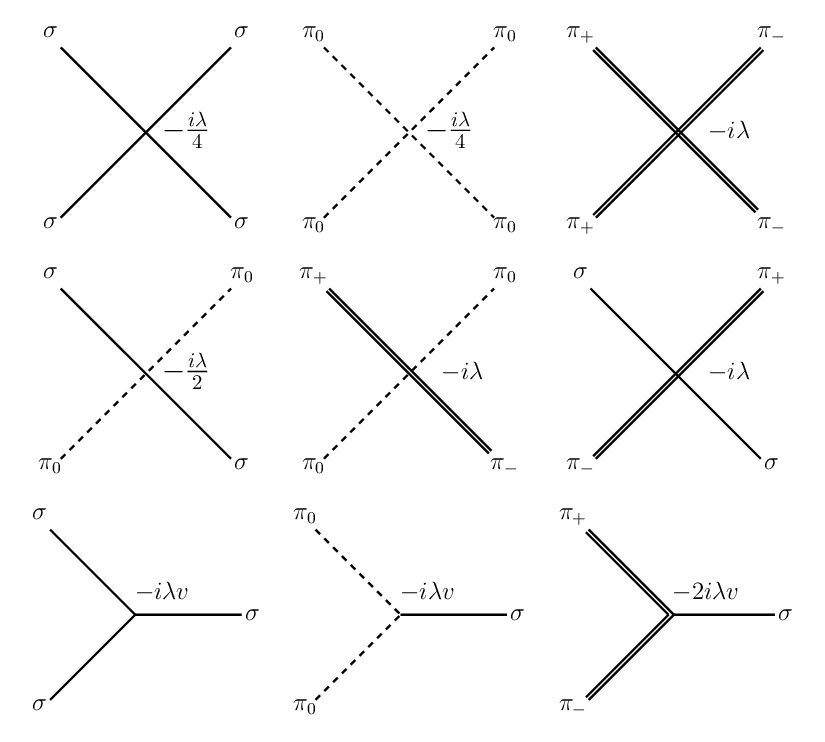}
    \caption{Feynman diagrams corresponding to the nine interaction vertices among the mesonic fields in the LSMq. Solid, dashed, and double solid lines denote the sigma meson, the neutral pion, and the charged pions, respectively.}
    \label{fig1}
\end{figure}
These masses are dynamically generated and depend on the vacuum expectation value $v$, which plays the role of the order parameter associated with the spontaneous breaking of chiral symmetry. The $\mathcal{L}_{int}$ contains the interaction terms among the fields and is defined as
\begin{align}
    \mathcal{L}_{int}&=-\frac{\lambda}{4}\sigma^{4}-\lambda v\sigma^{3}-\lambda v^{3}\sigma-\lambda\sigma^{2}\pi_{-}\pi_{+} -2\lambda v \sigma\pi_{-}\pi_{+} \nonumber \\
    &-\frac{\lambda}{2}\sigma^{2}\pi_{0}^{2}-\lambda v\sigma \pi_{0}^{2}-\lambda \pi_{-}^{2}\pi_{+}^{2}-\lambda\pi_{-}\pi_{+}\pi_{0}^{2}-\frac{\lambda}{4}\pi_{0}^{4} \nonumber \\ 
    &+a^{2}v\sigma -g\bar{\psi}\psi\sigma-ig\gamma^{5}\bar{\psi}\left(\tau_{+}\pi_{+}+\tau_{-}\pi_{-}+\tau_{3}\pi_{0}\right)\psi.
    \label{interactinglagrangian}
\end{align}

\begin{figure}[t]
    \centering
    \includegraphics[scale=0.65]{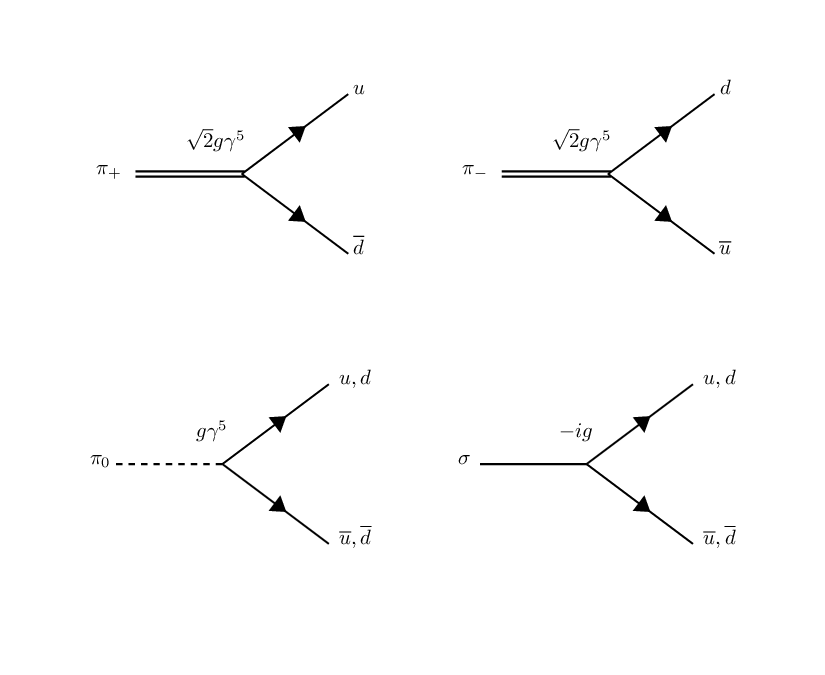}
    \caption{Feynman diagrams corresponding to the four interaction vertices between mesons and quarks fields in the LSMq. Solid, dashed, double solid lines and solid line with arrow denote the sigma meson, the neutral pion, the charged pions, and the quarks, respectively.}
    \label{fig2}
\end{figure}

The interaction terms contained in $\mathcal{L}_{int}$ determine the Feynman rules required for the perturbative expansion of the theory. The Feynman diagrams corresponding to all interaction vertices appearing in Eq.~(\ref{interactinglagrangian}) are shown in Figs.~\ref{fig1} and~\ref{fig2}. In the following section we compute the one-loop self-energy corrections to the propagators of all the fields of the model in the presence of a magnetic background and a thermal medium.

\section{One-loop self-energies \label{sec3}}

In this section, we compute the one-loop self-energies for all the fields of the LSMq. For each field, we first identify the Feynman diagrams contributing to the corresponding self-energy and write the associated analytical expressions. We then evaluate the different contributions and combine them to obtain the complete one-loop correction to the propagator. The same procedure is applied to all the fields of the model.

\subsection{Neutral pion one-loop self-energy\label{sec3.1}}

\begin{figure}[h!]
    \centering
    \includegraphics[scale=0.42]{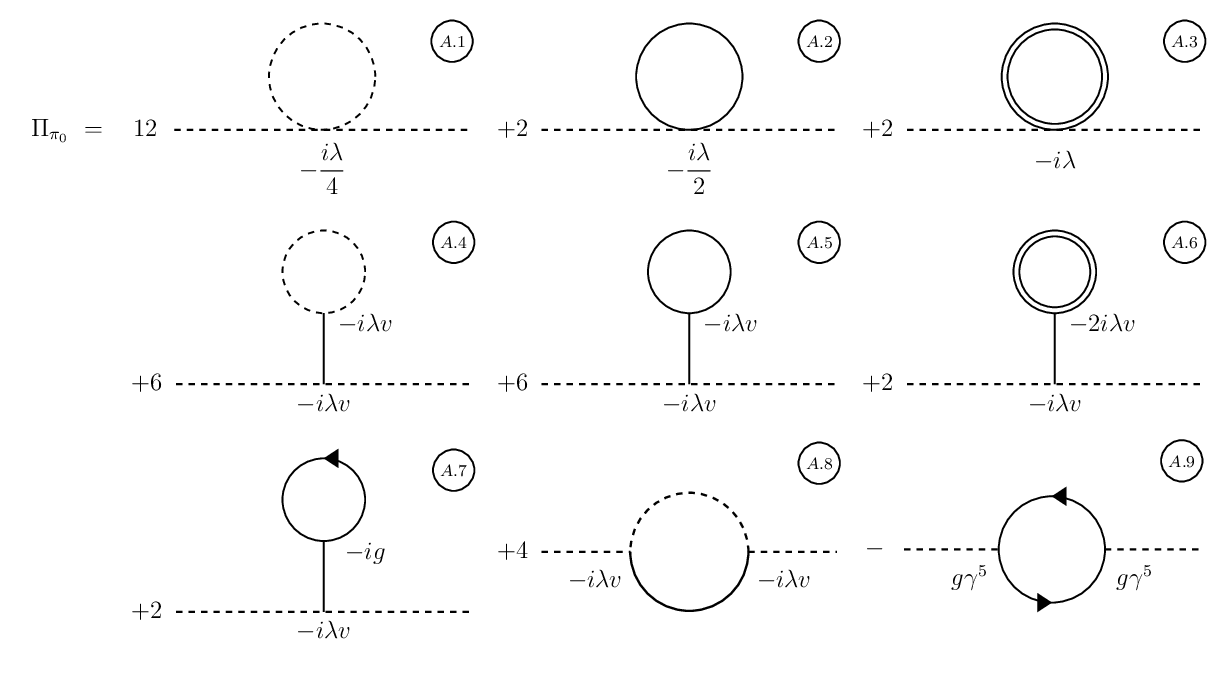}
    \caption{One-loop self-energy diagrams for the neutral pion field. Solid lines represent the sigma meson, dashed lines correspond to neutral pions, double solid lines denote charged pions, and solid lines with arrows represent quark fields. Each vertex includes the corresponding coupling coefficient, and the factors multiplying the diagrams account for the associated combinatorial factors.}
    \label{fig3}
\end{figure}
We begin by computing the one-loop self-energy of the neutral pion. In Fig.~\ref{fig3}, all the Feynman diagrams contributing to this quantity are shown. We label each contribution as $A.j$, with $j=1,2,\ldots,9$. The full neutral pion self-energy is then given by
\begin{align}
    -i\Pi_{\pi_{0}} = & -i\Pi_{\pi_{0}}^{(A.1)}  -i\Pi_{\pi_{0}}^{(A.2)} -i\Pi_{\pi_{0}}^{(A.3)} -i\Pi_{\pi_{0}}^{(A.4)} \nonumber \\
    &-i\Pi_{\pi_{0}}^{(A.5)}- i\Pi_{\pi_{0}}^{(A.6)} -i\Pi_{\pi_{0}}^{(A.7)} -i\Pi_{\pi_{0}}^{(A.8)} \nonumber \\
    &-i\Pi_{\pi_{0}}^{(A.9)}.
\end{align}

We evaluate each contribution separately. The first term corresponds to the tadpole diagram with a neutral pion circulating in the loop. In this case, only thermal effects need to be considered. The corresponding expression is
\begin{equation}
    - i \Pi_{\pi_{0}}^{(A.1)} = 12 \left( -\frac{i\lambda}{4} \right) \int \frac{d^{4}q}{(2\pi)^{4}}\frac{i}{q^{2}-m_{\pi}^{2}+i\epsilon},
    \label{A.1start}
\end{equation}
where $12$ is the symmetry factor, $(-i\lambda/4)$ is the vertex factor, and $q$ is the internal momentum. The same convention for constructing the initial expressions is used throughout the remaining contributions and self-energies. The symmetry and the vertex factors from the Feynman rules shown in Figs.~\ref{fig1} and~\ref{fig2}, together with the corresponding diagrams for each self-energy contribution. Therefore, in what follows we do not explicitly describe each element entering the initial expressions.

To evaluate this expression, we move to the imaginary-time formalism, performing a Wick rotation and replacing $q^0 \rightarrow i\omega_n$, where $\omega_n=2\pi n T$ are the bosonic Matsubara frequencies. Then, it becomes
\begin{equation}
    - i \Pi_{\pi_{0}}^{(A.1)} =\ 12 \left( -\frac{i\lambda}{4} \right) T \sum_{n=-\infty}^{\infty} \int \frac{d^{3}q}{(2\pi)^{3}} \frac{1}{\omega_{n}^{2} + \vec{q}^{2} + m_{\pi}^{2}}.
    \label{A.1inMatsubara}
\end{equation}
After performing the sum over Matsubara frequencies, we obtain
\begin{equation}
    -i\Pi_{\pi_{0}}^{(A.1)} = -3i\lambda \, \int \frac{d^{3}q}{(2\pi)^{3}} \frac{1}{2E_{\pi}} \Big( 1 + 2n_{B}(E_{\pi}) \Big),
    \label{A.1finalexpression}
\end{equation}
where $E_{\pi}=\sqrt{\vec{q}^2+m_\pi^2}$ and $n_B(X)$ is the Bose-Einstein distribution function. Since no restrictions are imposed on the temperature, the momentum integral cannot be evaluated analytically. However, the structure of the result allows us to identify its main contributions. The first term inside the brackets corresponds to the vacuum contribution, which can be regularized using dimensional regularization and subsequently renormalized. The second term represents the thermal contribution, which is ultraviolet finite and encodes the medium effects. This term gives the finite medium contribution to the neutral pion self-energy.

The next contribution corresponds to the tadpole diagram with a sigma meson circulating in the loop. This case is analogous to the previous one, with the appropriate replacement of the symmetry and vertex factors, as well as the pion mass by the sigma meson mass. The resulting expression is
\begin{equation}
    -i\Pi_{\pi_{0}}^{(A.2)} = -i\lambda \, \int \frac{d^{3}q}{(2\pi)^{3}} \frac{1}{2E_{\sigma}} \Big( 1 + 2n_{B}(E_{\sigma}) \Big),
\end{equation}
where $E_{\sigma}=\sqrt{\vec{q}^2+m_\sigma^2}$. As before, the first term inside the brackets corresponds to the vacuum contribution, which can be regularized and renormalized, while the second term represents the thermal contribution. The latter encodes the finite-temperature contribution to the neutral pion self-energy.

We next consider the first one-loop contribution involving a charged particle in the loop, namely the tadpole diagram with a charged pion. The corresponding expression is
\begin{align}
   -i\Pi_{\pi_{0}}^{(A.3)} &=\ 2\left(-i\lambda\right) \, iT \sum_{n} \int \frac{d^{3}q}{(2\pi)^{3}} \int_{0}^{\infty} \frac{ds}{\cos\left(\vert eB \vert s\right)}\nonumber \\
   &\times e^{is\left((i\omega_{n})^{2} - q_{3}^{2} - q_{\perp}^{2}\frac{\tan\left(\vert eB\vert s\right)}{\vert eB \vert s} - m_{\pi}^{2}\right)},
   \label{A.3initial}
\end{align}
where $q_\perp^2=q_1^2+q_2^2$ are the momentum components perpendicular to the magnetic field direction, $|eB|$ is the magnetic-field strength, and $s$ is the Schwinger proper-time. We rewrite Eq.~(\ref{A.3initial}) by separating the momentum into longitudinal and transverse components, $q_3$ and $q_\perp$, respectively. This gives
\begin{align}
    -i\Pi_{\pi_{0}}^{(A.3)}&= \ 2\lambda \ T \sum_{n} \int_{0}^{\infty} \frac{ds}{\cos\left(\vert eB \vert s\right)} \int_{-\infty}^{\infty} dq_{3} e^{-isq_{3}^{2}} \nonumber \\
    &\times \int_{-\infty}^{\infty} d^{2}q_{\perp} e^{-isq_{\perp}^{2}\frac{\tan\left(\vert eB\vert s\right)}{\vert eB \vert s}}  e^{-is\left(\omega_{n}^{2} + m_{\pi}^{2}\right)}.
    \label{A.3initialMatsubara}
\end{align}
Both momentum integrals are Gaussians and can be performed straightforwardly. The result is
\begin{equation}
    -i\Pi_{\pi_{0}}^{(A.3)}=-\frac{i\lambda |eB| T}{4\pi^{3/2}} \sum_{n} \int_{0}^{\infty} \frac{ds}{\sin\left(\vert eB \vert s\right)} \frac{1}{\sqrt{is}}e^{-is\left(\omega_{n}^{2} + m_{\pi}^{2}\right)}.
    \label{A.3MatsubaraIntds}
\end{equation}
We now perform the analytic continuation $s\rightarrow -i\tau$, obtaining
\begin{equation}
    -i\Pi_{\pi_{0}}^{(A.3)}=-\frac{i\lambda |eB| T}{4\pi^{3/2}}   \int_{0}^{\infty} \frac{d\tau}{\sinh\left(\vert eB \vert \tau\right)} \frac{e^{-\tau m_{\pi}^{2}}}{\sqrt{\tau}}\sum_{n}e^{-\tau \omega_{n}^{2}}.
    \label{A.3analyticcontinuation}
\end{equation}

To evaluate the sum over the Matsubara modes, we write explicitly
\begin{equation}
    \sum_{n} e^{-\tau \omega_{n}^{2}}=\sum_{n=-\infty}^{\infty} e^{-4\tau\pi^{2}T^{2}n^2},
    \label{ExpMatsubaraExplicit}
\end{equation}
which can be expressed in terms of the \textit{Jacobi theta} function of the third kind,
\begin{align}
    \vartheta_{3} (z\vert x)&=\sum_{n=-\infty}^{\infty} \text{exp}\left(i\pi xn^{2}+2i\pi z n\right) \nonumber \\
    &= 1 + 2\sum_{n=1}^{\infty} e^{i\pi x n^{2}} \cos\left(2n\pi z\right).
    \label{JacobiTheta3}
\end{align}
Using Eqs.~(\ref{ExpMatsubaraExplicit}) and~(\ref{JacobiTheta3}) in Eq.~(\ref{A.3analyticcontinuation}), we obtain
\begin{align}
    -i\Pi_{\pi_{0}}^{(A.3)}=-\frac{i\lambda |eB|}{8\pi^{2}}   &\int_{0}^{\infty} \frac{d\tau}{\sinh\left(\vert eB \vert \tau\right)} \frac{e^{-\tau m_{\pi}^{2}}} {\tau} \nonumber\\
    &\times \left(1 + 2\sum_{n=1}^{\infty} e^{-\frac{n^{2}}{4 T^{2}\tau}} \right).
    \label{A.3JustPropertime}
\end{align}
From this expression, we identify the purely magnetic contribution (first term inside the brackets) and the thermomagnetic contribution (second term). The purely magnetic term contains ultraviolet divergences, as expected from the vacuum contribution. In order to isolate this piece, we consider the limit $|eB|\to 0$, obtaining
\begin{equation}
    -i\Pi_{\pi_{0},B=0}^{(A.3)}=-\frac{i\lambda}{8\pi^{2}} \int_{0}^{\infty} \frac{d\tau}{\tau^{2}} e^{-\tau m_{\pi}^{2}},
    \label{A.3vacuum}
\end{equation}
which corresponds to the vacuum contribution. Therefore, in the full expression, we subtract the vacuum piece given in Eq.~(\ref{A.3vacuum}) and add back the corresponding renormalized vacuum term.

The next three contributions to the neutral pion self-energy correspond to tadpole-type diagrams with an additional internal line connecting the neutral pion propagator to the loop, as shown in Fig.~\ref{fig3} for diagrams $A.4$, $A.5$, and $A.6$. In all three cases, the extra line is a zero-momentum sigma meson propagator. Thus, they follow from diagrams $A.1$, $A.2$, and $A.3$, respectively, after replacing the vertex factors and including the zero-momentum sigma propagator. The resulting expressions are
\begin{equation}
    -i\Pi_{\pi_{0}}^{(A.4)} =\frac{6i \lambda^{2}v^{2}}{m_{\sigma}^{2}}\, T \sum_{n=-\infty}^{\infty} \int \frac{d^{3}q}{(2\pi)^{3}} \frac{1}{\omega_{n}^{2} + \vec{q}^{2} + m_{\pi}^{2}},
    \label{A.4initial}
\end{equation}
\begin{equation}
    -i\Pi_{\pi_{0}}^{(A.5)} =\frac{6i\ \lambda^{2}v^{2}}{m_{\sigma}^{2}}\, T \sum_{n=-\infty}^{\infty} \int \frac{d^{3}q}{(2\pi)^{3}} \frac{1}{\omega_{n}^{2} + \vec{q}^{2} + m_{\sigma}^{2}},
    \label{A.5initial}
\end{equation}
and
\begin{align}
    -i\Pi_{\pi_{0}}^{(A.6)} &=-\frac{1}{2\pi^{3}}\frac{\lambda^{2} v^{2}}{m_{\sigma}^{2}} T \sum_{n}\int d^{3}q \int_{0}^{\infty} \frac{ds}{\cos(|eB| s)}\nonumber \\
    &\times e^{-is\left( \omega_{n}^{2} + q_{3}^{2} + q_{\perp}^{2} \frac{\tan(|eB| s)}{|eB| s} + m_{\pi}^{2} \right)}.
    \label{A.6initial}
\end{align}
After carrying out the momentum integrations and Matsubara sums, these expressions become
\begin{equation}
    -i\Pi_{\pi_{0}}^{(A.4)} =\frac{6i\lambda^{2}v^{2}}{m_{\sigma}^{2}} \, \int \frac{d^{3}q}{(2\pi)^{3}} \frac{1}{2E_{\pi}} \Big( 1 + 2n_{B}(E_{\pi}) \Big),
    \label{A.4final}
\end{equation}
\begin{equation}
    -i\Pi_{\pi_{0}}^{(A.5)} =\frac{6i\ \lambda^{2}v^{2}}{m_{\sigma}^{2}} \, \int \frac{d^{3}q}{(2\pi)^{3}} \frac{1}{2E_{\sigma}} \Big( 1 + 2n_{B}(E_{\sigma}) \Big),
    \label{A.5final}
\end{equation}
and
\begin{align}
    -i\Pi_{\pi_{0}}^{(A.6)} =\frac{i}{4\pi^{2}}\frac{\lambda^{2} v^{2}}{m_{\sigma}^{2}} \vert eB \vert & \int_{0}^{\infty} \frac{d\tau}{\sinh(\vert eB \vert \tau)} \frac{e^{-\tau m_{\pi}^{2}}}{\tau} \nonumber \\
    &\times \left(1 + 2\sum_{n=1}^{\infty} e^{-\frac{n^{2}}{4 T^{2}\tau}} \right).
    \label{A.6final}
\end{align}

The seventh contribution to the neutral pion self-energy corresponds to the tadpole diagram labeled as $A.7$. Its initial expression is
\begin{align}
    -i\Pi_{\pi_{0}}^{(A.7)} &=-2(-i\lambda v) (-ig) \left( -\frac{i}{m_{\sigma}^{2}} \right) N_{c} i T \sum_{n} \sum_{f} \int \frac{d^{3}q}{(2\pi)^{3}} \nonumber \\
    &\times \text{Tr}\Bigg \{ \int_{0}^{\infty}\frac{ds}{\cos(|q_fB| s)} e^{-is\left( \tilde{\omega}_{n}^{2} + q_{3}^{2} + q_{\perp}^{2} \frac{\tan(|q_fB| s)}{|q_fB| s} + m_{f}^{2} \right)}  \nonumber \\
& \times  \Big [ \left(\cos(|q_fB| s) + \text{sgn}(q_{f}B) \gamma^{1} \gamma^{2} \sin(\vert q_{f} B \vert s)\right)\nonumber \\
&\times \left(m_{f} -\gamma^{0}\tilde{\omega}_{n} - \gamma^{3} q_{3}\right)  - \frac{\gamma^1q_1+\gamma^2q_2}{\cos(|q_fB| s)} \Big ] \Bigg \},
\end{align}
where $N_c$ is the number of colors, $\sum_f$ denotes the sum over quark flavors, $\tilde{\omega}_n=(2n+1)\pi T$ are the fermionic Matsubara frequencies, $q_f$ is the quark electric charge and $\text{sgn}(q_fB)$ is the sign function. To evaluate this contribution, we perform the trace over Dirac indices. Since the trace of an odd number of gamma matrices vanishes, only the term proportional to the identity in Dirac space, namely the one proportional to $m_f$, contributes. The expression then reduces to
\begin{align}
    -i\Pi_{\pi_{0}}^{(A.7)} &=\frac{2\lambda v g}{m_{\sigma}^{2}}  N_{c} \sum_{f}4m_{f} T \sum_{n}  \int_{0}^{\infty} ds \nonumber \\
    &\times \int \frac{d^{3}q}{(2\pi)^{3}}  \ e^{-is\left( \tilde{\omega}_{n}^{2} + q_{3}^{2} + q_{\perp}^{2} \frac{\tan(|q_fB| s)}{|q_fB| s} + m_{f}^{2} \right)}.
    \label{A:7afterTrace}
\end{align}

The momentum integrals are Gaussian and can be evaluated directly. The resulting expression is
\begin{align}
    -i\Pi_{\pi_{0}}^{(A.7)}&=-\frac{i\lambda v g}{\pi^{3/2}m_{\sigma}^{2}} |q_{f}B| N_{c} \sum_{f}m_{f} T \sum_{n} \nonumber \\
    &\times \int_{0}^{\infty} \frac{ds}{ \tan(|q_{f}B| s)} 
    \frac{1}{\sqrt{is}}  \ e^{-is\left( \tilde{\omega}_{n}^{2} + m_{f}^{2} \right)}.
\end{align}
We now perform the analytic continuation $s\rightarrow -i\tau$, obtaining
\begin{align}
    -i\Pi_{\pi_{0}}^{(A.7)}&= -\frac{i\lambda v g}{\pi^{3/2}m_{\sigma}^{2}} |q_{f}B| N_{c} \sum_{f}m_{f} T \sum_{n}  \nonumber \\
    &\times \int_{0}^{\infty} \frac{d\tau}{ \tanh(|q_{f}B| \tau)} \frac{1}{\sqrt{\tau}}  \ e^{-\tau\left( \tilde{\omega}_{n}^{2} + m_{f}^{2} \right)}.
    \label{A.7beforesumn}
\end{align}

We now evaluate the sum over fermionic Matsubara frequencies. To this end, we first note the relation
\begin{equation}
    \sum_{n} e^{-\tau \tilde{\omega}_{n}^{2}} =e^{-\tau \pi^{2} T^{2}} \sum_{n=-\infty}^{\infty} e^{-\tau(4n^{2}\pi^{2}T^{2}+4n\pi^{2}T^{2})},
    \label{explicitMatsubarafermions}
\end{equation}
which allows us to express the sum in terms of the Jacobi theta function. For the present case, we have
\begin{equation}
    \vartheta_{3} (2i\pi T^{2}\tau\vert 4i\pi T^{2} \tau)= \sum_{n=-\infty}^{\infty} e^{-\tau(4n^{2}\pi^{2}T^{2} + 4n\pi^{2}T^{2})}.
    \label{JacobiThetafermions}
\end{equation}
Using the modular property of the Jacobi theta function, Eq.~(\ref{JacobiThetafermions}) can be rewritten as
\begin{align}
    \vartheta_{3}(2i\pi T^{2}\tau \vert 4i\pi T^{2}\tau)&=\frac{e^{\pi^{2}T^{2}\tau}}{2T\sqrt{\pi\tau}}\nonumber \\
    &\times \left(1 + 2\sum_{n=1}^{\infty} (-1)^{n} e^{-\frac{n^{2}}{4 T^{2}\tau}} \right).
    \label{JacobiThetaFermionsfinal}
\end{align}
Hence, using Eqs.~(\ref{explicitMatsubarafermions}) and~(\ref{JacobiThetaFermionsfinal}) in Eq.~(\ref{A.7beforesumn}), the expression for this contribution to the neutral pion self-energy becomes
\begin{align}
    -i\Pi_{\pi_{0}}^{(A.7)} = &-\frac{i\lambda v g}{2\pi^{2}m_{\sigma}^{2}} |q_{f}B| N_{c} \sum_{f}m_{f} \int_{0}^{\infty} \frac{d\tau}{\tanh(|q_{f}B | \tau)}\nonumber \\
    &\times \frac{e^{-\tau m_{f}^{2}}}{\tau} \left(1 + 2\sum_{n=1}^{\infty} (-1)^{n} e^{-\frac{n^{2}}{4 T^{2}\tau}} \right).
    \label{A.7final}
\end{align}

In Eq.~(\ref{A.7final}), the first term inside the brackets corresponds to the purely magnetic contribution, whereas the second term represents the thermomagnetic contribution. The purely magnetic term contains ultraviolet divergences associated with the vacuum contribution. Taking the limit $|q_f B|\rightarrow 0$, we isolate the vacuum piece,
\begin{equation}
    -i\Pi_{\pi_{0},B=0}^{(A.7)} =-\frac{i\lambda v g}{2\pi^{2}m_{\sigma}^{2}} N_{c} \sum_{f}m_{f} \int_{0}^{\infty} d\tau \frac{e^{-\tau m_{f}^{2}}}{\tau^{2}}.
\end{equation}
In the full expression, we subtract this vacuum contribution and add back its renormalized counterpart, following the same procedure used in the previous contributions involving charged particles in the loop.

The next contribution, labeled $A.8$, involves two neutral fields in the loop, namely a neutral pion and a sigma meson. Therefore, this diagram contains only vacuum and finite-temperature contributions
\begin{align}
    -i\Pi_{\pi_{0}}^{(A.8)} &= 4(-i\lambda v)^{2} i T \sum_n \int \frac{d^{3}q}{(2\pi)^{3}}  \frac{i}{\left[\omega_n^{2}+\vec q^{\,2}+m_\sigma^{2}\right]}\nonumber \\
    &\times \frac{i} {\left[(\omega-\omega_n)^{2}+(\vec{k}-\vec q)^{2}+m_{\pi}^{2}\right]},
    \label{A.8initial}
\end{align}
where $\omega$ is the external Matsubara frequency of the neutral pion. Simplifying Eq.~(\ref{A.8initial}), we obtain
\begin{equation}
    -i\Pi_{\pi_{0}}^{(A.8)} = 4i\lambda^{2} v^{2} \int \frac{d^{3}q}{(2\pi)^{3}} T \sum_n \frac{1} {\left[(\omega-\omega_n)^{2}+E_{1}^{2}\right] \left[\omega_{n}^{2}+E_{2}^{2}\right]},
\end{equation}
where $E_1=\sqrt{(\vec{k}-\vec q)^{2}+m_{\pi}^{2}}$ and $E_2=\sqrt{\vec q^{\,2}+m_{\sigma}^{2}}$. After performing the sum over Matsubara frequencies, the expression becomes
\begin{widetext}
\begin{align}
    -i\Pi_{\pi_{0}}^{(A.8)} &=  \ i\lambda^{2} v^{2}  \int \frac{d^{3}q}{(2\pi)^{3}} \frac{1}{E_{1}E_{2}}\Bigg[\Big(1+n_{B}(E_{1})+n_{B}(E_{2})\Big) \left( \frac{1}{i\omega + E_{1} + E_{2}} - \frac{1}{i\omega - E_{1} - E_{2}}\right) \nonumber \\
&+ \Big(n_{B}(E_{1})-n_{B}(E_{2})\Big)  \left( \frac{1}{i\omega - E_{1}+E_{2}} - \frac{1}{i\omega + E_{1} - E_{2}} \right) \Bigg].
\label{A.8final}
\end{align}
From this expression, we identify two types of contributions. The first term inside the brackets contains a vacuum piece, corresponding to the term independent of the Bose–Einstein distribution functions, namely the constant term ``1''. The remaining terms represent the matter contributions and encode the thermal effects of the medium.

The last contribution to the neutral pion self-energy is the quark-antiquark loop. In this case, the magnetic-field dependence is encoded in the Schwinger proper-time representation of the quark propagators, while thermal effects are incorporated through the Matsubara formalism. The initial expression is

\begin{align}
    -i\Pi_{\pi_{0}}^{(A.9)} &= -i g^{2} N_{c} \sum_{f} T \sum_{n} \int \frac{d^{3}q}{(2\pi)^{3}} \text{Tr}\left\lbrace \left[ \int_{0}^{\infty}\frac{ds}{\cos(|q_fB| s)} e^{-is\left( (\omega - \tilde{\omega}_{n})^{2}+(k_{3}-q_{3})^{2} + (k-q)_{\perp}^{2} \frac{\tan(|q_fB| s)}{|q_fB| s} + m_{f}^{2} \right)} \right. \right. \nonumber \\
& \times \left[ \left(\cos(|q_fB| s) + \text{sgn}(q_{f}B) \gamma^{1} \gamma^{2} \sin(\vert q_{f} B \vert s)\right)\left(m_{f} - \gamma^{0}(\omega-\tilde{\omega}_{n})-\gamma^{3}(k_{3}-q_{3})\right)  \right. \nonumber \\
& \times \left. \left. - \frac{(\slashed{k}-\slashed{q})_{\perp}}{\cos(|q_fB| s)} \right] \right]\gamma^{5} \left[\int_{0}^{\infty} \frac{ds'}{\cos(|q_fB| s')} e^{-is'\left( \tilde{\omega}_{n}^{2} +q_{3}^{2} + q_{\perp}^{2} \frac{\tan(|q_fB| s')}{|q_fB| s'} + m_{f}^{2} \right)} \right. \nonumber \\
& \times \left.\left.\left[ \left(\cos(|q_fB| s') + \text{sgn}(-q_{f}B) \gamma^{1} \gamma^{2} \sin(\vert q_{f} B \vert s')\right)\left(m_{f} -\gamma^{0}\tilde{\omega}_{n}-\gamma^{3}q_{3}\right) - \frac{\slashed{q}_{\perp}}{\cos(|q_fB| s')} \right] \right] \gamma^{5} \right\rbrace,
\label{A.9initial}
\end{align}
where $s$ and $s'$ are the Schwinger proper times. We begin by computing the trace over Dirac indices 
\begin{align}
\text{Tr} & \left\lbrace  \left[ \left(\cos(|q_fB| s) + \text{sgn}(q_{f}B) \gamma^{1} \gamma^{2} \sin(\vert q_{f} B \vert s)\right)\left(m_{f} -\gamma^{0}(\omega-\tilde{\omega}_{n})-\gamma^{3}(k_{3}-q_{3})\right) - \frac{(\slashed{k}-\slashed{q})_{\perp}}{\cos(|q_fB| s)} \right] \gamma^{5} \right. \nonumber \\
& \times \left.\left[ \left(\cos(|q_fB| s') + \text{sgn}(-q_{f}B) \gamma^{1} \gamma^{2} \sin(\vert q_{f} B \vert s')\right)\left(m_{f} - \gamma^{0}\tilde{\omega}_{n}-\gamma^{3}q_{3}\right) - \frac{\slashed{q}_{\perp}}{\cos(|q_fB| s')} \right] \gamma^{5} \right\rbrace \nonumber \\
&=4\left[\left(\cos(|q_fB| s)\cos(|q_fB| s') + \sin(|q_fB| s)\sin(|q_fB| s')\right)\left(m_{f}^{2} + (\omega - \tilde{\omega}_{n})\tilde{\omega}_{n} + (k_{3}-q_{3})q_{3}\right)\right.\nonumber \\
& \ \ \ \ \left. + \frac{(k_{1}-q_{1})q_{1} + (k_{2}-q_{2})q_{2}}{\cos(|q_fB| s)\cos(|q_fB| s')} \right].
\label{A.9trace}
\end{align}
Substituting the result of the trace into Eq.~(\ref{A.9initial}) and performing the analytic continuations $s \rightarrow -i\tau$ and $s' \rightarrow -i \tau'$, we obtain
\begin{align}
    -i\Pi_{\pi_{0}}^{(A.9)} &= \ 4ig^{2} N_{c} \sum_{f} T \sum_{n} \int \frac{d^{3}q}{(2\pi)^{3}} \int_{0}^{\infty}\frac{d\tau}{\cosh(\vert q_{f}B \vert \tau)} e^{-\tau\left[ (\omega - \tilde{\omega}_{n})^{2}+(k_{3}-q_{3})^{2} + (k-q)_{\perp}^{2} \frac{\tanh(\vert q_{f}B \vert \tau)}{\vert q_{f}B \vert \tau} + m_{f}^{2} \right]} \nonumber \\
& \times \int_{0}^{\infty}\frac{d\tau'}{\cosh(\vert q_{f}B \vert \tau')} e^{-\tau'\left( \tilde{\omega}_{n}^{2} +q_{3}^{2} + q_{\perp}^{2} \frac{\tan(\vert q_{f}B \vert \tau')}{\vert q_{f}B \vert \tau'} + m_{f}^{2} \right)} \nonumber \\
& \times \left[\left(\cosh(\vert q_{f}B \vert \tau)\cosh(\vert q_{f}B \vert \tau') - \sinh(\vert q_{f}B \vert \tau)\sinh(\vert q_{f}B \vert \tau')\right)\left(m_{f}^{2} + (\omega - \tilde{\omega}_{n})\tilde{\omega}_{n} + (k_{3}-q_{3})q_{3}\right)\right.\nonumber \\
& \left. + \frac{(k_{1}-q_{1})q_{1} + (k_{2}-q_{2})q_{2}}{\cosh(\vert q_{f}B \vert \tau)\cosh(\vert q_{f}B \vert \tau')} \right].
\end{align}
The momentum integrals are Gaussian and can be evaluated analytically, and the expression becomes
\begin{align}
-i\Pi_{\pi_{0}}^{(A.9)} &= \frac{4ig^{2}}{(2\pi)^{3}} \pi^{3/2} N_{c} \sum_{f} T \int_{0}^{\infty}d\tau \int_{0}^{\infty}d\tau' \frac{\vert q_{f} B \vert}{\sinh(\vert q_{f}B \vert (\tau+\tau'))(\tau+\tau')^{1/2}} e^{-\frac{\tau \tau' k_{3}^{2}}{\tau + \tau'}} e^{-m_{f}^{2} (\tau + \tau')} e^{-\tau\omega^{2}}\nonumber \\
& \times e^{-\frac{k_{\perp}^{2}}{\vert q_{f}B \vert} \frac{\sinh(\vert q_{f}B \vert \tau)\sinh(\vert q_{f}B \vert \tau')}{\sinh(\vert q_{f}B \vert (\tau+\tau'))}} \left\lbrace \cosh(\vert q_{f}B \vert (\tau - \tau')) \left[ \sum_{n=-\infty}^{\infty} e^{-\left[ (\tau + \tau')\tilde{\omega}_{n}^{2} -2 \omega \tilde{\omega}_{n} \tau\right]} \right.\right. \nonumber \\
& \times \left. \left(m_{f}^{2} + \left(\tau + \tau'\right)^{-2}\left[(\tau +\tau')\left(k_{3}^{2}\tau -\frac{1}{2}  \right)- k_{3}^{2}\tau^{2} \right]\right) + \sum_{n=-\infty}^{\infty} (\omega - \tilde{\omega}_{n})\tilde{\omega}_{n} e^{-\left[ (\tau + \tau')\tilde{\omega}_{n}^{2} -2 \omega \tilde{\omega}_{n} \tau\right]}\right]\\
& +\frac{1}{\sinh^{2}(\vert q_{f}B \vert (\tau+\tau'))} \left[k_{\perp}^{2}\sinh(\vert q_{f}B \vert \tau)\sinh(\vert q_{f}B \vert \tau') - \vert q_{f} B\vert \sinh(\vert q_{f}B \vert (\tau+\tau'))\right] \nonumber \\
& \times \left. \sum_{n=-\infty}^{\infty} e^{-\left[ (\tau + \tau')\tilde{\omega}_{n}^{2} -2 \omega \tilde{\omega}_{n} \tau\right]} \right\rbrace.
\label{A.9aftermomentumint}
\end{align}
\end{widetext}

In order to express the sums over Matsubara frequencies in terms of Jacobi theta functions, we identify three different structures
\begin{align}
S_{0} &= \sum_{n=-\infty}^{\infty} e^{-\left[ (\tau + \tau')\tilde{\omega}_{n}^{2} -2 \omega \tilde{\omega}_{n} \tau\right]} ,\nonumber \\
S_{1} &= \sum_{n=-\infty}^{\infty} \tilde{\omega}_{n} \ e^{-\left[ (\tau + \tau')\tilde{\omega}_{n}^{2} -2 \omega \tilde{\omega}_{n} \tau\right]} ,\nonumber \\
S_{2} &= \sum_{n=-\infty}^{\infty} \tilde{\omega}_{n}^{2} \ e^{-\left[ (\tau + \tau')\tilde{\omega}_{n}^{2} -2 \omega \tilde{\omega}_{n} \tau\right]} .
\end{align}
We analyze these sums one by one. For the first one, $S_0$, we write
\begin{equation}
    S_{0} = \sum_{n=-\infty}^{\infty} e^{- 4\pi^{2}T^{2}(\tau + \tau')(n+\frac{1}{2})^{2} +2\tau\omega\pi T(2n+1) }.
\end{equation}
Using the Jacobi theta functions of the second and fourth kinds,
\begin{align}
    \vartheta_{2} (z\vert x) &= \sum_{n=-\infty}^{\infty} e^{i\pi(n+\frac{1}{2})^{2}x} e^{i(2n+1)z}, \nonumber \\
    &= (-ix)^{-1/2}e^{-\frac{iz^{2}}{\pi x}} \vartheta_{4} \left(-\frac{z}{x} \vert -\frac{1}{x}\right),
\end{align}
where
\begin{equation}
    \vartheta_{4}(z\vert x) = 1 + 2 \sum_{n=1}^{\infty} (-1)^{n} e^{i\pi x n^{2}} \cos(2nz),
\end{equation}
with $x=4i(\tau+\tau')\pi T^{2}$ and $z = -2i\tau \omega \pi T $. Then $S_0$ can be written as
\begin{align}
    S_{0} &=\frac{e^{\frac{\tau^{2}\omega^{2}}{\tau+\tau'}}}{2T\sqrt{\pi(\tau+\tau')}} \Big[1 \nonumber \\
    &+ 2\sum_{n=1}^{\infty} (-1)^{n} e^{-\frac{n^{2}}{4 T^{2}(\tau+\tau')}}\cos\left(\frac{n\tau\omega}{(\tau+\tau')T}\right) \Bigg].
    \label{S0final}
\end{align}
For $S_1$, we write
\begin{equation}
    S_{1} =\pi T \sum_{n=-\infty}^{\infty}(2n+1) e^{- 4\pi^{2}T^{2}(\tau + \tau')(n+\frac{1}{2})^{2} +2\tau\omega\pi T(2n+1)},
\end{equation}
and, in order to express it in terms of Jacobi theta functions, we write
\begin{equation}
S_{1} = -i\pi T (-ix)^{-1/2} \frac{\partial}{\partial z} \left[e^{-\frac{iz^{2}}{\pi x}} \vartheta_{4} \left(-\frac{z}{x} \vert -\frac{1}{x}\right)\right],
\end{equation}
where $z$ and $x$ are defined as in the case of $S_0$. After some algebra, we arrive at
\begin{align}
S_{1}&=\frac{e^{\frac{\tau^{2}\omega^{2}}{\tau+\tau'}}}{2T\sqrt{\pi(\tau+\tau')}}  \Bigg\{ \frac{\tau \omega}{\tau+\tau'} \nonumber \\
&+ 2\sum_{n=1}^{\infty} (-1)^{n} e^{-\frac{n^{2}}{4 T^{2}(\tau+\tau')}} \bigg[\frac{\tau \omega}{\tau+\tau'}\cos\left(\frac{n\tau\omega}{(\tau+\tau')T}\right) \nonumber \\
&- \frac{n}{2T(\tau+\tau')}\sin\left(\frac{n\tau\omega}{(\tau+\tau')T}\right) \bigg]\Bigg \}.
\label{S1final}
\end{align}
For $S_2$, the explicit expression is
\begin{equation}
    S_{2} = \pi^{2} T^{2} \sum_{n=-\infty}^{\infty} (2n+1)^{2} e^{- 4\pi^{2}T^{2}(\tau + \tau')(n+\frac{1}{2})^{2} +2\tau\omega\pi T(2n+1)},
\end{equation}
which can be written in terms of Jacobi theta functions as
\begin{equation}
S_{2} = -\pi^{2} T^{2} (-ix)^{-1/2} \frac{\partial^{2}}{\partial z^2{}} \left[e^{-\frac{iz^{2}}{\pi x}} \vartheta_{4} \left(-\frac{z}{x} \vert -\frac{1}{x}\right)\right],
\end{equation}
with $z$ and $x$ defined as before. After some algebra, we arrive at
\begin{align}
S_{2} &= \frac{1}{2T\sqrt{\pi(\tau+\tau')}} e^{\frac{\tau^{2}\omega^{2}}{\tau+\tau'}} \Bigg\{\frac{1}{2(\tau+\tau')}+\frac{\tau^{2} \omega^{2}}{(\tau+\tau')^{2}} \nonumber \\
&+ 2\sum_{n=1}^{\infty} (-1)^{n} e^{-\frac{n^{2}}{4 T^{2}(\tau+\tau')}}\bigg[ \left(\frac{1}{2(\tau+\tau')}+\frac{\tau^{2} \omega^{2}}{(\tau+\tau')^{2}} \right. \nonumber \\
&\left.- \frac{n^{2}}{4T^{2}(\tau+\tau')^{2}}\right)\cos\left(\frac{n\tau\omega}{(\tau+\tau')T}\right) \nonumber \\
&- \frac{n\tau\omega}{T(\tau+\tau')^2}\sin\left(\frac{n\tau\omega}{(\tau+\tau')T}\right) \bigg]\Bigg \}.
\label{S2final}
\end{align}

Once the final expressions for $S_0$, $S_1$, and $S_2$ have been obtained, and after performing the change of variables $\tau = u(1-v)$ and $\tau'=uv$, we substitute them into Eq.~(\ref{A.9aftermomentumint}) to obtain
\begin{widetext}
    \begin{align}
-i\Pi_{\pi_{0}}^{(A.9)} &= \frac{ig^{2}}{(2\pi)^{2}} N_{c} \sum_{f} \int_{0}^{\infty}du \int_{0}^{1}dv \frac{\vert q_{f} B \vert}{\sinh(\vert q_{f}B \vert u)} e^{-uv(1-v) k_{3}^{2}} e^{-m_{f}^{2} u} e^{(i\omega)^{2}uv(1-v)} \nonumber \\
& \times e^{-\frac{k_{\perp}^{2}}{\vert q_{f}B \vert} \frac{\sinh(\vert q_{f}B \vert u(1-v))\sinh(\vert q_{f}B \vert uv)}{\sinh(\vert q_{f}B \vert u)}} \left\lbrace \cosh(\vert q_{f}B \vert u(1-2v)) \left[ \left(1 + 2\sum_{n=1}^{\infty} (-1)^{n} e^{-\frac{n^{2}}{4 T^{2}u}}\right. \right.\right. \nonumber \\
& \times \left. \cosh\left(\frac{(1-v)(i\omega)n}{T}\right) \right)\left(m_{f}^{2} + \frac{1}{u} \left(uv(1-v)k_{3}^{2} - \frac{1}{2}\right)\right) \nonumber \\
& - i\omega \left[(1-v)(i\omega) +2\sum_{n=1}^{\infty} (-1)^{n} e^{-\frac{n^{2}}{4 T^{2}u}} \left((1-v)(i\omega)\cosh\left(\frac{(1-v)(i\omega)n}{T}\right) - \frac{n}{2Tu}\sinh\left(\frac{(1-v)(i\omega)n}{T}\right) \right)\right] \nonumber \\
& - \left(\frac{1}{2u} - (1-v)^{2}(i\omega)^{2} +2\sum_{n=1}^{\infty} (-1)^{n} e^{-\frac{n^{2}}{4 T^{2}u}} \left[ \left(\frac{1}{2u}-(1-v)^{2}(i\omega)^{2} - \frac{n^{2}}{4T^{2}u^{2}}\right)\cosh\left(\frac{(1-v)(i\omega)n}{T}\right)  \right. \right.\nonumber \\
& \left. \left. + \frac{n(1-v)(i\omega)}{T u}\sinh\left(\frac{(1-v)(i\omega)n}{T}\right) \right]\right)\Bigg] \nonumber \\
& + \frac{1}{\sinh^{2}(\vert q_{f}B \vert u)} \left[k_{\perp}^{2}\sinh(\vert q_{f}B \vert u(1-v))\sinh(\vert q_{f}B \vert uv) - \vert q_{f} B\vert \sinh(\vert q_{f}B \vert u)\right] \nonumber \\
& \times \left. \left[1 + 2\sum_{n=1}^{\infty} (-1)^{n} e^{-\frac{n^{2}}{4 T^{2}u}}\cosh\left(\frac{(1-v)(i\omega)n}{T}\right) \right] \right\rbrace.
\label{A.9final}
\end{align}

In Eq.~(\ref{A.9final}) we identify the purely magnetic terms as those that do not contain exponential factors depending on $T$, whereas the remaining ones correspond to thermomagnetic contributions. As expected, the purely magnetic contribution contains the vacuum piece, which is ultraviolet divergent. Taking the limit $|q_fB|\rightarrow 0$, we obtain
\begin{align}
-i\Pi_{\pi_{0},B=0}^{(A.9)} &= \frac{ig^{2}}{(2\pi)^{2}} N_{c} \sum_{f} \int_{0}^{\infty}du \int_{0}^{1}dv \ e^{-\left(k_{\perp}^{2}+k_{3}^{2}-(i\omega)^{2}\right)uv(1-v)}e^{-m_{f}^{2}u}\frac{1}{u}\Bigg(m_{f}^{2} + \frac{1}{u}\left(uv(1-v)k_{3}^{2}- \frac{1}{2}\right) \nonumber \\
&- (i\omega)^{2} (1-v)  - \frac{1}{2u} + (1-v)^{2}(i\omega)^{2} + \frac{1}{u}\left(uv(1-v)k_{\perp}^{2}-1\right)\Bigg).
\label{A.9vacuum}
\end{align}
\end{widetext}

In the full expression, we subtract this vacuum contribution and add back its renormalized counterpart, following the same procedure used in all the previous contributions involving charged particles in the loop. 

\subsection{Sigma meson one-loop self-energy\label{sec3.2}}

\begin{figure}[h]
    \centering
    \includegraphics[scale=0.42]{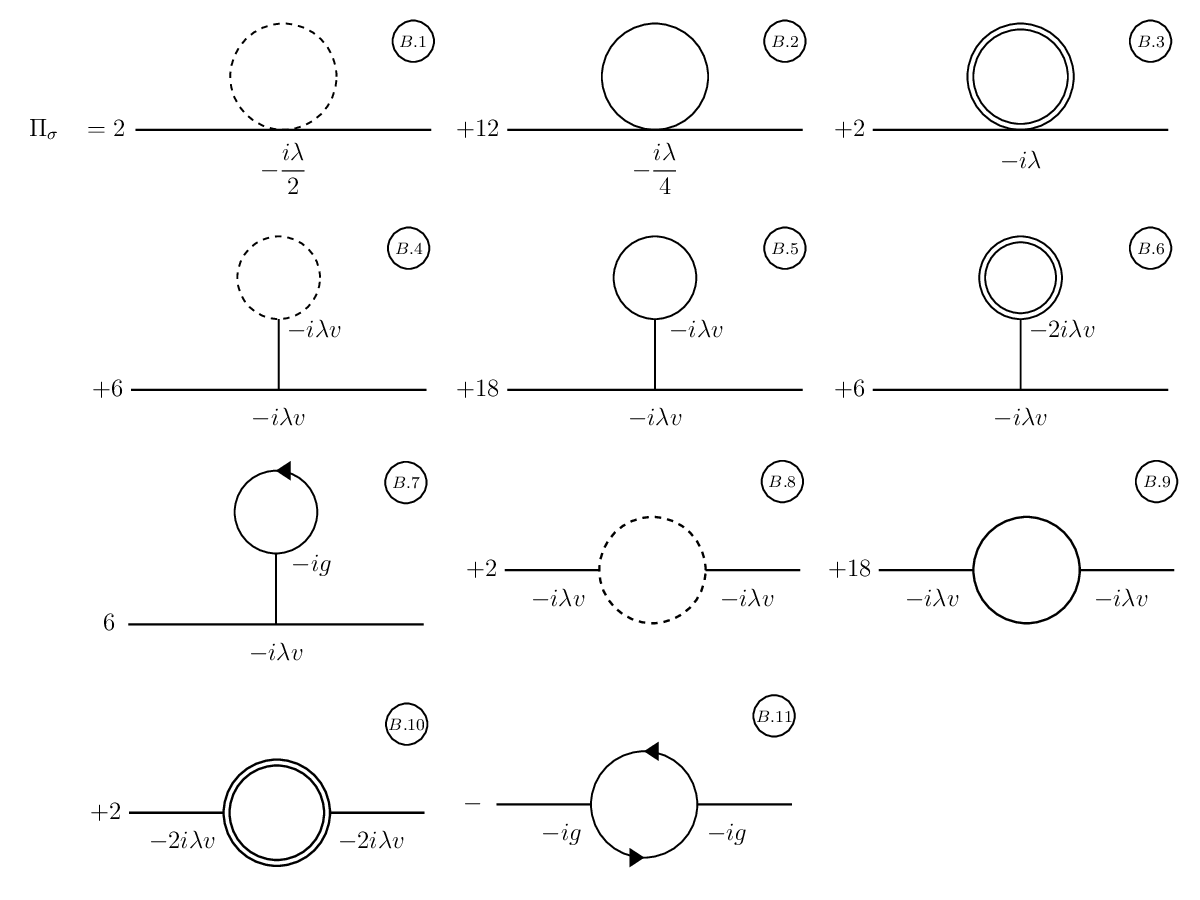}
    \caption{One-loop self-energy diagrams for the sigma meson field. Solid lines represent the sigma meson, dashed lines correspond to neutral pions, double solid lines denote charged pions, and solid lines with arrows represent quark fields. Each vertex explicitly displays the corresponding coupling coefficient, and the factors multiplying each diagram account for the associated combinatorial factors.}
    \label{fig4}
\end{figure}

We now turn to the one-loop self-energy of the sigma meson. The corresponding diagrams are shown in Fig.~\ref{fig4}. There are eleven contributions, labeled as $B.j$, with $j=1,2,\ldots,11$. The full sigma meson self-energy is given by
\begin{align}
    -i\Pi_{\sigma} =&-i\Pi_{\sigma}^{(B.1)}  -i\Pi_{\sigma}^{(B.2)}-i\Pi_{\sigma}^{(B.3)} \nonumber \\
    &-i\Pi_{\sigma}^{(B.4)} -i\Pi_{\sigma}^{(B.5)} -i\Pi_{\sigma}^{(B.6)} \\ 
    &-i\Pi_{\sigma}^{(B.7)} -i\Pi_{\sigma}^{(B.8)} -i\Pi_{\sigma}^{(B.9)} \nonumber \\
    &-i\Pi_{\sigma}^{(B.10)} -i\Pi_{\sigma}^{(B.11)},
    \label{structuresigmaselfenergy}
\end{align}

We evaluate them in turn. For this self-energy, however, we make use of the results previously obtained for the neutral pion self-energy. In fact, most of the terms labeled $B.j$ have a direct correspondence with those labeled $A.j$. Therefore, the terms from $B.1$ to $B.7$ can be obtained directly from $A.1$ to $A.7$, respectively, by replacing the corresponding vertex and symmetry factors. The remaining terms, which do not have a direct counterpart in the neutral pion case, are computed separately.

We present the final expressions for $B.1$ to $B.7$. These results are obtained within the Matsubara formalism; when charged particles circulate in the loop, the corresponding propagators are written in the Schwinger proper-time representation. The first two contributions are given by
\begin{equation}
-i\Pi_{\sigma}^{(B.1)} = -i\lambda \, \int \frac{d^{3}q}{(2\pi)^{3}} \frac{1}{2E_{\pi}} \left( 1 + 2n_{B}(E_{\pi}) \right),
\label{B.1final}
\end{equation}
and
\begin{equation}
-i\Pi_{\sigma}^{(B.2)} = -3i\lambda \, \int \frac{d^{3}q}{(2\pi)^{3}} \frac{1}{2E_{\sigma}} \left( 1 + 2n_{B}(E_{\sigma}) \right),
\label{B.2final}
\end{equation}
where $E_\pi=\sqrt{\vec{q}^{\,2}+m_\pi^2}$ and $E_\sigma=\sqrt{\vec{q}^{\,2}+m_\sigma^2}$. For the loop with charged pions, one finds
\begin{align}
    -i\Pi_{\sigma}^{(B.3)} = -\frac{i\lambda}{8\pi^{2}} \ \vert eB \vert  & \int_{0}^{\infty} \frac{d\tau}{\sinh\left(\vert eB \vert \tau\right)} \frac{e^{-\tau m_{\pi}^{2}}} {\tau} \nonumber \\
    &\times \left(1 + 2\sum_{n=1}^{\infty} e^{-\frac{n^{2}}{4 T^{2}\tau}} \right),
\end{align}
whose vacuum contribution is given by
\begin{equation}
    -i\Pi_{\sigma,B=0}^{(B.3)} =-\frac{i\lambda}{8\pi^{2}} \int_{0}^{\infty} \frac{d\tau}{\tau^{2}} e^{-\tau m_{\pi}^{2}}.
\end{equation}
The following two terms are
\begin{equation}
-i\Pi_{\sigma}^{(B.4)} = \frac{6i\ \lambda^{2}v^{2}}{m_{\sigma}^{2}} \, \int \frac{d^{3}q}{(2\pi)^{3}} \frac{1}{2E_{\pi}} \left( 1 + 2n_{B}(E_{\pi}) \right),
\label{B.4final}
\end{equation}
and
\begin{equation}
-i\Pi_{\sigma}^{(B.5)} = \frac{18i \lambda^{2}v^{2}}{m_{\sigma}^{2}} \, \int \frac{d^{3}q}{(2\pi)^{3}} \frac{1}{2E_{\sigma}} \left( 1 + 2n_{B}(E_{\sigma}) \right).
\label{B.5final}
\end{equation}
For the charged-pion tadpole with an additional zero-momentum sigma propagator, we obtain
\begin{align}
    -i\Pi_{\sigma}^{(B.6)} =\frac{3i\lambda^{2}v^{2}}{4\pi^{2}m_{\sigma}^{2}} \ \vert eB \vert & \int_{0}^{\infty} \frac{d\tau}{\sinh\left(\vert eB \vert \tau\right)} \frac{e^{-\tau m_{\pi}^{2}}} {\tau} \nonumber \\
    &\times \left(1 + 2\sum_{n=1}^{\infty} e^{-\frac{n^{2}}{4 T^{2}\tau}} \right),
    \label{B.6final}
\end{align}
where $\tau$ denotes the Schwinger proper time. The fermionic tadpole contribution is
\begin{align} 
-i\Pi_{\sigma}^{(B.7)} = & - \frac{3 i \lambda v g}{2\pi^{2}m_{\sigma}^{2}} \vert q_{f}B \vert N_{c} \sum_{f}m_{f} \int_{0}^{\infty} \frac{d\tau}{\tanh(\vert q_{f}B \vert \tau)} \nonumber \\
\times &\frac{e^{-\tau m_{f}^{2}}}{\tau} \left(1 + 2\sum_{n=1}^{\infty} (-1)^{n} e^{-\frac{n^{2}}{4 T^{2}\tau}} \right).
\label{B.7final}
\end{align}
The vacuum contributions associated with Eqs.~(\ref{B.6final}) and~(\ref{B.7final}) are
\begin{equation}
    -i\Pi_{\sigma,B=0}^{(B.6)}=\frac{3i\lambda^{2}v^{2}}{4\pi^{2}m_{\sigma}^{2}} \int_{0}^{\infty} \frac{d\tau}{\tau^{2}} e^{-\tau m_{\pi}^{2}},
\end{equation}
and
\begin{equation}
   -i\Pi_{\sigma,B=0}^{(B.7)} =-\frac{3 i \lambda v g}{2\pi^{2}m_{\sigma}^{2}} N_{c} \sum_{f}m_{f} \int_{0}^{\infty} d\tau \frac{e^{-\tau m_{f}^{2}}}{\tau^{2}},
\end{equation}
respectively.

The contributions $B.8$ and $B.9$ do not have a direct counterpart among the $A.j$ terms. Both diagrams share the same structure as $A.8$. Therefore, their expressions can be obtained by appropriately replacing the fields and the corresponding vertex and symmetry factors. Thus, we can write
\begin{align}
-i\Pi_{\pi_{0}}^{(B.8)} &= \frac{i\lambda^{2} v^{2}}{2} \int \frac{d^{3}q}{(2\pi)^{3}} \frac{1}{E_{3}E_{4}} \nonumber \\
&\times \left[(1+n_{B}(E_{3})+n_{B}(E_{4})) \right. \nonumber \\
&\left. \times \left( \frac{1}{i\omega + E_{3} + E_{4}} - \frac{1}{i\omega - E_{3} - E_{4}}\right)\right. \nonumber \\
& + \left. (n_{B}(E_{3})-n_{B}(E_{4})) \right.\nonumber \\
&\left. \times \left( \frac{1}{i\omega - E_{3}+E_{4}} - \frac{1}{i\omega + E_{3} - E_{4}} \right) \right],
\label{B.8final}
\end{align}
where $E_{3}=\sqrt{(\vec{k}-\vec q)^{2}+m_{\pi}^{2}}$ and $E_{4}=\sqrt{\vec q^{\,2}+m_{\pi}^{2}}$, and
\begin{align}
-i\Pi_{\sigma}^{(B.9)} &= \frac{9i\lambda^{2} v^{2}}{2} \int \frac{d^{3}q}{(2\pi)^{3}} \frac{1}{E_{5}E_{6}}\nonumber \\
&\times \left[(1+n_{B}(E_{5})+n_{B}(E_{6})) \right. \nonumber \\
&\left. \times \left( \frac{1}{i\omega + E_{5} + E_{6}} - \frac{1}{i\omega - E_{5} - E_{6}}\right)\right. \nonumber \\
& + \left. (n_{B}(E_{5})-n_{B}(E_{6})) \right. \nonumber \\
&\left. \times\left( \frac{1}{i\omega - E_{5}+E_{6}} - \frac{1}{i\omega + E_{5} - E_{6}} \right) \right],
\label{B.9final}
\end{align}
where $E_{5}=\sqrt{(\vec{k}-\vec q)^{2}+m_{\sigma}^{2}}$ and $E_{6}=\sqrt{\vec q^{2}+m_{\sigma}^{2}}$. From Eqs.~(\ref{B.1final}), (\ref{B.2final}), (\ref{B.4final}), (\ref{B.5final}), (\ref{B.8final}), and~(\ref{B.9final}), we identify two types of contributions. The terms independent of the Bose-Einstein distribution functions correspond to the vacuum contribution, while those containing the distribution functions encode the thermal (matter) effects of the medium.

We next consider the contribution involving a charged-pion pair, corresponding to diagram $B.10$. The initial expression is
\begin{align}
-i\Pi_{\sigma}^{(B.10)} &= 2 \left( -2i\lambda v \right)^{2} \int \frac{d^{4}q}{(2\pi)^{4}} \int_{0}^{\infty} \frac{ds}{\cos(\vert eB \vert s)} \nonumber \\
&\times e^{is\left((k-q)_{\parallel}^{2} - (k-q)_{\perp}^{2} \frac{\tan(\vert eB \vert s)}{\vert eB \vert s} - m_{\pi}^{2} + i \epsilon\right)} \nonumber \\
& \times \int_{0}^{\infty}\frac{ds'}{\cos(\vert eB \vert s')} e^{is' \left(q_{\parallel}^{2} - q_{\perp}^{2} \frac{\tan(\vert eB \vert s')}{\vert eB \vert s'} - m_{\pi}^{2} + i \epsilon\right)},
\end{align}
where $s$ and $s'$ are Schwinger proper times. 

We next perform the momentum integrations. To this end, we first implement the analytic continuations $s\rightarrow -i\tau$ and $s'\rightarrow -i\tau'$, and use the Gaussian integrals
\begin{align}
    \int d^{2}q_{\perp} \ e^{-(k-q)_{\perp}^{2}\alpha} e^{-q_{\perp}^{2}\alpha'} &= \left(\frac{\pi}{\alpha + \alpha'}\right) e^{-\frac{\alpha \alpha' k_{\perp}^{2}}{\alpha + \alpha'}}, \nonumber \\
    \int dq_{3} \ e^{-\tau (k_{3}-q_{3})^{2}} e^{-\tau' q_{3}^{2}} &= \sqrt{\frac{\pi}{\tau + \tau'}}e^{-\frac{\tau \tau' k_{3}^{2}}{\tau + \tau'}},
\end{align}
with 
\begin{equation*}
\alpha = \frac{\tanh(\vert eB \vert \tau)}{\vert eB \vert} , \quad \alpha' = \frac{\tanh(\vert eB \vert \tau')}{\vert eB \vert}.
\end{equation*}
After these integrations, one obtains
\begin{align}
-i\Pi_{\sigma}^{(B.10)} &= \frac{i\lambda^{2} v^{2}}{\pi^{3/2}} T \int_{0}^{\infty} d\tau \int_{0}^{\infty} d\tau' \frac{1}{\sqrt{\tau + \tau'}}\nonumber \\
&\times \frac{\vert eB \vert}{\sinh(\vert eB \vert (\tau + \tau'))} e^{-\frac{\tau \tau' k_{3}^{2}}{\tau + \tau'}} e^{-\tau\omega^{2}} \nonumber \\
&\times e^{-\frac{k_{\perp}^{2}}{\vert eB \vert} \frac{\sinh(\vert eB \vert \tau)\sinh(\vert eB \vert \tau')}{\sinh(\vert eB \vert (\tau+\tau'))}} e^{-m_{\pi}^{2}(\tau+\tau')}  \nonumber \\
&\times \sum_{n} e^{-\left[(\tau+\tau')\omega_{n}^{2} - 2\omega \omega_{n} \tau\right]}.
\label{B.10aftermomentumintegration}
\end{align}

We now evaluate the sum over bosonic Matsubara frequencies, which can be written as
\begin{align}
    e^{-\omega^2\tau}\sum_{n} e^{-\left[(\tau+\tau')\omega_{n}^{2} - 2\omega \omega_{n} \tau\right]} &= \frac{e^{-\frac{\omega^{2}\tau\tau'}{\tau+\tau'}}}{2T\sqrt{\pi(\tau+\tau')}} \left[  1 \right. \nonumber \\
    &\left. + 2\sum_{n=1}^{\infty} e^{-\frac{n^{2}}{4 (\tau + \tau')T^{2}}}  \right. \nonumber \\
    &\left. \times \cos\left(\frac{n\omega \tau}{(\tau + \tau')T}\right) \right].
    \label{sumB.10}
\end{align}
Substituting Eq.~(\ref{sumB.10}) into Eq.~(\ref{B.10aftermomentumintegration}), and performing the change of variables $\tau = u (1-v)$ and $\tau'=uv$, we obtain

\begin{align}
    -i\Pi_{\sigma}^{(B.10)} &= \frac{i\lambda^{2} v^{2}}{2\pi^{2}} \int_{0}^{\infty} du \int_{0}^{1} dv \frac{\vert eB \vert}{\sinh(\vert eB \vert u)} \nonumber \\
    &\times e^{-uv(1-v) k_{3}^{2}}e^{-u m_{\pi}^{2}} e^{uv(1-v)(i\omega)^{2}}  \nonumber \\
    & \times e^{-\frac{k_{\perp}^{2}}{\vert eB \vert} \frac{\sinh(\vert eB \vert u(1-v))\sinh(\vert eB \vert uv)}{\sinh(\vert eB \vert u)}}\nonumber \\
    &\times \bigg[1 + 2\sum_{n=1}^{\infty} e^{-\frac{n^{2}}{4 T^{2}u}} \cosh\left(\frac{(1-v)(i\omega)n}{T}\right) \bigg].
    \label{B.10final}
\end{align}
The corresponding vacuum contribution is obtained in the limit $|eB|\rightarrow 0$, yielding
\begin{align}
-i\Pi_{\sigma,B=0}^{(B.10)} =\frac{i\lambda^{2} v^{2}}{2\pi^{2}} &\int_{0}^{\infty} du \int_{0}^{1} dv  \frac{e^{-u m_{\pi}^{2}}}{u}\nonumber \\
&\times e^{-\left[k_{3}^{2} + k_{\perp}^{2} - (i\omega)^2\right]uv(1-v)}.
\label{B.10vacuum}
\end{align}

The final contribution to the sigma meson self-energy corresponds to diagram $B.11$, where the loop is formed by a quark-antiquark pair. This contribution therefore incorporates both thermal and magnetic-field effects. The starting expression is 
\begin{widetext}
\begin{align} 
-i\Pi_{\sigma}^{(B.11)} = & \ - (-ig)^{2} N_{c} \sum_{f} \int \frac{d^{4}q}{(2\pi)^{4}} \text{Tr}\Bigg\{ \int_{0}^{\infty} \frac{ds}{\cos(\vert q_{f}B \vert s)} e^{is\big( (k-q)_{\parallel}^{2} - (k-q)_{\perp}^{2} \frac{\tan(\vert q_{f}B \vert s)}{\vert q_{f}B \vert s} - m_{f}^{2} + i \epsilon \big)} \nonumber \\
\times & \bigg[ \left(\cos(\vert q_{f}B \vert s) + \text{sgn}(q_{f}B) \gamma^{1} \gamma^{2} \sin(\vert q_{f} B \vert s)\right)\left(m_{f} + (\slashed{k}-\slashed{q})_{\parallel}\right) - \frac{(\slashed{k}-\slashed{q})_{\perp}}{\cos(\vert q_{f}B \vert s)} \bigg] \nonumber \\
\times & \int_{0}^{\infty} \frac{ds'}{\cos(\vert q_{f}B \vert s')} e^{is'\big( q_{\parallel}^{2} - q_{\perp}^{2} \frac{\tan(\vert q_{f}B \vert s')}{\vert q_{f}B \vert s'} - m_{f}^{2} + i \epsilon \big)} \nonumber \\
\times & \bigg[ \left(\cos(\vert q_{f}B \vert s') + \text{sgn}(-q_{f}B) \gamma^{1} \gamma^{2} \sin(\vert q_{f} B \vert s')\right)\left(m_{f} + \slashed{q}_{\parallel}\right) - \frac{\slashed{q}_{\perp}}{\cos(\vert q_{f}B \vert s')} \bigg] \Bigg\}.
\label{B.11initial}
\end{align}
We first compute the trace over Dirac indices, obtaining
\begin{align}
\text{Tr} &\left \{ \bigg[ \left(\cos(\vert q_{f}B \vert s) + \text{sgn}(q_{f}B) \gamma^{1} \gamma^{2} \sin(\vert q_{f} B \vert s)\right)\left(m_{f} + (\slashed{k}-\slashed{q})_{\parallel}\right) - \frac{(\slashed{k}-\slashed{q})_{\perp}}{\cos(\vert q_{f}B \vert s)} \bigg] \right.\nonumber \\
&\left. \times \bigg[ \left(\cos(\vert q_{f}B \vert s') + \text{sgn}(-q_{f}B) \gamma^{1} \gamma^{2} \sin(\vert q_{f} B \vert s')\right)\left(m_{f} + \slashed{q}_{\parallel}\right) - \frac{\slashed{q}_{\perp}}{\cos(\vert q_{f}B \vert s')} \bigg]  \right \} \nonumber \\
&= 4 \bigg[\big(\cos(\vert q_{f}B \vert s)\cos(\vert q_{f}B \vert s') + \sin(\vert q_{f}B \vert s)\sin(\vert q_{f}B \vert s')\big)\big(m_{f}^{2} - (\omega - \tilde{\omega}_{n})\tilde{\omega}_{n} - (k_{3}-q_{3})q_{3}\big) \nonumber \\
&\ \ \ \ - \frac{(k_{1}-q_{1})q_{1} + (k_{2}-q_{2})q_{2}}{\cos(\vert q_{f}B \vert s)\cos(\vert q_{f}B \vert s')} \bigg].
\label{B.11trace}
\end{align}
Substituting Eq.~(\ref{B.11trace}) into Eq.~(\ref{B.11initial}), we get
\begin{align} 
-i\Pi_{\sigma}^{(B.11)} &= 4ig^{2} N_{c} \sum_{f} T \sum_{n} \int \frac{d^{3}q}{(2\pi)^{3}} \int_{0}^{\infty}\frac{ds}{\cos(\vert q_{f}B \vert s)} e^{-is\big( (\omega - \tilde{\omega}_{n})^{2}+(k_{3}-q_{3})^{2} + (k-q)_{\perp}^{2} \frac{\tan(\vert q_{f}B \vert s)}{\vert q_{f}B \vert s} + m_{f}^{2} \big)} \nonumber \\
\times & \int_{0}^{\infty}\frac{ds'}{\cos(\vert q_{f}B \vert s')} e^{-is'\big( \tilde{\omega}_{n}^{2} +q_{3}^{2} + q_{\perp}^{2} \frac{\tan(\vert q_{f}B \vert s')}{\vert q_{f}B \vert s'} + m_{f}^{2} \big)} \bigg[\big(\cos(\vert q_{f}B \vert s)\cos(\vert q_{f}B \vert s') + \sin(\vert q_{f}B \vert s)\sin(\vert q_{f}B \vert s')\big) \nonumber \\
& \times \big(m_{f}^{2} - (\omega - \tilde{\omega}_{n})\tilde{\omega}_{n} - (k_{3}-q_{3})q_{3}\big) - \frac{(k_{1}-q_{1})q_{1} + (k_{2}-q_{2})q_{2}}{\cos(\vert q_{f}B \vert s)\cos(\vert q_{f}B \vert s')} \bigg].
\label{B.11aftertrace}
\end{align}
After implementing the analytic continuations $s \rightarrow -i\tau$ and $s' \rightarrow -i\tau'$, and perform the momentum integrations. For brevity, we omit the intermediate steps, as they closely follow those used in the evaluation of the fermionic contribution to the neutral pion self-energy. The resulting expression, after the sum over the Matsubara modes and the change of variables $\tau=u(1-v)$ and $\tau'=uv$, is
\begin{align}
-i\Pi_{\sigma}^{(B.11)} &=- \frac{ig^{2}}{(2\pi)^{2}} N_{c} \sum_{f} \int_{0}^{\infty}du \int_{0}^{1}dv \frac{\vert q_{f} B \vert}{\sinh(\vert q_{f}B \vert u)} e^{-uv(1-v) k_{3}^{2}} e^{-um_{f}^{2}} e^{(i\omega)^{2}uv(1-v)}\nonumber \\
&\times e^{-\frac{k_{\perp}^{2}}{\vert q_{f}B \vert} \frac{\sinh(\vert q_{f}B \vert u(1-v))\sinh(\vert q_{f}B \vert uv)}{\sinh(\vert q_{f}B \vert u)}} \Bigg\{ \cosh(\vert q_{f}B \vert u(1-2v)) \Bigg[ \bigg(1 + 2\sum_{n=1}^{\infty} (-1)^{n} e^{-\frac{n^{2}}{4 T^{2}u}} \nonumber \\
&\times \cosh\Big(\frac{(1-v)n(i\omega)}{T}\Big) \bigg) \bigg(m_{f}^{2} - \frac{1}{u} \left(uv(1-v)k_{3}^{2} - \frac{1}{2}\right)\bigg) + (i \omega) \bigg[(1-v)(i\omega) \nonumber \\
&+ 2\sum_{n=1}^{\infty} (-1)^{n} e^{-\frac{n^{2}}{4 T^{2}u}} \bigg((1-v)(i\omega)\cosh\Big(\frac{(1-v)n(i\omega)}{T}\Big) - \frac{n}{2Tu}\sinh\Big(\frac{(1-v)n(i\omega)}{T}\Big) \bigg)\bigg] \nonumber \\
&+ \frac{1}{2u} - (1-v)^{2}(i\omega)^{2} +2\sum_{n=1}^{\infty} (-1)^{n} e^{-\frac{n^{2}}{4 T^{2}u}} \bigg[ \Big(\frac{1}{2u}-(1-v)^{2}(i\omega)^{2} - \frac{n^{2}}{4T^{2}u^{2}}\Big)\cosh\Big(\frac{(1-v)n(i\omega)}{T}\Big) \nonumber \\
&+ \frac{(1-v)n(i\omega)}{T u}\sinh\Big(\frac{(1-v)n(i\omega)}{T}\Big) \bigg]\Bigg] - \frac{k_{\perp}^{2}\sinh(\vert q_{f}B \vert u(1-v))\sinh(\vert q_{f}B \vert uv) - \vert q_{f} B\vert \sinh(\vert q_{f}B \vert u)}{\sinh^{2}(\vert q_{f}B \vert u)}  \nonumber \\
&\times \bigg[1 + 2\sum_{n=1}^{\infty} (-1)^{n} e^{-\frac{n^{2}}{4 T^{2}u}}\cosh\Big(\frac{(1-v)n(i\omega)}{T}\Big) \bigg] \Bigg\}.
\label{B.11final}
\end{align}
The corresponding vacuum contribution, obtained in the limit where both the temperature and the magnetic field vanish, is given by
\begin{align}
    -i\Pi_{\sigma,B=0}^{(B.11)}=&- \frac{ig^{2}}{(2\pi)^{2}} N_{c} \sum_{f} \int_{0}^{\infty}du \int_{0}^{1}dv e^{-\big(k_{\perp}^{2}+k_{3}^{2}-(i\omega)^{2}\big)uv(1-v)}e^{-m_{f}^{2}u}\frac{1}{u}\bigg(m_{f}^{2}- \frac{1}{u} \Big(uv(1-v)k_{3}^{2} - \frac{1}{2}\Big) \nonumber \\
    &+(i\omega)^{2} (1-v)+\frac{1}{2u}- (1-v)^{2}(i\omega)^{2} - \frac{1}{u}\left(uv(1-v)k_{\perp}^{2}-1\right)\bigg).
    \label{B.11vacuum}
\end{align}
\end{widetext}
This completes the one-loop calculation of the sigma meson self-energy.

\subsection{Charged pion one-loop self-energy\label{sec3.3}}

\begin{figure}[h]
    \centering
    \includegraphics[scale=0.42]{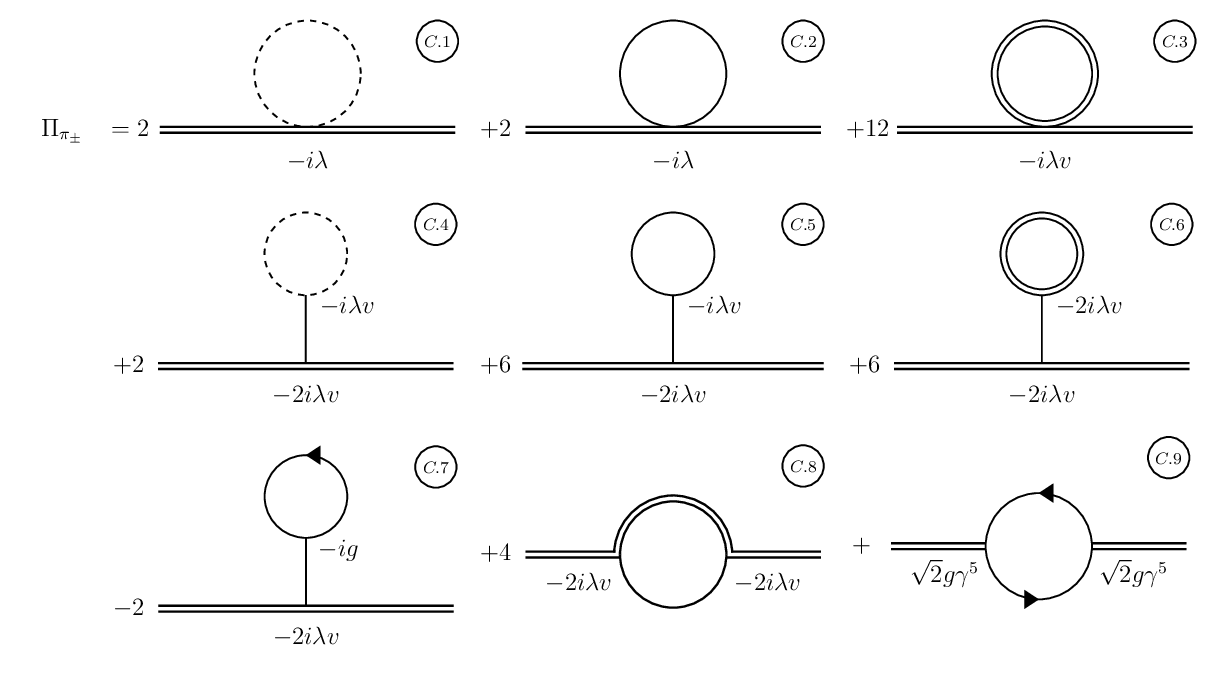}
    \caption{One-loop self-energy diagrams for the charged pion fields. Solid lines represent the sigma meson, dashed lines correspond to neutral pions, double solid lines denote charged pions, and solid lines with arrows represent quark fields. Each vertex explicitly displays the corresponding coupling coefficient. The coefficients multiplying each diagram are the corresponding combinatorial factors.}
    \label{fig5}
\end{figure}

We now turn to the one-loop self-energy of the charged pions. The corresponding Feynman diagrams are shown in Fig.~\ref{fig5} and are labeled as $C.j$, with $j=1,\ldots,9$. The full one-loop self-energy can be written as 
\begin{align}
-i\Pi_{\pi_{\pm}} = & -i\Pi_{\pi_{\pm}}^{(C.1)}  -i\Pi_{\pi_{\pm}}^{(C.2)} -i\Pi_{\pi_{\pm}}^{(C.3)} \nonumber \\
&-i\Pi_{\pi_{\pm}}^{(C.4)} -i\Pi_{\pi_{\pm}}^{(C.5)} -i\Pi_{\pi_{\pm}}^{(C.6)}\nonumber \\
&-i\Pi_{\pi_{\pm}}^{(C.7)} -i\Pi_{\pi_{\pm}}^{(C.8)} -i\Pi_{\pi_{\pm}}^{(C.9)}.
\end{align}

We evaluate them in turn. As in the previous cases, we note that the first seven contributions are closely related to those of the neutral pion self-energy. Therefore, the corresponding expressions can be obtained by appropriately replacing the symmetry and coupling factors according to the Feynman rules shown in Figs.~\ref{fig1} and~\ref{fig2}. The resulting expressions are
\begin{equation}
-i\Pi_{\pi_{\pm}}^{(C.1)} = -2i\lambda \int \frac{d^{3}q}{(2\pi)^{3}} \frac{1}{2E_{\pi}} \big( 1 + 2n_{B}(E_{\pi}) \big),
\label{C.1final}
\end{equation}
\begin{equation}
-i\Pi_{\pi_{\pm}}^{(C.2)} = -2i\lambda \int \frac{d^{3}q}{(2\pi)^{3}} \frac{1}{2E_{\sigma}} \big( 1 + 2n_{B}(E_{\sigma}) \big),
\label{C.2final}
\end{equation}
\begin{align}
-i\Pi_{\pi_\pm}^{(C.3)}&= -\frac{3i\lambda}{4\pi^{2}} \ \vert eB \vert  \int_{0}^{\infty} \frac{d\tau}{\sinh\left(\vert eB \vert \tau\right)} \frac{e^{-\tau m_{\pi}^{2}}} {\tau} \nonumber \\
&\times \left(1 + 2\sum_{n=1}^{\infty} e^{-\frac{n^{2}}{4 T^{2}\tau}} \right),
\label{C.3final}
\end{align}
where the corresponding vacuum contribution is
\begin{equation}
    -i\Pi_{\pi_\pm,B=0}^{(C.3)}=-\frac{3i\lambda}{4\pi^{2}} \int_{0}^{\infty} \frac{d\tau}{\tau^{2}} e^{-\tau m_{\pi}^{2}},
\end{equation}
\begin{equation}
-i\Pi_{\pi_{\pm}}^{(C.4)} = \frac{4i \lambda^{2}v^{2}}{m_{\sigma}^{2}} \int \frac{d^{3}q}{(2\pi)^{3}} \frac{1}{2E_{\pi}} \big( 1 + 2n_{B}(E_{\pi}) \big),
\end{equation}
\begin{equation}
-i\Pi_{\pi_{\pm}}^{(C.5)} = \frac{12i \lambda^{2}v^{2}}{m_{\sigma}^{2}} \int \frac{d^{3}q}{(2\pi)^{3}} \frac{1}{2E_{\sigma}} \big( 1 + 2n_{B}(E_{\sigma}) \big),
\end{equation}
\begin{align}
-i\Pi_{\pi_{\pm}}^{(C.6)} &=  \frac{3i\lambda^{2}v^{2}}{2\pi^{2}m_{\sigma}^{2}} \ \vert eB \vert  \int_{0}^{\infty} \frac{d\tau}{\sinh\left(\vert eB \vert \tau\right)} \frac{e^{-\tau m_{\pi}^{2}}} {\tau}\nonumber \\
&\times \bigg(1 + 2\sum_{n=1}^{\infty} e^{-\frac{n^{2}}{4 T^{2}\tau}} \bigg),
\end{align}
with the corresponding vacuum term
\begin{equation}
    -i\Pi_{\pi_{\pm},B=0}^{(C.6)}= \frac{3i\lambda^{2}v^{2}}{2\pi^{2}m_{\sigma}^{2}} \int_{0}^{\infty} \frac{d\tau}{\tau^{2}} e^{-\tau m_{\pi}^{2}},
\end{equation}
and
\begin{align} 
-i\Pi_{\pi_\pm}^{(C.7)} &= - \frac{i \lambda v g}{\pi^{2}m_{\sigma}^{2}} \vert q_{f}B \vert N_{c} \sum_{f}m_{f} \int_{0}^{\infty} \frac{d\tau}{\tanh(\vert q_{f}B \vert \tau)} \nonumber \\
&\times \frac{e^{-\tau m_{f}^{2}}}{\tau} \bigg(1 + 2\sum_{n=1}^{\infty} (-1)^{n} e^{-\frac{n^{2}}{4 T^{2}\tau}} \bigg),
\end{align}
where the corresponding vacuum contribution is
\begin{equation}
    -i\Pi_{\pi_\pm,B=0}^{(C.7)}=-\frac{i \lambda v g}{\pi^{2}m_{\sigma}^{2}} N_{c} \sum_{f}m_{f} \int_{0}^{\infty} d\tau \frac{e^{-\tau m_{f}^{2}}}{\tau^{2}}.
\end{equation}
As before, we define $E_\pi=\sqrt{\vec{q}^{\,2}+m_\pi^2}$ and $E_\sigma=\sqrt{\vec{q}^{\,2}+m_\sigma^2}$.

The remaining two contributions, labeled $C.8$ and $C.9$, require a different treatment. These correspond, respectively, to a loop involving a sigma meson and a charged pion, and to a quark-antiquark loop with quarks of different flavor and, consequently, different electric charges. In these cases, the Schwinger phases associated with the propagators do not cancel, preventing a straightforward evaluation in momentum space. As a result, the computation must be carried out in coordinate space.

The evaluation of these contributions is technically more involved and constitutes a central part of the present work. To the best of our knowledge, such loops have not been explicitly computed within the LSMq, either in the presence of a magnetic field alone or including both magnetic and thermal effects. We therefore present the derivation in detail, as it provides both new results and a systematic procedure that may be useful for future studies of thermomagnetic effects in effective descriptions of QCD, as well as in QCD itself.

We begin with the expression for the self-energy contribution corresponding to a loop formed by a charged pion and a sigma meson, which must be evaluated in space--time coordinates
\begin{widetext}
\begin{equation}
    -i\Pi_{\pi_{\pm}}^{(C.8)} = 4(-2i\lambda v)^{2}\int d^{4}x \int d^{4}y  \; F(x,k_{1}) D^{B}(x-y)D(y-x) F(y,k_{2}),
    \label{C.8initial}
\end{equation}
where the propagators for the charged pion and the sigma meson in space-time coordinates are
\begin{align}
    D^{B}(x-y) &= e^{i\phi(x,y)} \int \frac{d^{4}p_{1}}{(2\pi)^{4}} \int_{0}^{\infty} \frac{ds}{\cos(|eB|s)} e^{is\big(p_{1\parallel}^{2} - p_{1\perp}^{2} \frac{\tan(|eB|s)}{|eB|s} - m_{\pi}^{2} + i\epsilon \big)} e^{-ip_{1}\cdot(x-y)}, \nonumber \\
    D(y-x) &= \int \frac{d^{4}p_{2}}{(2\pi)^{4}} \frac{i}{p_{2}^{2}-m_{\sigma}^{2} + i\epsilon}e^{-ip_{2}\cdot(y-x)},
    \label{propagadoresneutralcharged}
\end{align}
respectively. The Schwinger phase in $D^{B}(x-y)$ is
\begin{equation}
    \phi(x,y) = e \int_{x}^{y} d\xi_{\mu} \Big[ A^{\mu}(\xi) + \frac{1}{2} F^{\mu\nu}(\xi - y)_{\nu} \Big].
    \label{Schwingerphase}
\end{equation}
The functions $F(X,K)$ are the Ritus functions for scalar fields~\cite{Ritus:1972ky}, defined as follows
\begin{equation}
    F(x,k_{1}) = \frac{1}{\sqrt{2\pi}} e^{-i(k_{1}^{0}x^{0} - k_{1}^{3}x^{3})}e^{-i \text{sgn}(eB)(k_{1}^{1} -k_{1}^{2})\theta} N_{k_{1}^{1},k_{1}^{2}}\, \xi^{(k_{1}^{1} -k_{1}^{2})/2} e^{-\frac{\xi}{2}}L_{k_{1}^{2}}^{k_{1}^{1} -k_{1}^{2}} (\xi),
    \label{Ritusf1}
\end{equation}
and
\begin{equation}
   F(y,k_{2}) = \frac{1}{\sqrt{2\pi}} e^{i(k_{2}^{0}y^{0} - k_{2}^{3}y^{3})}e^{i \text{sgn}(eB)(k_{2}^{1} -k_{2}^{2})\theta'} N_{k_{2}^{1},k_{2}^{2}}\, \xi'^{(k_{2}^{1} -k_{2}^{2})/2} e^{-\frac{\xi'}{2}}L_{k_{2}^{2}}^{k_{2}^{1} -k_{2}^{2}} (\xi'), 
   \label{Ritusf2}
\end{equation}
with
\begin{align}
    \xi &= \frac{|eB|}{2}r^{2}, \quad N_{k_{1}^{1},k_{1}^{2}}= \bigg( |eB|\frac{k_{1}^{2}!}{k_{1}^{1}!} \bigg)^{1/2}, \nonumber \\
    \xi' &= \frac{|eB|}{2}r'^{2}, \quad N_{k_{2}^{1},k_{2}^{2}}= \bigg( |eB|\frac{k_{2}^{2}!}{k_{2}^{1}!} \bigg)^{1/2},
    \label{notationRitusfunc}
\end{align}
and $\text{sgn}(eB)$ the sign function. Therefore, substituting Eqs.~(\ref{propagadoresneutralcharged})-(\ref{Ritusf2}) into Eq.~(\ref{C.8initial}), we can write
\begin{align}
-i\Pi_{\pi_{\pm}}^{(C.8)} = & - \frac{16 \lambda^{2}v^{2}}{2\pi} \int d^{4}x \int d^{4}y \ e^{-i(k_{1}^{0}x^{0} - k_{1}^{3}x^{3})}e^{-i \text{sgn}(eB)(k_{1}^{1} -k_{1}^{2})\theta} N_{k_{1}^{1},k_{1}^{2}}\, \xi^{(k_{1}^{1} -k_{1}^{2})/2} e^{-\frac{\xi}{2}}L_{k_{1}^{2}}^{k_{1}^{1} -k_{1}^{2}} (\xi) \nonumber \\
\times & e^{ie \int_{x}^{y} d\xi_{\mu} \Big[ A^{\mu}(\xi) + \frac{1}{2} F^{\mu\nu}(\xi - y)_{\nu} \Big]} \int \frac{d^{4}p_{1}}{(2\pi)^{4}} \int_{0}^{\infty} \frac{ds}{\cos(|eB|s)} e^{is\big(p_{1\parallel}^{2} - p_{1\perp}^{2} \frac{\tan(|eB|s)}{|eB|s} - m_{\pi}^{2} + i\epsilon \big)} e^{-ip_{1}\cdot(x-y)} \nonumber \\
\times & \int \frac{d^{4}p_{2}}{(2\pi)^{4}} \frac{i}{p_{2}^{2}-m_{\sigma}^{2} + i\epsilon} e^{-ip_{2}\cdot(y-x)} e^{i(k_{2}^{0}y^{0} - k_{2}^{3}y^{3})}e^{i \text{sgn}(eB)(k_{2}^{1} -k_{2}^{2})\theta'} N_{k_{2}^{1},k_{2}^{2}}\, \xi'^{(k_{2}^{1} -k_{2}^{2})/2} e^{-\frac{\xi'}{2}}L_{k_{2}^{2}}^{k_{2}^{1} -k_{2}^{2}} \xi').
\end{align}
We first evaluate the Schwinger phase. We choose the gauge $A^{\mu}(\xi)=(0,-\frac{B}{2}\xi^{2},\frac{B}{2}\xi^{1},0)$, in such a way that $F^{\mu\nu}$ has only two non-zero components $F^{12}=-F^{21}=B$; the path in the phase chosen is $\xi^{\mu}(\lambda)=x^{\mu}+\lambda(y^{\mu}-x^{\mu}), \quad 0\le\lambda\le1$. This gives
\begin{equation}
    e^{ie \int_{x}^{y} d\xi_{\mu} \Big[ A^{\mu}(\xi) + \frac{1}{2} F^{\mu\nu}(\xi - y)_{\nu} \Big]}=e^{ie\int_{0}^{1} d\lambda\, \Delta_{\mu}\Big[A^{\mu}(\xi(\lambda))+\frac{1}{2}F^{\mu\nu}(\xi(\lambda)-y)_{\nu}\Big]},
\end{equation}
with $\Delta_{\mu}=y_{\mu}-x_{\mu}$. After integrating over $\lambda$, we get
\begin{equation}
    e^{ie \int_{x}^{y} d\xi_{\mu} \Big[ A^{\mu}(\xi) + \frac{1}{2} F^{\mu\nu}(\xi - y)_{\nu} \Big]}=e^{ie\frac{B}{2}(y^2x^1-x^2y^1)}.
\end{equation}
The integrals over $x_\parallel$ and $y_\parallel$ can be performed straightforwardly, using the integral representation of the Dirac delta functions. Then, we write
\begin{align}
-i\Pi_{\pi_{\pm}}^{(C.8)} = & - \frac{16 \lambda^{2}v^{2}}{2\pi} (2\pi)^{4} \int d^{2}x_{\perp} \int d^{2}y_{\perp} \ e^{-i \text{sgn}(eB)(k_{1}^{1} -k_{1}^{2})\theta} N_{k_{1}^{1},k_{1}^{2}}\, \xi^{(k_{1}^{1} -k_{1}^{2})/2} e^{-\frac{\xi}{2}} L_{k_{1}^{2}}^{k_{1}^{1} -k_{1}^{2}} (\xi) \nonumber \\
\times & e^{i\frac{eB}{2}(y^2x^1-x^2y^1)} \int \frac{d^{4}p_{1}}{(2\pi)^{4}} \int_{0}^{\infty} \frac{ds}{\cos(|eB|s)} e^{is\big(p_{1\parallel}^{2} - p_{1\perp}^{2} \frac{\tan(|eB|s)}{|eB|s} - m_{\pi}^{2} + i\epsilon \big)} e^{i(p_{1}^{1}(x^{1}-y^{1}) + p_{1}^{2}(x^{2}-y^{2}))} \nonumber \\
\times & \int \frac{d^{4}p_{2}}{(2\pi)^{4}} \frac{i}{p_{2}^{2}-m_{\sigma}^{2} + i\epsilon} e^{i(p_{2}^{1}(x^{1}-y^{1}) + p_{2}^{2}(x^{2}-y^{2}))} e^{i \text{sgn}(eB)(k_{2}^{1} -k_{2}^{2})\theta'} N_{k_{2}^{1},k_{2}^{2}}\, \xi'^{(k_{2}^{1} -k_{2}^{2})/2} e^{-\frac{\xi'}{2}}L_{k_{2}^{2}}^{k_{2}^{1} -k_{2}^{2}}(\xi') \nonumber \\
\times & \delta^{(2)}(k_{1\parallel}+p_{1\parallel}-p_{2\parallel})\,\delta^{(2)}(p_{2\parallel}-k_{2\parallel}-p_{1\parallel}).
\end{align}
The next step is to compute the integrals over the perpendicular coordinates \(x_\perp\) and \(y_\perp\). We identify that
\begin{equation}
    \xi=\frac{|eB|}{2}r^{2}=\frac{|eB|}{2}x_{\perp}^{2}=\frac{|eB|}{2}(x_{1}^{2}+x_{2}^{2}),
\qquad
\xi'=\frac{|eB|}{2}r'^{2}=\frac{|eB|}{2}y_{\perp}^{2}=\frac{|eB|}{2}(y_{1}^{2}+y_{2}^{2}).
\end{equation}
Hence, we choose polar coordinates with $x_{1}=r\cos\theta, \ x_{2}=r\sin\theta, \ y_{1}=r'\cos\theta', \ \text{and} \ y_{2}=r'\sin\theta'$ and the integral over the angles $\theta$ and $\theta'$ can be performed, yielding
\begin{align}
-i\Pi_{\pi_{\pm}}^{(C.8)}&= - \frac{16\lambda^{2}v^{2}}{2\pi} (2\pi)^{4} \int\frac{d^{4}p_{1}}{(2\pi)^{4}} \int_{0}^{\infty}\frac{ds}{\cos(|eB|s)} e^{is\big(p_{1\parallel}^{2}-p_{1\perp}^{2}\frac{\tan(|eB|s)}{|eB|s}-m_{\pi}^{2}+i\epsilon\big)} \nonumber \\
\times & \int\frac{d^{4}p_{2}}{(2\pi)^{4}}\frac{i}{p_{2}^{2}-m_{\sigma}^{2}+i\epsilon} N_{k_{1}^{1},k_{1}^{2}} N_{k_{2}^{1},k_{2}^{2}} \Big(\frac{|eB|}{2}\Big)^{\Delta k_{1}/2} \Big(\frac{|eB|}{2}\Big)^{\Delta k_{2}/2} \nonumber \\
\times & \sum_{n,m,q} \frac{i^{\,m-1}e^{-im\varphi_{0}}}{m-n+\Delta k_{2}} \frac{(-i)^{q+1}e^{-iq\varphi_{0}}}{q+n-\Delta k_{1}} \Big(e^{i2\pi(m-n+\Delta k_{2})}-1\Big) \Big(e^{i2\pi(q+n-\Delta k_{1})}-1\Big) \nonumber \nonumber \\
\times & \int_{0}^{\infty} dr\, r^{\Delta k_{1} +1} e^{-\frac{|eB|}{4}r^{2}} L_{k_{1}^{2}}^{\Delta k_{1}} \Big(\frac{|eB|}{2}r^{2}\Big) J_{q}\big(r\sqrt{(\Delta p^{1})^{2} +(\Delta p^{2})^{2}}\big) \nonumber \\
\times & \int_{0}^{\infty} dr'\, r'^{\Delta k_{2} +1} e^{-\frac{|eB|}{4}r'^{2}} L_{k_{2}^{2}}^{\Delta k_{2}} \Big(\frac{|eB|}{2}r'^{2}\Big) J_{m}\big(r'\sqrt{(\Delta p^{1})^{2} +(\Delta p^{2})^{2}}\big) J_{n}\Big(\frac{|eB|}{2}rr'\Big),
\label{C.8full}
\end{align}
where $J_l(X)$ are the Bessel functions of first kind, $\Delta K_{j}=K_{j}^{1}-K_{j}^{2}$ is the generic expression for the difference between the components 1 and 2 of a given momentum, and $\varphi_0=\tan^{-1}(\Delta p^2/\Delta p^1)$. The Bessel functions arise from the angular integrations, where the Jacobi--Anger identity is used to expand plane waves in terms of cylindrical harmonics. Equation~(\ref{C.8full}) gives the complete expression, which involves several numerical integrations and a triple sum. To further simplify the expression, we consider the inertial frame in which the external particle is at rest, namely $k_{j}^{1}=k_{j}^{2}=k_{j}^{3}=0$ and $k_{j}^{2}=(k_{j}^{0})^{2}$, with $j=1,2$. Thus, the expression we need to compute becomes
\begin{align}
-i\Pi_{\pi_{\pm}}^{(C.8)} &= - \frac{8 \lambda^{2}v^{2}}{\pi} |eB|(2\pi)^{4} \int d^{2}x_{\perp} \int d^{2}y_{\perp} \  e^{-\frac{\xi}{2}} e^{i\frac{eB}{2}(y^2x^1-x^2y^1)} \nonumber \\
& \times \int \frac{d^{4}p_{1}}{(2\pi)^{4}} \int_{0}^{\infty} \frac{ds}{\cos(|eB|s)} e^{is\big(p_{1\parallel}^{2} - p_{1\perp}^{2} \frac{\tan(|eB|s)}{|eB|s} - m_{\pi}^{2} + i\epsilon \big)} e^{i(p_{1}^{1}(x^{1}-y^{1}) + p_{1}^{2}(x^{2}-y^{2}))}\nonumber \\
&\times \int \frac{d^{4}p_{2}}{(2\pi)^{4}} \frac{i}{p_{2}^{2}-m_{\sigma}^{2} + i\epsilon} e^{i(p_{2}^{1}(x^{1}-y^{1}) + p_{2}^{2}(x^{2}-y^{2}))} e^{-\frac{\xi'}{2}}\nonumber \\
&\times \delta(k_{1}^{0}+p_{1}^{0}-p_{2}^{0})\,\delta(-k_{2}^{0}-p_{1}^{0}+p_{2}^{0})\,\delta^{(2)}(p_{2\parallel}^{3}-p_{1\parallel}^{3}).
\label{C.8zerospacialcomponents}
\end{align}
We compute the integrals over $x_\perp$ and $y_\perp$ component by component, using
\begin{align}
    \int dy^{2} e^{\big\{ i\big( i\frac{|eB|}{4}(y^{2})^{2} + y^{2}[(p_{2}^{2} - p_{1}^{2})+\frac{eB}{2}x^{1}] \big) \big\}} &= \sqrt{\frac{4\pi}{|eB|}} \exp{\bigg\{-\frac{\big[(p_{2}^{2} - p_{1}^{2}) + \frac{eB}{2} x^{1}\big]^{2}}{|eB|}\bigg\}}, \nonumber \\
    \int dy^{1} e^{\big\{ i\big( -\frac{|eB|}{4}(y^{1})^{2} + y^{1}[(p_{2}^{1} - p_{1}^{1}) -\frac{eB}{2}x^{2}] \big) \big\}} &= \sqrt{\frac{4\pi}{|eB|}} \exp{\bigg\{-\frac{\big[(p_{2}^{1} - p_{1}^{1}) - \frac{eB}{2} x^{2}\big]^{2}}{|eB|}\bigg\}}, \nonumber \\
    e^{-\frac{(p_{2}^{1} - p_{1}^{1})^{2}}{|eB|}}\int dx^{2} e^{\big\{ - \frac{|eB|}{2}(x^{2})^{2} + x^{2} \big[ \frac{eB(p_{2}^{1} - p_{1}^{1})}{|eB|}+ i(p_{1}^{2} - p_{2}^{2})\big]  \big\}} &= \sqrt{\frac{2\pi}{|eB|}} e^{-\frac{(p_{2}^{1} - p_{1}^{1})^{2}}{|eB|}} e^{\bigg\{\frac{2}{|eB|} \Big[ \frac{eB(p_{2}^{1} - p_{1}^{1})}{|eB|}+ i(p_{1}^{2} - p_{2}^{2})\Big]^{2} \bigg\}}, \nonumber \\
    e^{-\frac{(p_{2}^{2} - p_{1}^{2})^{2}}{|eB|}} \int dx^{1} e^{\big\{ -\frac{|eB|}{2}(x^{1})^{2} +x^{1} \big[ i(p_{1}^{1} - p_{2}^{1}) -\frac{eB(p_{2}^{2}-p_{1}^{2})}{|eB|}\big] \big\}} &= \sqrt{\frac{2\pi}{|eB|}} e^{-\frac{(p_{2}^{2} - p_{1}^{2})^{2}}{|eB|}} e^{\bigg\{ \frac{2}{|eB|} \Big[ i(p_{1}^{1} - p_{2}^{1}) -\frac{eB(p_{2}^{2}-p_{1}^{2})}{|eB|}\Big]^{2}  \bigg\}}.
    \label{integralscomponents}
\end{align}
We substitute the expressions in Eq.~(\ref{integralscomponents}) into Eq.~(\ref{C.8zerospacialcomponents}) and obtain
\begin{align}
-i\Pi_{\pi_{\pm}}^{(C.8)} &=- \frac{8 i\lambda^{2}v^{2}}{\pi} |eB|(2\pi)^{4} \frac{8\pi^{2}}{|eB|^{2}} \int \frac{d^{4}p_{1}}{(2\pi)^{4}} \int_{0}^{\infty} \frac{ds}{\cos(|eB|s)} e^{is\big(p_{1\parallel}^{2} - p_{1\perp}^{2} \frac{\tan(|eB|s)}{|eB|s} - m_{\pi}^{2} + i\epsilon \big)} \nonumber \\
& \times \int \frac{d^{4}p_{2}}{(2\pi)^{4}} \frac{1}{p_{2}^{2}-m_{\sigma}^{2} + i\epsilon} e^{-\frac{(p_{2}^{1} - p_{1}^{1})^{2}}{|eB|}} e^{-\frac{(p_{2}^{2} - p_{1}^{2})^{2}}{|eB|}} \delta(k_{1}^{0}+p_{1}^{0}-p_{2}^{0})\,\delta(-k_{2}^{0}-p_{1}^{0}+p_{2}^{0})\,\delta^{(2)}(p_{2}^{3}-p_{1}^{3}).
\end{align}
Using the Dirac delta functions, we can perform one of the momentum integrals and obtain the expression
\begin{align}
-i\Pi_{\pi_{\pm}}^{(C.8)} &= -16 i\lambda^{2}v^{2}\mathcal{V}_\parallel \int_{0}^{\infty} \frac{ds}{\cos(|eB|s)} \frac{1}{i\tan(|eB|s) + 1} e^{is\big[- m_{\pi}^{2} + i\epsilon \big]} \nonumber \\
& \times \int \frac{d^{4}p_{2}}{(2\pi)^{4}}  e^{is\big[(p_{2}^{0} - k_{2}^{0})^{2} - (p_{2}^{3})^2\big]} \frac{1}{p_{2}^{2}-m_{\sigma}^{2} + i\epsilon} e^{-\frac{(p_{2\perp})^{2}}{|eB|}}e^{\frac{(p_{2\perp})^{2}}{|eB|(i\tan(|eB|s)+1)}},
\label{C.8justintegralp2}
\end{align}
where $\mathcal{V}_\parallel$ denotes the volume associated with the parallel coordinates. To compute the integral over $p_2$, we first rewrite the Feynman propagator as
\begin{equation}
\frac{i}{p_{2}^{2}-m_{\sigma}^{2}+i\epsilon} = \int_{0}^{\infty}ds' e^{is'\left(p_{2}^{2}-m_{\sigma}^{2}+i\epsilon\right)}.
\end{equation}
Equation~(\ref{C.8justintegralp2}) is written as follows
\begin{align}
-i\Pi_{\pi_{\pm}}^{(C.8)} &= -16 \lambda^{2}v^{2} \mathcal{V}_\parallel \int_{0}^{\infty} \frac{ds}{\cos(|eB|s)} \frac{1}{i\tan(|eB|s) + 1} e^{is\big[- m_{\pi}^{2} + i\epsilon \big]} \nonumber \\
&\times \int_{0}^{\infty}ds' \int \frac{d^{4}p_{2}}{(2\pi)^{4}}  e^{is\big[(p_{2}^{0} - k_{2}^{0})^{2} - (p_{2}^{3})^2\big]} e^{is'\big((p_{2}^{0})^{2} - (p_{2}^{1})^{2} - (p_{2\perp})^{2} + m_{\sigma}^{2}+i\epsilon\big)} e^{-\frac{(p_{2\perp})^{2}}{|eB|}}e^{\frac{(p_{2\perp})^{2}}{|eB|(i\tan(|eB|s)+1)}}.
\end{align}
We now integrate over the three spatial components of $p_2$, using first the integral over $p_2^3$
\begin{equation}
\int \frac{dp_{2}^{3}}{2\pi} e^{-i(p_{2}^{3})^{2}(s+s')} = \frac{1}{2\pi}\sqrt{\frac{\pi}{i(s+s')}},
\end{equation}
and then over $p_{2\perp}$
\begin{equation}
\int \frac{d^{2}p_{2\perp}}{(2\pi)^2} e^{-p_{2\perp}^{2}\big[ \frac{1}{|eB|}\big(1 - \frac{1}{i\tan(|eB|s)+1}\big) + is'\big]} = \frac{1}{4\pi} \bigg[\frac{1}{|eB|}\Big(\frac{i\tan(|eB|s)}{i\tan(|eB|s)+1}\Big) + is' \bigg]^{-1}.
\end{equation}
Then, we have
\begin{align}
-i\Pi_{\pi_{\pm}}^{(C.8)} &= - \frac{2 \lambda^{2}v^{2}\mathcal{V}_\parallel}{\pi^2} \int_{0}^{\infty} \frac{ds}{\cos(|eB|s)} \frac{1}{i\tan(|eB|s) + 1} e^{is\big[- m_{\pi}^{2} + i\epsilon \big]} \\
&\times \int_{0}^{\infty}ds'  \sqrt{\frac{\pi}{i(s+s')}} e^{is\big[- m_{\sigma}^{2} + i\epsilon \big]} \bigg[\frac{1}{|eB|}\Big(\frac{i\tan(|eB|s)}{i\tan(|eB|s)+1}\Big) + is' \bigg]^{-1}\int \frac{dp_{2}^{0}}{2\pi}e^{is(p_{2}^{0} - k_{2}^{0})^{2}} e^{is'(p_{2}^{0})^{2}} .
\end{align}
We implement the analytic continuations $s \rightarrow -i\tau$ and $s' \rightarrow -i\tau'$, and the temperature dependence is incorporated through the Matsubara formalism. This gives
\begin{align}
-i\Pi_{\pi_{\pm}}^{(C.8)} &= \frac{2 i\lambda^{2} v^{2} \mathcal{V}_\parallel}{\pi^2} \int_{0}^{\infty} \frac{d\tau}{\cosh(|eB|\tau)} \frac{1}{\tanh(|eB|\tau) + 1} e^{-\tau m_{\pi}^{2}} \nonumber \\
&\times \int_{0}^{\infty}d\tau'  \sqrt{\frac{\pi}{\tau+\tau'}} e^{- \tau' m_{\sigma}^{2}}\bigg[\frac{1}{|eB|}\Big(\frac{\tanh(|eB|\tau)}{\tanh(|eB|\tau)+1}\Big) + \tau' \bigg]^{-1} T\sum_{n} e^{\tau(i\omega_{n} - i\omega)^{2}} e^{\tau'(i\omega_{n})^{2}}
\end{align}
The sum over the Matsubara frequencies is similar to the expression in Eq.~(\ref{sumB.10}), where we now have
\begin{equation}
\sum_{n=-\infty}^{\infty} e^{-\tau(\omega_{n} - \omega)^{2}} e^{-\tau'\omega_{n}^{2}}= \frac{1}{2T\sqrt{\pi(\tau+\tau')}} e^{-\frac{\tau\tau' \omega^{2}}{\tau + \tau'}}\bigg[ 1 + 2\sum_{n=1}^{\infty} e^{-\frac{n^{2}}{4(\tau + \tau')T^{2}}} \cos\Big(\frac{n\omega\tau}{(\tau+\tau')T}\Big) \bigg].
\end{equation}
Then, after the sum, we obtain
\begin{align}
-i\Pi_{\pi_{\pm}}^{(C.8)} &= \frac{i\lambda^{2} v^{2} \mathcal{V}_\parallel}{\pi^2} \int_{0}^{\infty} \frac{d\tau}{\cosh(|eB|\tau)} \frac{1}{\tanh(|eB|\tau) + 1} e^{-\tau m_{\pi}^{2}}\int_{0}^{\infty}d\tau' \frac{1}{\tau+\tau'} e^{- \tau' m_{\sigma}^{2}} \nonumber \\
&\times e^{-\frac{\tau\tau' \omega^{2}}{\tau + \tau'}}\bigg[ 1 + 2\sum_{n=1}^{\infty} e^{-\frac{n^{2}}{4(\tau + \tau')T^{2}}} \cos\Big(\frac{n\omega\tau}{(\tau+\tau')T}\Big) \bigg] \bigg[\frac{1}{|eB|}\big(\frac{\tanh(|eB|\tau)}{\tanh(|eB|\tau)+1}\Big) + \tau' \bigg]^{-1}.
\end{align}
Finally, performing the change of variables \(\tau = u(1-v)\) and \(\tau' = uv\), we obtain the final expression in momentum space
\begin{align}
-i\Pi_{\pi_{\pm}}^{(C.8)} &= \frac{i\lambda^{2} v^{2}}{\pi^{2}} \int_{0}^{\infty} du \int_{0}^{1}dv \frac{1}{\cosh(|eB|u(1-v))} \frac{1}{\tanh(|eB|u(1-v)) + 1} e^{-u(1-v) m_{\pi}^{2}} e^{- uv m_{\sigma}^{2}}\nonumber \nonumber \\
\times & e^{(i\omega)^2uv(1-v)}\bigg[ 1 + 2\sum_{n=1}^{\infty} e^{-\frac{n^{2}}{4T^{2}u}} \cosh\Big(\frac{(1-v)n(i\omega)}{T}\Big) \bigg] \bigg[\frac{1}{|eB|}\Big(\frac{\tanh(|eB|u(1-v))}{\tanh(|eB|u(1-v))+1}\Big) + uv \bigg]^{-1},
\end{align}
\end{widetext}
where the vacuum piece is
\begin{align}
    -i\Pi_{\pi_{\pm},B=0}^{(C.8)}&=\frac{i\lambda^{2} v^{2}}{\pi^{2}} \int_{0}^{\infty} \frac{du}{u} \int_{0}^{1}dv e^{-u(1-v) m_{\pi}^{2}} \nonumber \\
    &\times e^{- uv m_{\sigma}^{2}} e^{uv(1-v)(i\omega)^{2}}.
\end{align}

The resulting expression involves a set of coupled integrals over proper times and momenta, which must be evaluated numerically. This structure reflects the nontrivial interplay between the magnetic background, the thermal medium, and the nonvanishing Schwinger phase associated with the charged pion.

The last contribution to the charged pion self-energy, labeled as $C.9$, corresponds to a quark--antiquark loop involving quarks of different flavor and, consequently, different electric charges. This makes the calculation particularly challenging, since the Schwinger phases associated with the two fermion propagators do not cancel. As in $C.8$, the calculation must therefore be carried out in coordinate space. This is one of the most technically involved one-loop terms considered in this work. For this reason, and in order to make the derivation transparent and reproducible, we present the main steps of the calculation in detail. We begin with
\begin{widetext}
\begin{equation}
-i\Pi_{\pi_{\pm}}^{(C.9)} = - N_{c} \sum_{f} \int d^{4}x \int d^{4}y  \; F(x,k_{1})  \text{Tr}\Big[ S^{B}(x-y)(\sqrt{2}g\gamma^{5}) S^{B}(y-x) (\sqrt{2}g\gamma^{5}) \Big] F(y,k_{2}),
\end{equation}   
where the fermion propagator is
\begin{align}
S^{B}(x-y)&= e^{i\phi(x,y)} \int \frac{d^{4}p}{(2\pi)^4} e^{-ip\cdot(x-y)}\int_{0}^{\infty}\frac{ds}{\cos(|q_fB| s)} e^{is\left( p_0^2- p_{3}^{2} - p_{\perp}^{2} \frac{\tan(|q_fB| s)}{|q_fB| s} - m_{f}^{2} \right)} \nonumber \\
&\times \Big [ \left(\cos(|q_fB| s) + \text{sgn}(q_{f}B) \gamma^{1} \gamma^{2} \sin(\vert q_{f} B \vert s)\right) \left(m_{f} +\gamma^{0}p_0 - \gamma^{3} p_{3}\right)  - \frac{\gamma^1p_1+\gamma^2p_2}{\cos(|q_fB| s)} \Big ],
\label{fermionpropagatorspacecoordinates}
\end{align}
the corresponding Schwinger phase is given by
\begin{equation}
\phi(x,y) = q_f \int_{x}^{y} d\xi_{\mu} \Big[ A^{\mu}(\xi) + \frac{1}{2} F^{\mu\nu}(\xi - y)_{\nu} \Big],
\end{equation}
and the Ritus functions are defined in Eqs.~(\ref{Ritusf1}) and~(\ref{Ritusf2}), with their respective notation defined there. Substituting these expressions, the correction $C.9$ becomes
\begin{align}
-i\Pi_{\pi_{\pm}}^{(C.9)} = & - 2 g^{2} N_{c} \sum_{f} \int d^{4}x \int d^{4}y \ F(x,k_{1}) \ e^{iq_{u} \int_{x}^{y} d\xi_{\mu} \big[ A^{\mu}(\xi) + \frac{1}{2} F^{\mu\nu}(\xi - y)_{\nu} \big]} \int \frac{d^{4}p_{1}}{(2\pi)^{4}} \int \frac{d^{4}p_{2}}{(2\pi)^{4}} e^{-ip_{1}\cdot(x-y)} \nonumber \\
\times & \int_{0}^{\infty} \frac{ds}{\cos(|q_{u}B|s)} e^{is\big(p_{1\parallel}^{2} - p_{1\perp}^{2} \frac{\tan(|q_{u}B|s)}{|q_{u}B|s} - m_{f}^{2} + i\epsilon \big)} \int_{0}^{\infty} \frac{ds'}{\cos(|q_{d}B|s')} e^{is'\big(p_{2\parallel}^{2} - p_{2\perp}^{2} \frac{\tan(|q_{d}B|s')}{|q_{d}B|s'} - m_{f}^{2} + i\epsilon \big)} \nonumber \\
\times & \text{Tr} \bigg\{ \bigg[ \big(\cos(|q_{u}B|s) + \text{sgn}(q_{u}B) \gamma^{1} \gamma^{2} \sin(|q_{u}B|s)\big)\big(m_{f} + \slashed{p}_{1\parallel}\big) - \frac{\slashed{p}_{1\perp}}{\cos(|q_{u}B|s)} \bigg]\gamma^{5}  \nonumber \\
\times & \bigg[ \big(\cos(|q_{d}B|s') + \text{sgn}(q_{d}B) \gamma^{1} \gamma^{2} \sin(|q_{d}B|s')\big)\big(m_{f} + \slashed{p}_{2\parallel}\big) - \frac{\slashed{p}_{2\perp}}{\cos(|q_{d}B|s')} \bigg]\gamma^{5}\bigg\} \nonumber \\
\times & F(y,k_{2}) \ e^{iq_{d} \int_{y}^{x} d\xi'_{\mu} \big[ A^{\mu}(\xi') + \frac{1}{2} F^{\mu\nu}(\xi' - x)_{\nu} \big]} e^{-ip_{2}\cdot(y-x)},
\end{align}
where $N_c$ is the number of colors, $\text{sgn}(q_fB)$ is the sign function, the charges of quarks up and down are $q_u$ and $q_d$, respectively. The two Schwinger phases are treated in the same way as in the $C.8$ contribution. Hence, we have
\begin{align}
e^{iq_{u} \int_{x}^{y} d\xi_{\mu} \big[ A^{\mu}(\xi) + \frac{1}{2} F^{\mu\nu}(\xi - y)_{\nu} \big]}&= e^{i\frac{q_{u}B}{2}(y^{2}x^{1}-x^{2}y^{1})}, \nonumber \\
e^{iq_{d} \int_{y}^{x} d\xi_{\mu} \big[ A^{\mu}(\xi') + \frac{1}{2} F^{\mu\nu}(\xi' - x)_{\nu} \big]} &= e^{-i\frac{q_{d}B}{2}(y^{2}x^{1}-x^{2}y^{1})}.
\label{Schwingerphases}
\end{align}
Clearly, Eq.~(\ref{Schwingerphases}) shows that the two Schwinger phases do not cancel. We then compute the trace over Dirac indices. The trace has the same structure as the one obtained for contribution $A.9$, given in Eq.~(\ref{A.9trace}), with the appropriate replacement of the electric charges. After inserting the explicit form of the Ritus functions, the contribution $C.9$ becomes
\begin{align}
-i\Pi_{\pi_{\pm}}^{(C.9)} = & - \frac{8}{2\pi} g^{2} N_{k_{1}^{1},k_{1}^{2}} N_{k_{2}^{1},k_{2}^{2}} N_{c} \sum_{f} \int d^{4}x \int d^{4}y e^{-i(k_{1}^{0}x^{0} - k_{1}^{3}x^{3})}e^{-i \text{sgn}(q_{u}B)(k_{1}^{1} -k_{1}^{2})\theta} \,\xi^{(k_{1}^{1} -k_{1}^{2})/2} \nonumber \\
\times & e^{-\frac{\xi}{2}}L_{k_{1}^{2}}^{k_{1}^{1} -k_{1}^{2}} (\xi) e^{i\frac{q_{u}B}{2}(y^{2}x^{1}-x^{2}y^{1})} e^{i(k_{2}^{0}y^{0} - k_{2}^{3}y^{3})} e^{i\text{sgn}(q_{d}B)(k_{2}^{1} -k_{2}^{2})\theta'} \xi'^{(k_{2}^{1} -k_{2}^{2})/2} e^{-\frac{\xi'}{2}} L_{k_{2}^{2}}^{k_{2}^{1} -k_{2}^{2}} (\xi') \nonumber \\
\times & e^{-i\frac{q_{d}B}{2}(y^{2}x^{1}-x^{2}y^{1})} \int \frac{d^{4}p_{1}}{(2\pi)^{4}} \int_{0}^{\infty} \frac{ds}{\cos(|q_{u}B|s)} e^{-ip_{1}\cdot(x-y)} e^{is\big(p_{1\parallel}^{2} - p_{1\perp}^{2} \frac{\tan(|q_{u}B|s)}{|q_{u}B|s} - m_{f}^{2} + i\epsilon \big)} \nonumber \\
\times & \int \frac{d^{4}p_{2}}{(2\pi)^{4}} \int_{0}^{\infty} \frac{ds'}{\cos(|q_{d}B|s')} e^{-ip_{2}\cdot(y-x)} e^{is'\big(p_{2\parallel}^{2} - p_{2\perp}^{2} \frac{\tan(|q_{d}B|s')}{|q_{d}B|s'} - m_{f}^{2} + i\epsilon \big)} \nonumber \\
\times & \bigg[\big(m_{f}^{2} + p_{1\parallel}\cdot p_{2\parallel}\big)\big(\cos(|q_{u}B|s)\cos(|q_{d}B|s') + \sin(|q_{u}B|s)\sin(|q_{d}B|s')\big) + \frac{p_{1\perp}\cdot p_{2\perp}}{\cos(|q_{u}B|s)\cos(|q_{d}B|s')} \bigg].
\end{align}

The integrals over $x_\parallel$ and $y_\parallel$ can be performed straightforwardly and give two Dirac delta functions, as in the $C.8$ contribution. We now write contribution $C.9$ as
\begin{align}
-i\Pi_{\pi_{\pm}}^{(C.9)} = & - \frac{8}{2\pi} g^{2} (2\pi)^{4} N_{k_{1}^{1},k_{1}^{2}} N_{k_{2}^{1},k_{2}^{2}} N_{c} \sum_{f} \int d^{2}x_{\perp} \int d^{2}y_{\perp} e^{-i \text{sgn}(q_{u}B)(k_{1}^{1} -k_{1}^{2})\theta} \,\xi^{(k_{1}^{1} -k_{1}^{2})/2} e^{-\frac{\xi}{2}}L_{k_{1}^{2}}^{k_{1}^{1} -k_{1}^{2}} (\xi) \nonumber \\
\times & e^{i\frac{q_{u}B}{2}(y^{2}x^{1}-x^{2}y^{1})} e^{i\text{sgn}(q_{d}B)(k_{2}^{1} -k_{2}^{2})\theta'} \xi'^{(k_{2}^{1} -k_{2}^{2})/2} e^{-\frac{\xi'}{2}} L_{k_{2}^{2}}^{k_{2}^{1} -k_{2}^{2}} (\xi') e^{-i\frac{q_{d}B}{2}(y^{2}x^{1}-x^{2}y^{1})}\nonumber \\
\times & \int \frac{d^{4}p_{1}}{(2\pi)^{4}} \int_{0}^{\infty} \frac{ds}{\cos(|q_{u}B|s)} e^{i\big(p_{1\perp}\cdot(x-y)_{\perp}\big)} e^{is\big(p_{1\parallel}^{2} - p_{1\perp}^{2} \frac{\tan(|q_{u}B|s)}{|q_{u}B|s} - m_{f}^{2} + i\epsilon \big)} \nonumber \\
\times & \int \frac{d^{4}p_{2}}{(2\pi)^{4}} \int_{0}^{\infty} \frac{ds'}{\cos(|q_{d}B|s')} e^{i \big(p_{2\perp}\cdot(y-x)_{\perp}\big)
} e^{is'\big(p_{2\parallel}^{2} - p_{2\perp}^{2} \frac{\tan(|q_{d}B|s')}{|q_{d}B|s'} - m_{f}^{2} + i\epsilon \big)} \nonumber \\
\times & \bigg[\big(m_{f}^{2} + p_{1\parallel}\cdot p_{2\parallel}\big)\big(\cos(|q_{u}B|s)\cos(|q_{d}B|s') + \sin(|q_{u}B|s)\sin(|q_{d}B|s')\big)\nonumber \\
+ & \frac{p_{1\perp} \cdot p_{2\perp}}{\cos(|q_{u}B|s)\cos(|q_{d}B|s')} \bigg]\delta^{(2)}(k_{1\parallel}+p_{1\parallel}-p_{2\parallel})\,\delta^{(2)}(p_{2\parallel}-k_{2\parallel}-p_{1\parallel}).
\end{align}
As we did in the previous contribution, the next step is to compute the integrals over the perpendicular components, $x_\perp$ and $y_\perp$, which can be written in terms of polar coordinates, $r$ and $\theta$. Following the same procedure used for $C.8$, the result after the angular integrations is
\begin{align}
-i\Pi_{\pi_{\pm}}^{(C.9)} = & -\frac{8}{2\pi} g^{2} (2\pi)^{4}N_{c} \sum_{f} \int \frac{d^{4}p_{1}}{(2\pi)^{4}} \int_{0}^{\infty} \frac{ds}{\cos(|q_{u}B|s)} e^{is\big(p_{1\parallel}^{2} - p_{1\perp}^{2} \frac{\tan(|q_{u}B|s)}{|q_{u}B|s} - m_{f}^{2} + i\epsilon \big)} \nonumber \\
\times & \int \frac{d^{4}p_{2}}{(2\pi)^{4}} \int_{0}^{\infty} \frac{ds'}{\cos(|q_{d}B|s')} e^{i \big(p_{2\perp}\cdot(y-x)_{\perp}\big)} e^{is'\big(p_{2\parallel}^{2} - p_{2\perp}^{2} \frac{\tan(|q_{d}B|s')}{|q_{d}B|s'} - m_{f}^{2} + i\epsilon \big)} N_{k_{1}^{1},k_{1}^{2}} N_{k_{2}^{1},k_{2}^{2}} \nonumber \\
\times & \bigg[\big(m_{f}^{2} + p_{1\parallel}\cdot p_{2\parallel}\big)\big(\cos(|q_{u}B|s)\cos(|q_{d}B|s') + \sin(|q_{u}B|s)\sin(|q_{d}B|s')\big)\nonumber \\
+ & \frac{p_{1\perp} \cdot p_{2\perp}}{\cos(|q_{u}B|s)\cos(|q_{d}B|s')} \bigg] \bigg(\frac{|q_{u}B|}{2}\bigg)^{(k_{1}^{1}-k_{1}^{2})/2}\bigg(\frac{|q_{d}B|}{2}\bigg)^{(k_{2}^{1}-k_{2}^{2})/2} \sum_{n,m,q} \frac{i^{\,m-1}e^{-im\varphi_{0}}}{m-n+(k_{2}^{1}-k_{2}^{2})} \nonumber \\
\times & \Big(e^{i2\pi(m-n+(k_{2}^{1}-k_{2}^{2})}-1\Big)  \frac{(-i)^{\,q+1}e^{-iq\varphi_{0}}}{q+n-(k_{1}^{1}-k_{1}^{2})} \Big(e^{i2\pi(q+n-(k_{1}^{1}-k_{1}^{2}))}-1\Big) \int_{0}^{\infty} dr \ r^{(k_{1}^{1}-k_{1}^{2})+1} e^{-\frac{|q_{u}B| r^{2}}{4}} \nonumber \\
\times & L_{k_{1}^{2}}^{k_{1}^{1} -k_{1}^{2}} \bigg(\frac{|q_{u}B| r}{2}\bigg) J_{q}\bigg(r\sqrt{(p_{2}^{1}-p_{1}^{1})^{2} + (p_{2}^{2}-p_{1}^{2})^{2}}\bigg) \int_{0}^{\infty} dr' \ r'^{(k_{2}^{1}-k_{2}^{2})+1} e^{-\frac{|q_{u}B| r'^{2}}{4}} L_{k_{2}^{2}}^{k_{2}^{1} -k_{2}^{2}} \bigg(\frac{|q_{u}B| r'}{2}\bigg) \nonumber \\
\times & J_{m}\bigg(r'\sqrt{(p_{2}^{1}-p_{1}^{1})^{2} + (p_{2}^{2}-p_{1}^{2})^{2}}\bigg) J_{n}\bigg(\frac{(q_{u} - q_{d}) B r r'}{2}\bigg)\nonumber \\
\times & \delta^{(2)}(k_{1\parallel}+p_{1\parallel}-p_{2\parallel})\,\delta^{(2)}(p_{2\parallel}-k_{2\parallel}-p_{1\parallel}).
\label{C.9full}
\end{align}

Since Eq.~(\ref{C.9full}) does not seem to admit further analytical simplification, we consider the inertial frame in which the external charged pion is at rest, namely $k_j^1=k_j^2=k_j^3=0$ and $k_j^2=(k_j^0)^2$, with $j=1,2$. In this frame, the expression to be evaluated becomes
\begin{align}
-i\Pi_{\pi_{\pm}}^{(C.9)} = & - \frac{8}{2\pi} g^{2} \sqrt{|q_{u}B| |q_{d}B|} (2\pi)^{4} N_{c} \sum_{f} \int d^{2}x_{\perp} \int d^{2}y_{\perp} e^{-\frac{\xi}{2}} e^{i\frac{B}{2}(y^{2}x^{1}-x^{2}y^{1})(q_{u}-q_{d})} e^{-\frac{\xi'}{2}} \nonumber \\
\times & \int \frac{d^{4}p_{1}}{(2\pi)^{4}} \int_{0}^{\infty} \frac{ds}{\cos(|q_{u}B|s)} e^{i\big(p_{1\perp}\cdot(x-y)_{\perp}\big)} e^{is\big(p_{1\parallel}^{2} - p_{1\perp}^{2} \frac{\tan(|q_{u}B|s)}{|q_{u}B|s} - m_{f}^{2} + i\epsilon\big)} \nonumber \\
\times & \int \frac{d^{4}p_{2}}{(2\pi)^{4}} \int_{0}^{\infty} \frac{ds'}{\cos(|q_{d}B|s')} e^{i \left[p_{2\perp}\cdot(y-x)_{\perp}\right]
} e^{is'\left[p_{2\parallel}^{2} - p_{2\perp}^{2} \frac{\tan(|q_{d}B|s')}{|q_{d}B|s'} - m_{f}^{2} + i\epsilon \right]} \nonumber \\
\times & \bigg[\big(m_{f}^{2} + p_{1\parallel}\cdot p_{2\parallel}\big)\big(\cos(|q_{u}B|s)\cos(|q_{d}B|s') + \sin(|q_{u}B|s)\sin(|q_{d}B|s')\big) \nonumber \\
+ & \frac{p_{1\perp}\cdot p_{2\perp}}{\cos(|q_{u}B|s)\cos(|q_{d}B|s')} \bigg] \delta(k_{1}^{0} + p_{1}^{0} -p_{2}^{0}) \delta(p_{2}^{0} - k_{2}^{0} -p_{1}^{0}) \delta(p_{1}^{3} - p_{2}^{3}) \delta(p_{2}^{3} - p_{1}^{3}).
\label{C.9zeroexternalmomentum}
\end{align}
From Eq.~(\ref{C.9zeroexternalmomentum}), the integrals over $x_\perp$ and $y_\perp$ can be computed component by component. The required expressions are
\begin{align}
    \int dy^{2} e^{ i\big( i\frac{|q_{d}B|}{4}(y^{2})^{2} + y^{2}\big[(p_{2}^{2} - p_{1}^{2})+\frac{q'B}{2}x^{1}\big]\big)} &= \sqrt{\frac{4\pi}{|q_{d}B|}} e^{-\frac{(p_{2}^{2}-p_{1}^{2})^{2}}{|q_{d}B|}} e^{-\frac{q'B}{|q_{d}B|} \big[ \frac{q'B}{4}(x^{1})^{2} + (p_{2}^{2} - p_{1}^{2})x^{1}\big]}, \nonumber \\
    \int dy^{1} e^{ i\big(i\frac{|q_{d}B|}{4}(y^{1})^{2} + y^{1}\big[(p_{2}^{1} - p_{1}^{1})+\frac{q'B}{2}x^{2}\big]\big)} &= \sqrt{\frac{4\pi}{|q_{d}B|}} e^{-\frac{(p_{2}^{1}-p_{1}^{1})^{2}}{|q_{d}B|}}e^{-\frac{q'B}{|q_{d}B|} \big[(p_{2}^{1} - p_{1}^{1})x^{2} - \frac{q'B}{4}(x^{2})^{2}\big]}, \nonumber \\
    \int dx^{2} e^{x^{2} \big[ \frac{q'B(p_{2}^{1} - p_{1}^{1})}{|q_{d}B|}+ i(p_{1}^{2} - p_{1}^{2})\big]} e^{-\frac{(x^{2})^{2}}{4}\big(|q_{u}B| + \frac{(q'B)^{2}}{|q_{d}B|}\big)} &= \sqrt{\frac{4\pi}{M}} e^{\frac{1}{M} \big[\frac{q'B(p_{2}^{1} - p_{1}^{1})}{|q_{d}B|}+ i(p_{1}^{2} - p_{1}^{2})\big]^{2}}, \nonumber \\
    \int dx^{1} e^{x^{1} \big[ i(p_{1}^{1} - p_{2}^{1}) - \frac{q'B(p_{2}^{2} - p_{1}^{2})}{|q_{d}B|}\big]} e^{-\frac{(x^{1})^{2}}{4}\big(|q_{u}B| + \frac{(q'B)^{2}}{|q_{d}B|}\big)} &= \sqrt{\frac{4\pi}{M}} e^{\frac{1}{M} \big[i(p_{1}^{1} - p_{2}^{1}) - \frac{q'B(p_{2}^{2} - p_{1}^{2})}{|q_{d}B|}\big]^{2}},
\end{align}
with $q'=q_u-q_d$ and $M = |q_{u}B| + \frac{(q'B)^{2}}{|q_{d}B|}$. After straightforward algebra, we obtain
\begin{align}
-i\Pi_{\pi_{\pm}}^{(C.9)} = & - \frac{8}{2\pi} g^{2} (2\pi)^{4} \sqrt{|q_{u}B| |q_{d}B|} N_{c} \sum_{f} \int \frac{d^{4}p_{1}}{(2\pi)^{4}} \int_{0}^{\infty} \frac{ds}{\cos(|q_{u}B|s)} e^{is\big(p_{1\parallel}^{2} - p_{1\perp}^{2} \frac{\tan(|q_{u}B|s)}{|q_{u}B|s} - m_{f}^{2} + i\epsilon \big)} \nonumber \\
\times & \int \frac{d^{4}p_{2}}{(2\pi)^{4}} \int_{0}^{\infty} \frac{ds'}{\cos(|q_{d}B|s')} e^{is'\big(p_{2\parallel}^{2} - p_{2\perp}^{2} \frac{\tan(|q_{d}B|s')}{|q_{d}B|s'} - m_{f}^{2} + i\epsilon\big)} \frac{4\pi}{|q_{d}B|} \frac{4\pi}{M} \nonumber \\
\times & e^{-\frac{(p_{2}^{1}-p_{1}^{1})^{2}}{|q_{d}B|}} e^{-\frac{(p_{2}^{2}-p_{1}^{2})^{2}}{|q_{d}B|}} e^{-\frac{1}{M} \left[\left(p_{2}^{1} - p_{1}^{1}\right)^2\left(1 - \frac{(q'B)^{2}}{|q_{d}B|^2}\right) + \left(p_{2}^{2} - p_{1}^{2}\right)^2\left(1 - \frac{(q'B)^{2}}{|q_{d}B|^2}\right)\right]} \nonumber \\
\times & \bigg[ \left(m_{f}^{2} + p_{1\parallel}\cdot p_{2\parallel}\right)\left(\cos(|q_{u}B|s)\cos(|q_{d}B|s') + \sin(|q_{u}B|s)\sin(|q_{d}B|s')\right) \nonumber \\
+ & \frac{p_{1\perp}p_{2\perp}}{\cos(|q_{u}B|s)\cos(|q_{d}B|s')} \bigg] \delta(k_{1}^{0} + p_{1}^{0} -p_{2}^{0}) \delta(p_{2}^{0} - k_{2}^{0} -p_{1}^{0}) \delta(p_{1}^{3} - p_{2}^{3}) \delta(p_{2}^{3} - p_{1}^{3}).
\end{align}
We next perform the integrations over the parallel components of one of the internal momenta. Choosing $p_1$, these integrations are direct because of the Dirac delta functions. We obtain
\begin{align}
-i\Pi_{\pi_{\pm}}^{(C.9)} = & -16g^{2} \sqrt{|q_{u}B| |q_{d}B|}\mathcal{V}_\parallel N_{c} \sum_{f} \int \frac{d^{4}p_{2}}{(2\pi)^{4}} \int d^{2}p_{1\perp} \int_{0}^{\infty} \frac{ds}{\cos(|q_{u}B|s)} e^{is\big((p_{2}^{0}-k_{2}^{0})^{2} - (p_{2}^{3})^{2} \big)} \nonumber \\
\times & e^{is\big(- p_{1\perp}^{2} \frac{\tan(|q_{u}B|s)}{|q_{u}B|s} - m_{f}^{2} + i\epsilon\big)} \int_{0}^{\infty} \frac{ds'}{\cos(|q_{d}B|s')} e^{is'\big(p_{2\parallel}^{2} - p_{2\perp}^{2} \frac{\tan(|q_{d}B|s')}{|q_{d}B|s'} - m_{f}^{2} + i\epsilon\big)} \frac{1}{|q_{d}B|M} \nonumber \\
\times & e^{-\frac{(p_{2}^{1}-p_{1}^{1})^{2}}{|q_{d}B|}} e^{-\frac{(p_{2}^{2}-p_{1}^{2})^{2}}{|q_{d}B|}} e^{-\frac{1}{M} \big[(p_{2}^{1} - p_{1}^{1})^2\big(1 - \frac{(q'B)^{2}}{|q_{d}B|^2}\big) + (p_{2}^{2} - p_{1}^{2})^2\big(1 - \frac{(q'B)^{2}}{|q_{d}B|^2}\big)\big]} \nonumber \\
\times & \bigg[ \big(m_{f}^{2} + (p_{2}^{0}-k_{2}^{0})p_{2}^{0} - (p_{2}^{3})^{2}\big)\big(\cos(|q_{u}B|s)\cos(|q_{d}B|s') + \sin(|q_{u}B|s)\sin(|q_{d}B|s')\big) \nonumber \\
+ & \frac{p_{1\perp}\cdot p_{2\perp}}{\cos(|q_{u}B|s)\cos(|q_{d}B|s')} \bigg] .
\end{align}
We compute the integral over $p_{1\perp}$ using the two master integrals
\begin{align}
    \int d^{2}p_{1\perp} e^{-i (p_{1\perp})^{2} \frac{\tan(|q_{u}B|s)}{|q_{u}B|}} e^{-(p_{2\perp} - p_{1\perp})^{2}Q} &= \frac{\pi}{R} e^{-(p_{2\perp})^{2}Q} e^{\frac{(p_{2\perp})^{2}Q^{2}}{R}}, \nonumber \\
    e^{-(p_{2\perp})^{2}Q} \int d^{2}p_{1\perp} \ p_{1\perp}\cdot p_{2\perp}e^{-(p_{1\perp})^{2}R} e^{2p_{1\perp}\cdot p_{2\perp}Q} &= \frac{\pi Q}{R^{2}} e^{-(p_{2\perp})^{2}Q} e^{\frac{(p_{2\perp})^{2}Q^{2}}{R}}(p_{2\perp})^{2},
\end{align}
with $R = \frac{i\tan(|q_{u}B|s)}{|q_{u}B|s} + Q$, $Q = \frac{1}{|q_{d}B|} + \frac{N}{M}$, $N = 1 - \frac{(q'B)^{2}}{|q_{d}B|^2}$ and $M = |q_{u}B| + \frac{(q'B)^{2}}{|q_{d}B|}$. The contribution $C.9$ becomes
\begin{align}
-i\Pi_{\pi_{\pm}}^{(C.9)} = & - 16 g^{2} \mathcal{V}_\parallel N_c \frac{\sqrt{|q_{u}B||q_{d}B|}}{|q_{d}B|M} \sum_{f} \int_{0}^{\infty} \frac{ds}{\cos(|q_{u}B|s)} \frac{\pi}{R} e^{is\big(- m_{f}^{2} + i\epsilon\big)}  \nonumber \\
\times & \int_{0}^{\infty} \frac{ds'}{\cos(|q_{d}B|s')} \int \frac{d^{4}p_{2}}{(2\pi)^{4}} e^{is'\big((p_{2}^{0})^{2} - (p_{2}^{3})^{2} - p_{2\perp}^{2} \frac{\tan(|q_{d}B|s')}{|q_{d}B|s'} - m_{f}^{2} + i\epsilon\big)} e^{-(p_{2\perp})^{2}Q} e^{\frac{(p_{2\perp})^{2}Q^{2}}{R}} e^{is\big((p_{2}^{0}-k_{2}^{0})^{2} - (p_{2}^{3})^{2} \big)} \nonumber \\
\times & \bigg[\big(m_{f}^{2} + (p_{2}^{0}-k_{2}^{0})p_{2}^{0} - (p_{2}^{3})^{2}\big)\big(\cos(|q_{u}B|s)\cos(|q_{d}B|s') + \sin(|q_{u}B|s)\sin(|q_{d}B|s')\big) \nonumber \\
-& \frac{Q}{R} \frac{(p_{2\perp})^{2}}{\cos(|q_{u}B|s)\cos(|q_{d}B|s')}\bigg].
\label{C.9justp_2integrals}
\end{align}
We proceed to compute the integral over the three-momentum $\vec{p}_2$, where we use the following results
\begin{align}
    \int d^{2}p_{2\perp} e^{-i (p_{2\perp})^{2} \frac{\tan(|q_{d}B|s')}{|q_{d}B|}} e^{-(p_{2\perp})^{2}Q} e^{\frac{(p_{2\perp})^{2}Q^{2}}{R}} &= \int d^{2}p_{1\perp} e^{- (p_{2\perp})^{2} Z}= \frac{\pi}{Z}, \nonumber \\
    \int d^{2}p_{2\perp} e^{-(p_{2\perp})^{2}Z} (p_{2\perp})^{2} &= \frac{\pi}{Z^{2}}, \nonumber \\
    \int dp_{2}^{3} e^{-i(p_{2}^{3})^{2}(s+s')} &= \sqrt{\frac{\pi}{i(s+s')}}, \nonumber \\
    \int dp_{2}^{3} e^{-i(p_{2}^{3})^{2}(s+s')} (p_{2}^{3})^{2} &= \frac{1}{2}\frac{\sqrt{\pi}}{(i(s+s'))^{3/2}},
    \label{masterintegralsforC.9}
\end{align}
with $Z=Q -\frac{Q^{2}}{R} +i \frac{\tan(|q_{d}B|s')}{|q_{d}B|}$. Using Eq.~(\ref{masterintegralsforC.9}) in Eq.~(\ref{C.9justp_2integrals}), we obtain
\begin{align}
-i\Pi_{\pi_{\pm}}^{(C.9)} = & - 16 g^{2} \mathcal{V}_\parallel N_c \frac{\sqrt{|q_{u}B||q_{d}B|}}{|q_{d}B|M} \frac{1}{(2\pi)^{3}}  \sum_{f} \int_{0}^{\infty} \frac{ds}{\cos(|q_{u}B|s)} \frac{\pi}{R}  e^{is\big( - m_{f}^{2} + i\epsilon \big)}  \nonumber \\
\times & \int_{0}^{\infty} \frac{ds'}{\cos(|q_{d}B|s')} \int \frac{dp_2^0}{2\pi} e^{is'\big( - m_{f}^{2} + i\epsilon \big)} e^{is'(p_{2}^{0})^{2}} e^{is(p_{2}^{0}-k_{2}^{0})^{2}}\frac{\pi}{Z}\sqrt{\frac{\pi}{i(s+s')}} \nonumber \\
\times & \bigg[ \Big(m_{f}^{2} + (p_{2}^{0}-k_{2}^{0})p_{2}^{0} - \frac{1}{2}\frac{1}{(i(s+s'))^{3/2}}\Big)\big(\cos(|q_{u}B|s)\cos(|q_{d}B|s') + \sin(|q_{u}B|s)\sin(|q_{d}B|s')\big) \nonumber \\
- & \frac{Q}{ZR} \frac{1}{\cos(|q_{u}B|s)\cos(|q_{d}B|s')} \bigg].
\end{align}
We now implement the analytic continuations \(s \rightarrow -i\tau\) and \(s' \rightarrow -i\tau'\), and through a Wick rotation in $p_2^0$, we write this contribution to the charged pion self-energy in the Matsubara formalism
\begin{align}
-i\Pi_{\pi_{\pm}}^{(C.9)} = &i \frac{\sqrt{\pi} g^{2} \mathcal{V}_\parallel N_c}{\pi^2} \frac{\sqrt{|q_{u}B||q_{d}B|}}{|q_{d}B|M} \sum_{f} T \int_{0}^{\infty} \frac{d\tau}{\cosh(|q_{u}B|\tau)}\int_{0}^{\infty} \frac{ds'}{\cos(|q_{d}B|s')} \nonumber \\
\times & \frac{1}{ZR} \frac{1}{\sqrt{\tau + \tau'}} e^{-m_{f}^{2}(\tau+\tau')} e^{-\tau\omega^{2}} \bigg\{ \Big(\cosh(|q_{u}B|\tau)\cosh(|q_{d}B|\tau') - \sinh(|q_{u}B|\tau)\sinh(|q_{d}B|\tau')\Big) \nonumber \\
\times & \bigg[ \bigg(m_{f}^{2} - \frac{1}{2(\tau + \tau')}\bigg) \sum_{n} e^{-[(\tau + \tau')\tilde{\omega}_{n}^{2} - 2\omega\tilde{\omega}_{n}\tau]} + \sum_{n} (\tilde{\omega}_{n} - \omega)\tilde{\omega}_{n} e^{-[(\tau + \tau')\tilde{\omega}_{n}^{2} - 2\omega\tilde{\omega}_{n}\tau]} \bigg] \nonumber \\
- & \frac{Q}{ZR} \frac{1}{\cosh(|q_{u}B|\tau)\cosh(|q_{d}B|\tau')} \sum_{n} e^{-[(\tau + \tau') \tilde{\omega}_{n}^{2} - 2\omega\tilde{\omega}_{n}\tau]} \bigg\}.
\label{C.9justsum}
\end{align}
The required sums over Matsubara frequencies are
\begin{align}
    e^{-\tau \omega^{2}}\sum_{n=-\infty}^{\infty} e^{-[(\tau + \tau')\tilde{\omega}_{n}^{2} - 2\omega\tilde{\omega}_{n}\tau]} = & \frac{1}{2T\sqrt{\pi(\tau+\tau')}} e^{-\frac{\tau\tau' \omega^{2}}{\tau + \tau'}}\bigg[ 1 + 2\sum_{n=1}^{\infty}(-1)^n e^{-\frac{n^{2}}{4(\tau + \tau')T^{2}}} \cos\left(\frac{n\omega\tau}{(\tau+\tau')T}\right)\bigg], \nonumber \\
    e^{-\tau\omega^{2}}\sum_{n=-\infty}^{\infty} \tilde{\omega}_{n} e^{-[(\tau + \tau')\tilde{\omega}_{n}^{2} - 2\omega\tilde{\omega}_{n}\tau]} = & \frac{1}{2T\sqrt{\pi(\tau+\tau')}} e^{-\frac{\tau\tau' \omega^{2}}{\tau + \tau'}} \bigg\{ \frac{\tau \omega}{\tau + \tau'} + 2\sum_{n=1}^{\infty} (-1)^{n}e^{-\frac{n^{2}}{4(\tau + \tau')T^{2}}} \nonumber \\
& \bigg[\frac{\tau \omega}{\tau + \tau'} \cos\Big(\frac{n\omega\tau}{(\tau+\tau')T}\Big) - \frac{n}{2(\tau+\tau')T}\sin\Big(\frac{n\omega\tau}{(\tau+\tau')T}\Big)\bigg]\bigg\},\nonumber \\
e^{-\tau\omega^{2}}\sum_{n=-\infty}^{\infty} \tilde{\omega}_{n}^{2} e^{-\left[(\tau + \tau')\tilde{\omega}_{n}^{2} - 2\omega\tilde{\omega}_{n}\tau\right]} = & \frac{1}{2T\sqrt{\pi(\tau+\tau')}} e^{-\frac{\tau\tau' \omega^{2}}{\tau + \tau'}} \bigg\{ \frac{1}{2(\tau + \tau')} + \frac{\tau^{2} \omega^{2}}{(\tau + \tau')^{2}} + 2\sum_{n=1}^{\infty} (-1)^{n}e^{-\frac{n^{2}}{4(\tau + \tau')T^{2}}} \nonumber\\
& \bigg[\Big(\frac{1}{2(\tau + \tau')} + \frac{\tau^{2} \omega^{2}}{(\tau + \tau')^{2}} -\frac{n^{2}}{4(\tau + \tau')^{2}T^{2}}\Big) \cos\Big(\frac{n\omega\tau}{(\tau+\tau')T}\Big) \nonumber \\
&  - \frac{n\tau \omega}{(\tau+\tau')T}\sin\Big(\frac{n\omega\tau}{(\tau+\tau')T}\Big)\bigg]\bigg\}.
\label{C.9sums}
\end{align}
Substituting Eq.~(\ref{C.9sums}) into Eq.~(\ref{C.9justsum}) and performing the change of variables \(\tau = u(1-v)\) and \(\tau' = uv\), we obtain the final expression in momentum space
\begin{align}
-i\Pi_{\pi_{\pm}}^{(C.9)} = & \frac{ig^{2}}{\pi^{2}} \frac{\sqrt{|q_{u}B||q_{d}B|}}{|q_{d}B|M}  N_{c} \sum_{f} \int_{0}^{\infty} du \int_{0}^{1} dv \frac{1}{\cosh(|q_{u}B|u(1-v))} \nonumber \\
\times &  \frac{1}{\cosh(|q_{d}B|uv)}\frac{1}{ZR} e^{-m_{f}^{2}u} e^{uv(1-v)(i\omega)^{2}} \Biggl\{ \big(\cosh(|q_{u}B|u(1-v))\cosh(|q_{d}B|uv)  \nonumber \\
- & \ \sinh(|q_{u}B|u(1-v))\sinh(|q_{d}B|uv)\big)\Biggl[ \biggl(m_{f}^{2} - \frac{1}{2u}\biggr) \biggl[ 1 + 2\sum_{n=1}^{\infty}(-1)^n e^{-\frac{n^{2}}{4T^{2}u}} \cosh\Big(\frac{(1-v)n(i\omega)}{T}\Big)\biggr] \nonumber \\
- & \ (i\omega) \biggl[(1-v)(i\omega) + 2\sum_{n=1}^{\infty} (-1)^{n}e^{-\frac{n^{2}}{4T^{2}u}} \biggl((1-v)(i\omega) \cosh\Big(\frac{(1-v)n (i\omega) }{T}\Big) \nonumber \\
- & \ \frac{n}{2Tu}\sinh\Big(\frac{(1-v)n(i\omega) }{T}\Big)\bigg) \biggr] -\Biggl( \frac{1}{2u} - (1-v)^{2}(i\omega)^{2} + 2 \sum_{n=1}^{\infty} (-1)^{n} e^{-\frac{n^{2}}{4T^{2}u}} \nonumber \\
\times & \biggl[\left(\frac{1}{2u} - (1-v)^{2}(i\omega)^{2} -\frac{n^{2}}{4u^{2}T^{2}}\right) \cosh\Big(\frac{(1-v)n(i\omega)}{T}\Big) + \frac{(1-v)n(i\omega)}{uT} \sinh\Big(\frac{(1-v)n (i\omega) }{T}\Big) \biggr] \Biggr) \Biggr] \nonumber \\
- & \frac{Q}{ZR} \frac{1}{\cosh(|q_{u}B|u(1-v))\cosh(|q_{d}B|uv)} \biggl[ 1 + 2\sum_{n=1}^{\infty} (-1)^n e^{-\frac{n^{2}}{4T^{2}u}} \cosh\Big(\frac{(1-v)n(i\omega)}{T}\Big)\biggr] \Biggr\},
\end{align}
with
\begin{align}
    ZR &= \frac{\tanh(\vert q_{d}B\vert uv)\tanh(\vert q_{u}B\vert u(1-v))}{\vert q_{d}B\vert \vert q_{u} B\vert} + \left(\frac{\tanh(\vert q_{d}B\vert uv)}{\vert q_{d}B\vert}+ \frac{\tanh(\vert q_{u}B\vert u(1-v)}{\vert q_{u} B\vert}\right) Q, \nonumber \\
    Q&= \frac{1}{\vert q_{d}B\vert} \left(1+ \frac{\vert q_{d}B\vert^{2}-[(q_{u}-q_{d})B]^{2}}{\vert q_{u} B\vert \vert q_{d}B\vert +[(q_{u}-q_{d})B]^{2}} \right).
\end{align}
The temperature-independent terms contain the vacuum contribution and are ultraviolet divergent. In order to isolate this piece, we take the limit $q_fB\rightarrow 0$, obtaining
\begin{equation}
-i\Pi_{\pi_{\pm},B=0}^{(C.9)} = -\frac{\sqrt{2}}{3\pi^{2}} ig^{2} N_{c} \sum_{f} \int_{0}^{\infty} du \int_{0}^{1} dv e^{-m_{f}^{2}u} e^{uv(1-v)(i\omega)^{2}} \frac{1}{u} \biggl(m_{f}^{2} -(i\omega)^{2}v(1-v) -\frac{2}{u} \biggr).
\end{equation}

This completes the one-loop calculation of the charged pion self-energy. 
\end{widetext}

\subsection{Quark one-loop self-energy\label{sec3.4}}

\begin{figure}[h]
    \centering
    \includegraphics[scale=0.47]{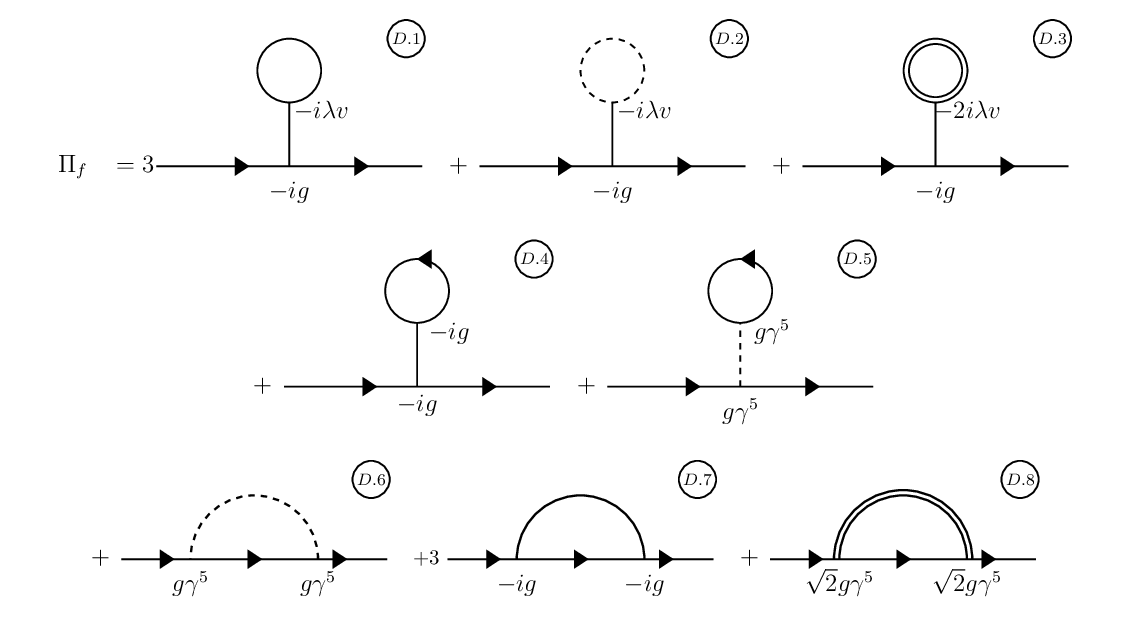}
    \caption{One-loop self-energy diagrams for the quark fields (up and down). Solid lines represent the sigma meson, dashed lines correspond to neutral pions, double solid lines denote charged pions, and solid lines with arrows represent quark fields. Each vertex explicitly displays the corresponding coupling coefficient. The coefficients multiplying each diagram are the corresponding combinatorial factors.}
    \label{fig6}
\end{figure}

We now turn to the one-loop self-energy of the quark fields. The corresponding Feynman diagrams are shown in Fig.~\ref{fig6} and labeled $D.j$, with $j=1,\ldots,8$. The full one-loop self-energy can be written as
\begin{align}
-i\Pi_{f} &= -i\Pi_{f}^{(D.1)}  -i\Pi_{f}^{(D.2)} -i\Pi_{f}^{(D.3)} \nonumber \\
&-i\Pi_{f}^{(D.4)} -i\Pi_{f}^{(D.5)} -i\Pi_{f}^{(D.6)} \nonumber \\
&-i\Pi_{f}^{(D.7)} -i\Pi_{f}^{(D.8)}.
\end{align}

The first five contributions correspond to tadpole-type diagrams with an additional internal line evaluated at zero momentum. These contributions need not be recomputed explicitly, since they follow directly from the corresponding results in previous subsections (see contributions $A.4$--$A.7$, $B.4$--$B.7$, and $C.4$--$C.7$) upon the appropriate replacement of symmetry factors, couplings, and internal fields. We thus obtain the following contributions
\begin{equation}
-i\Pi_{f}^{(D.1)} = \frac{3i\lambda v g}{m_{\sigma}^{2}}\,\int \frac{d^{3}q}{(2\pi)^{3}} \frac{1}{2E_{\sigma}} \big(1 + 2n_{B}(E_{\sigma})\big),
\label{D.1final}
\end{equation}
\begin{equation}
-i\Pi_{f}^{(D.2)} = \frac{i\lambda v g}{m_{\sigma}^{2}}\,\int \frac{d^{3}q}{(2\pi)^{3}} \frac{1}{2E_{\pi}} \big( 1 + 2n_{B}(E_{\pi}) \big),
\label{D.2final}
\end{equation}
\begin{align}
-i\Pi_{f}^{(D.3)} = & \frac{ig\lambda v}{8\pi^{2}m_{\sigma}^{2}} \ |eB|  \int_{0}^{\infty} \frac{d\tau}{\sinh\left(|eB| \tau\right)} \frac{e^{-\tau m_{\pi}^{2}}} {\tau} \nonumber \\
\times & \bigg(1 + 2\sum_{n=1}^{\infty} e^{-\frac{n^{2}}{4 T^{2}\tau}} \bigg),
\label{D.3final}
\end{align}
\begin{align}
-i\Pi_{f}^{(D.4)} = & - \frac{ig^{2}}{4\pi^{2}m_{\sigma}^{2}} |q_{f}B| N_{c} \sum_{f}m_{f}  \int_{0}^{\infty} \frac{d\tau}{\tanh(|q_{f}B| \tau)} \nonumber \\
\times &\frac{e^{-\tau m_{f}^{2}}}{\tau} \bigg(1 + 2\sum_{n=1}^{\infty} (-1)^{n} e^{-\frac{n^{2}}{4 T^{2}\tau}} \bigg),
\label{D.4final}
\end{align}
and 
\begin{align}
-i\Pi_{f}^{(D.5)} = & \frac{i g^{2}}{4\pi^{2}m_{\pi}^{2}} |q_{f}B| N_{c} \sum_{f}m_{f}  \int_{0}^{\infty} \frac{d\tau}{\tanh(|q_{f}B| \tau)} \nonumber \\
\times & \frac{e^{-\tau m_{f}^{2}}}{\tau} \bigg(1 + 2\sum_{n=1}^{\infty} (-1)^{n} e^{-\frac{n^{2}}{4 T^{2}\tau}} \bigg).
\label{D.5final}
\end{align}

In Eqs.~(\ref{D.1final}) and~(\ref{D.2final}) $n_B(X)$ is the Bose-Einstein distribution function, with $E_\sigma=\sqrt{q^2+m_\sigma^2}$ and $E_\pi=\sqrt{q^2+m_\pi^2}$. In Eqs.~(\ref{D.3final})--(\ref{D.5final}), the purely magnetic terms contain ultraviolet-divergent vacuum contributions, which can be identified by taking the limit $eB \rightarrow 0$ 
\begin{equation}
    -i\Pi_{f,B=0}^{(D.3)} =\frac{ig\lambda v}{8\pi^{2}m_{\sigma}^{2}}   \int_{0}^{\infty} \frac{d\tau}{\tau^{2}} e^{-\tau m_{\pi}^{2}},
    \label{D.3vacuum}
\end{equation}
\begin{equation}
    -i\Pi_{f,B=0}^{(D.4)} =-\frac{ig^{2}}{4\pi^{2}m_{\sigma}^{2}}  N_{c} \sum_{f}m_{f} \int_{0}^{\infty} d\tau \frac{e^{-\tau m_{f}^{2}}}{\tau^{2}},
\end{equation}
and
\begin{equation}
    -i\Pi_{f,B=0}^{(D.5)} =\frac{i g^{2}}{4\pi^{2}m_{\pi}^{2}} N_{c} \sum_{f}m_{f} \int_{0}^{\infty} d\tau \frac{e^{-\tau m_{f}^{2}}}{\tau^{2}}.
\end{equation}

The last three contributions, $D.6$, $D.7$, and $D.8$, require a different treatment. In these cases, the Schwinger phases associated with the propagators do not cancel. Therefore, the computation must be carried out in coordinate space using the Ritus formalism. To the best of our knowledge, this is the first time that these contributions have been computed within the LSMq including thermomagnetic effects without imposing any restriction on $T$ or $|eB|$. We therefore present the derivation in detail.

We begin with the expression for the self-energy contribution corresponding to a loop formed by a neutral pion and a quark, which must be evaluated in space-time coordinates
\begin{align}
-i\Pi_{f}^{(D.6)} = g^{2} \int d^{4}x \int & d^{4}y  \; \overline{U}(x,k_{1}) \gamma^{5} S^{B}(x-y) \nonumber \\
& \times \gamma^{5} D(y-x) U(y,k_{2}),
\label{D.6initial}
\end{align}
where the Ritus functions for fermions are
\begin{equation}
U(y,k_{2}) = E(y,k_{2}) u(k_{2}), \quad \bar{U}(x,k_{1})=U^{\dagger}(x,k_{1})\gamma^{0},
\label{spinorsRitus}
\end{equation}
with
\begin{align}
E(x,k_{1}) &= \sum_{\lambda=\pm} \Gamma^{\lambda} F(x,k_{1\lambda}), \nonumber \\ 
u(k_{1}) &= \frac{1}{\sqrt{2(E_{f}+m_{f})}} \Big[ \slashed{\Pi}_{s} (E_{f}, k_{1}) + m_{f} \mathbb{I}\Big] 
\begin{pmatrix}
    1 \\ 0 \\ -1 \\ 0
\end{pmatrix} , \nonumber \\
\Pi_{s}^{\mu} &= \bigg( k_{1}^{0}, \frac{\text{s}_{f}}{2}\sqrt{2\ell q_{f} B}, -\frac{\text{s}_{f}}{2}\sqrt{2\ell q_{f} B},k_{1}^{3}\bigg), \nonumber \\
E_{f} &= \sqrt{m_{f}^{2} +2\ell q_{f} B + (k_{1}^{3})^{2}},
\end{align}
and 
\begin{equation}
k_{1\lambda} = ( k_{1}^{0},k_{1\text{s}_{f}\lambda}^{1},k_{1}^{2},k_{1}^{3} ).
\end{equation}
In this notation $\lambda$ can have the values $\pm1$, therefore the component one of $k_{1\lambda}$ and the factor $\Gamma^{\lambda}$ can be written as
\begin{align}
    k^1_{1\text{s}_{f}\pm} &= k_{1}^{1} - \frac{(1\mp \text{s}_{f})}{2}, \nonumber \\
    \Gamma^{\lambda} &= \frac{1}{2} \Big(1 \pm i \gamma^{1}\gamma^{2}\Big),
\end{align}
respectively; with  $s_f=\text{sgn}(q_fB)$.
The scalar Ritus functions in terms of these dependencies is
\begin{widetext}
\begin{equation}
F(x,k_{1\lambda}) = \frac{1}{\sqrt{2\pi}} e^{-i(k_{1}^{0}x^{0} - k_{1}^{3}x^{3})}e^{-i \text{s}_{f}(k_{1\text{s}_{f}\pm}^{1} -k_{1}^{2})\theta} N_{k_{1\text{s}_{f}\pm}^{1},k_{1}^{2}}\, \xi^{(k_{1\text{s}_{f}\pm}^{1} -k_{1}^{2})/2} e^{-\frac{\xi}{2}}L_{k_{1}^{2}}^{k_{1\text{s}_{f}\pm}^{1} -k_{1}^{2}} (\xi),
\end{equation}
where
\begin{equation}
\xi = \frac{|q_{f}B|}{2}r^{2}, \quad N_{k_{1\text{s}_{f}\pm}^{1},k_{1}^{2}}= \bigg( |q_{f}B| \frac{k_{1}^{2}!}{k_{1\text{s}_{f}\pm}^{1}!} \bigg)^{1/2}.
\label{notationspinorsRitus}
\end{equation}
The coordinate-space propagators for the neutral pion and the quark are
\begin{equation}
D(y-x)  = \int \frac{d^{4}p_{1}}{(2\pi)^{4}} e^{-ip_{1}\cdot(y-x)} \frac{i}{p_{1}^{2}-m_{\pi}^{2}+i\epsilon},
\end{equation}
\begin{align}
S^{B}(x-y) = & \ e^{iq_{f}\int_{x}^{y} d\xi'_{\mu} \big[A^{\mu}(\xi')+\frac{1}{2}F^{\mu\,\nu}(\xi'-y)_{\nu}\big]} \int \frac{d^{4}p_{2}}{(2\pi)^{4}} e^{-ip_{2}\cdot(x-y)}  \int_{0}^{\infty} \frac{ds}{\cos(|q_{f}B| s)} e^{is\big(p_{2 \parallel}^{2}-p_{2 \perp}^{2} \frac{\tan(|q_{f}B| s)}{|q_{f}B|} -m_{f}^{2} +i\epsilon\big)} \nonumber \\
\times & \bigg[ \big(\cos(|q_{f}B| s) + \text{s}_{f}\gamma^{1}\gamma^{2}\sin(|q_{f}B| s)\big)\big(m_{f}+\slashed{p}_{2\parallel}\big) - \frac{\slashed{p}_{2\perp}}{\cos(|q_{f}B| s)} \bigg],
\end{align}
respectively. With all these elements specified, the explicit initial expression for the contribution $D.6$ can be written as
\begin{align}
-i\Pi_{f}^{(D.6)} = & \ g^{2} \int d^{4}x \int d^{4}y  \bigg[ \frac{1}{2} \Big(1 - i \gamma^{1}\gamma^{2}\Big) \frac{1}{\sqrt{2\pi}} e^{i(k_{1}^{0}x^{0} - k_{1}^{3}x^{3})}e^{i\text{s}_{f}(k_{1\text{s}_{f}+}^{1} -k_{1}^{2})\theta} N_{k_{1\text{s}_{f}+}^{1},k_{1}^{2}}\, \xi^{(k_{1\text{s}_{f}+}^{1} -k_{1}^{2})/2} e^{-\frac{\xi}{2}}L_{k_{1}^{2}}^{k_{1\text{s}_{f}+}^{1} -k_{1}^{2}} (\xi)\nonumber \\
+ & \frac{1}{2} \Big(1 + i \gamma^{1}\gamma^{2}\Big) \frac{1}{\sqrt{2\pi}} e^{i(k_{1}^{0}x^{0} - k_{1}^{3}x^{3})}e^{i\text{s}_{f}(k_{1\text{s}_{f}-}^{1} -k_{1}^{2})\theta} N_{k_{1\text{s}_{f}-}^{1},k_{1}^{2}}\, \xi^{(k_{1\text{s}_{f}-}^{1} -k_{1}^{2})/2} e^{-\frac{\xi}{2}}L_{k_{1}^{2}}^{k_{1\text{s}_{f}-}^{1} -k_{1}^{2}} (\xi) \bigg] \nonumber \\
\times & \frac{1}{\sqrt{2(E_{f}+m_{f})}} 
\begin{pmatrix}
    1, \ 0, \ 1, \ 0
\end{pmatrix} 
\bigg( \gamma^{0}k_{1}^{0} + \frac{\gamma^{1}\text{s}_{f}}{2}\sqrt{2\ell q_{f}B} - \frac{\gamma^{2}\text{s}_{f}}{2}\sqrt{2\ell q_{f}B} + \gamma^{3}k_{1}^{3} + m_{f} \mathbb{I}\bigg)\nonumber \\
\times & \gamma^{5} e^{iq_{f}\int_{x}^{y} d\xi'_{\mu} \big[A^{\mu}(\xi')+\frac{1}{2}F^{\mu\,\nu}(\xi'-y)_{\nu}\big]}  \int\frac{d^{4}p_{2}}{(2\pi)^{4}} e^{-i p_{2} \cdot (x-y)} \int_{0}^{\infty} \frac{ds}{\cos(|q_{f}B|s)}e^{is\big(p_{2 \parallel}^{2}-p_{2 \perp}^{2} \frac{\tan(|q_{f}B| s)}{|q_{f}B|} -m_{f}^{2} +i\epsilon\big)} \nonumber \\
\times & \bigg[\big(\cos(|q_{f}B|s)+ \text{s}_{f}\gamma^{1}\gamma^{2} \sin (|q_{f}B| s)\big)\big(m_{f} + \slashed{p}_{2 \parallel}\big)-\frac{\slashed{p}_{2 \perp}}{\cos(|q_{f}B| s)} \bigg] \gamma^{5} \int\frac{d^{4}p_{1}}{(2\pi)^{4}} \frac{ie^{-ip_{1}\cdot(y-x)}}{p_{1}^2-m_{\pi}^{2}+i\epsilon} \nonumber \\
\times & \bigg[ \frac{1}{2} \Big(1 + i \gamma^{1}\gamma^{2}\Big) \frac{1}{\sqrt{2\pi}} e^{-i(k_{2}^{0}y^{0} - k_{2}^{3}y^{3})}e^{-i\text{s}_{f}(k_{2s+}^{1} -k_{2}^{2})\theta} N_{k_{2\text{s}_{f}+}^{1},k_{2}^{2}}\, \xi'^{(k_{2\text{s}_{f}+}^{1} -k_{2}^{2})/2} e^{-\frac{\xi'}{2}}L_{k_{1}^{2}}^{k_{1\text{s}_{f}+}^{1} -k_{1}^{2}} (\xi')\nonumber \\
+ & \frac{1}{2} \Big(1 - i \gamma^{1}\gamma^{2}\Big) \frac{1}{\sqrt{2\pi}} e^{-i(k_{2}^{0}y^{0} - k_{2}^{3}y^{3})}e^{-i\text{s}_{f}(k_{2s-}^{1} -k_{2}^{2})\theta} N_{k_{2\text{s}_{f}-}^{1},k_{2}^{2}}\, \xi'^{(k_{2\text{s}_{f}-}^{1} -k_{2}^{2})/2} e^{-\frac{\xi'}{2}}L_{k_{2}^{2}}^{k_{2\text{s}_{f}-}^{1} -k_{2}^{2}}(\xi') \bigg] \nonumber \\
\times & \frac{1}{\sqrt{2(E_{f}+m_{f})}} \bigg( \gamma^{0}k_{2}^{0} - \frac{\gamma^{1}\text{s}_{f}}{2}\sqrt{2\ell q_{f}B} + \frac{\gamma^{2}\text{s}_{f}}{2}\sqrt{2\ell q_{f}B} - \gamma^{3}k_{2}^{3} + m_{f} \mathbb{I}\bigg)
\begin{pmatrix}
    1 \\ 0 \\ -1 \\ 0
\end{pmatrix}.
\label{D.6explicitinitial}
\end{align}
We now simplify this expression. First, we commute one of the $\gamma^5$ matrices through the Dirac matrices appearing in the quark propagator and then use $(\gamma^5)^2=\mathbb{I}$. Then, we compute the integral within the Schwinger phase in the same fashion we did before, getting
\begin{equation}
e^{iq_{f}\int_{x}^{y} d\xi'_{\mu} \big[ A^{\mu}(\xi') + \frac{1}{2}F^{\mu\nu}(\xi'-y)_{\nu} \big]} = e^{i\frac{q_{f}B}{2}(x^{1}y^{2}-x^{2}y^{1})}.
\label{Schwingerphaseevaluation}
\end{equation}
We compute the integrals over $x_\parallel$ and $y_\parallel$
\begin{align}
&\int d^{4}x \int d^{4}y \ e^{i(k_{1}^{0}x^{0} - k_{1}^{3}x^{3})} e^{-ip_{1}\cdot(x-y)} e^{-ip_{2}\cdot(y-x)} e^{-i(k_{2}^{0}y^{0} - k_{2}^{3}y^{3})}= \nonumber \\
&\int d^{2}x_{\perp} \int d^{2}y_{\perp} e^{-ip_{1\perp}\cdot(y-x)_{\perp}} e^{-ip_{2\perp}\cdot(x-y)_{\perp}} (2\pi)^{4} \delta^{(2)}(p_{2\parallel} - k_{1\parallel} - p_{1\parallel}) \delta^{(2)}(k_{2\parallel} + p_{1\parallel} - p_{2\parallel}).
\end{align}
After performing these steps, the expression for $D.6$ can be written
\begin{align}
-i\Pi_{f}^{(D.6)} = & \frac{g^{2}}{2\pi} \int d^{2}x_{\perp} \int d^{2}y_{\perp}  \bigg[ \frac{1}{2} \Big(1 - i \gamma^{1}\gamma^{2}\Big) e^{i\text{s}_{f}(k_{1\text{s}_{f}+}^{1} -k_{1}^{2})\theta} N_{k_{1\text{s}_{f}+}^{1},k_{1}^{2}}\, \xi^{(k_{1\text{s}_{f}+}^{1} -k_{1}^{2})/2} e^{-\frac{\xi}{2}}L_{k_{1}^{2}}^{k_{1\text{s}_{f}+}^{1} -k_{1}^{2}} (\xi)\nonumber \\
+ & \frac{1}{2} \Big(1 + i \gamma^{1}\gamma^{2}\Big) e^{i\text{s}_{f}(k_{1\text{s}_{f}-}^{1} -k_{1}^{2})\theta} N_{k_{1\text{s}_{f}-}^{1},k_{1}^{2}}\, \xi^{(k_{1\text{s}_{f}-}^{1} -k_{1}^{2})/2} e^{-\frac{\xi}{2}}L_{k_{1}^{2}}^{k_{1\text{s}_{f}-}^{1} -k_{1}^{2}} (\xi) \bigg] \nonumber \\
\times & \frac{1}{\sqrt{2(E_{f}+m_{f})}} 
\begin{pmatrix}
    1, \ 0, \ 1, \ 0
\end{pmatrix} 
\bigg( \gamma^{0}k_{1}^{0} + \frac{\gamma^{1}\text{s}_{f}}{2}\sqrt{2\ell q_{f}B} - \frac{\gamma^{2}\text{s}_{f}}{2}\sqrt{2\ell q_{f}B} + \gamma^{3}k_{1}^{3} + m_{f} \mathbb{I}\bigg)\nonumber \\
\times & e^{\frac{iq_{f}}{2}(x^{1}y^{2}-x^{2}y^{1})}  \int\frac{d^{4}p_{2}}{(2\pi)^{4}} e^{-i p_{2\perp} \cdot (x-y)_{\perp}} \int_{0}^{\infty} \frac{ds}{\cos(|q_{f}B|s)}e^{is\big(p_{2 \parallel}^{2}-p_{2 \perp}^{2} \frac{\tan(|q_{f}B| s)}{|q_{f}B|} -m_{f}^{2} +i\epsilon\big)} \nonumber \\
\times & \bigg[\big(\cos(|q_{f}B|s)+ \text{s}_{f}\gamma^{1}\gamma^{2} \sin (|q_{f}B| s)\big)\big(m_{f} - \slashed{p}_{2 \parallel}\big) +\frac{\slashed{p}_{2 \perp}}{\cos(|q_{f}B| s)} \bigg] \int\frac{d^{4}p_{1}}{(2\pi)^{4}} \frac{ie^{-ip_{1\perp}\cdot(y-x)_{\perp}}}{p_{1}^2-m_{\pi}^{2}+i\epsilon} \nonumber \\
\times & \bigg[ \frac{1}{2} \Big(1 + i \gamma^{1}\gamma^{2}\Big) e^{-i\text{s}_{f}(k_{2s+}^{1} -k_{2}^{2})\theta'} N_{k_{2\text{s}_{f}+}^{1},k_{2}^{2}}\, \xi'^{(k_{2\text{s}_{f}+}^{1} -k_{2}^{2})/2} e^{-\frac{\xi'}{2}}L_{k_{1}^{2}}^{k_{1\text{s}_{f}+}^{1} -k_{1}^{2}} (\xi')\nonumber \\
+ & \frac{1}{2} \Big(1 - i \gamma^{1}\gamma^{2}\Big) e^{-i\text{s}_{f}(k_{2s-}^{1} -k_{2}^{2})\theta'} N_{k_{2\text{s}_{f}-}^{1},k_{2}^{2}}\, \xi'^{(k_{2\text{s}_{f}-}^{1} -k_{2}^{2})/2} e^{-\frac{\xi'}{2}}L_{k_{2}^{2}}^{k_{2\text{s}_{f}-}^{1} -k_{2}^{2}}(\xi') \bigg] \nonumber \\
\times & \frac{1}{\sqrt{2(E_{f}+m_{f})}} \bigg(\gamma^{0}k_{2}^{0} - \frac{\gamma^{1}\text{s}_{f}}{2}\sqrt{2\ell q_{f}B} + \frac{\gamma^{2}\text{s}_{f}}{2}\sqrt{2\ell q_{f}B} - \gamma^{3}k_{2}^{3} + m_{f} \mathbb{I}\bigg) 
\begin{pmatrix}
    1 \\ 0 \\ -1 \\ 0
\end{pmatrix}\nonumber \\
\times & (2\pi)^{4} \delta^{(2)}(p_{2\parallel} - k_{1\parallel} - p_{1\parallel}) \delta^{(2)}(k_{2\parallel} + p_{1\parallel} - p_{2\parallel}).
\end{align}
We now focus on the integrations over $x_\perp$ and $y_\perp$. For this purpose, we use
\begin{align}
\xi &= \frac{|q_{f}B|}{2} r^{2} =\frac{|q_{f}B|}{2} x_\perp^{2}= \frac{|q_{f}B|}{2} ((x^{1})^{2} + (x^{2})^{2}), \nonumber \\
\xi' &= \frac{|q_{f}B|}{2}r'^{2}=\frac{|q_{f}B|}{2}y_\perp^{2} = \frac{|q_{f}B|}{2}((y^{1})^{2} + (y^{2})^{2}).
\end{align}
Introducing polar coordinates, $x^{1}=r\cos\theta$, $x^{2}=r\sin\theta$, $y^{1}=r'\cos\theta'$, and $y^{2}=r'\sin\theta'$, the angular integrations can be performed. The result for $D.6$ is
\begin{align}
-i\Pi_{f}^{(D.6)} = & \ \frac{g^{2}}{2\pi} \int d^{4}x_{\perp} \int d^{4}y_{\perp} \bigg[ \frac{1}{2} \Big(1 - i \gamma^{1}\gamma^{2}\Big) N_{k_{1\text{s}_{f}+}^{1},k_{1}^{2}}\, + \frac{1}{2} N_{k_{1\text{s}_{f}-}^{1},k_{1}^{2}}\, \bigg] \nonumber \\
\times & \frac{1}{\sqrt{2(E_{f}+m_{f})}} 
\begin{pmatrix}
    1, \ 0, \ 1, \ 0
\end{pmatrix} 
\bigg( \gamma^{0}k_{1}^{0} + \frac{\gamma^{1}\text{s}_{f}}{2}\sqrt{2\ell q_{f}B} - \frac{\gamma^{2}\text{s}_{f}}{2}\sqrt{2\ell q_{f}B} + \gamma^{3}k_{1}^{3} + m_{f} \mathbb{I}\bigg) \nonumber \\
\times & \sum_{n,m,q}  J_{q}\bigg(r\sqrt{(p_{2}^{1}-p_{1}^{1})^{2} + (p_{2}^{2}-p_{1}^{2})^{2}}\bigg) J_{m}\bigg(r'\sqrt{(p_{2}^{1}-p_{1}^{1})^{2} + (p_{2}^{2}-p_{1}^{2})^{2}}\bigg) J_{n}\bigg(\frac{|q_{f}B|}{2} r r'\bigg) \nonumber \\
\times & \Bigg[ \bigg(\frac{|q_{f}B|}{2}\bigg)^{(k_{1\text{s}_{f}+}^{1} - k_{1}^{2})/2} \bigg(\frac{|q_{f}B|}{2}\bigg)^{(k_{2\text{s}_{f}+}^{1} - k_{2}^{2})/2} \frac{i^{\,m-1}e^{-im\varphi_{0}}}{m-n +(k_{2\text{s}_{f}+}^{1}-k_{2}^{2})} \Big(e^{i2\pi\big(m-n+(k_{2\text{s}_{f}+}^{1}-k_{2}^{2})\big)}-1\Big) \nonumber \\
\times & \frac{(-i)^{\,q+1}e^{-iq\varphi_{0}}}{q+n-(k_{1\text{s}_{f}+}^{1}-k_{1}^{2})} \bigg(e^{i2\pi\big(q+ n- (k_{1\text{s}_{f}+}^{1}-k_{1}^{2})\big)}-1\bigg) \int_{0}^{\infty} dr \ r^{k_{1\text{s}_{f}+}^{1}-k_{1}^{2}+1} e^{-\frac{|q_{f}B|r^{2}}{4}} L_{k_{1}^{1}}^{k_{1\text{s}_{f}+}^{1} -k_{1}^{2}}\bigg(\frac{|q_{f}B|r^{2}}{2}\bigg) \nonumber \\
\times & \int_{0}^{\infty} dr' \ r'^{k_{2\text{s}_{f}+}^{1}-k_{1}^{2}+1} e^{-\frac{|q_{f}B| r'^{2}}{4}} L_{k_{2}^{1}}^{k_{2\text{s}_{f}+}^{1} -k_{2}^{2}} \bigg(\frac{|q_{f}B| r'^{2}}{2}\bigg) + \bigg(\frac{|q_{f}B|}{2}\bigg)^{(k_{1\text{s}_{f}+}^{1} - k_{1}^{2})/2} \bigg(\frac{|q_{f}B|}{2}\bigg)^{(k_{2\text{s}_{f}-}^{1} - k_{2}^{2})/2}  \nonumber \\
\times & \frac{i^{\,m-1}e^{-im\varphi_{0}}}{m-n+(k_{2\text{s}_{f}-}^{1}-k_{2}^{2})} \Big(e^{i2\pi\big(m-n+(k_{2\text{s}_{f}-}^{1}-k_{2}^{2})\big)}-1\Big)\frac{(-i)^{\,q+1}e^{-iq\varphi_{0}}}{q+n-(k_{1\text{s}_{f}+}^{1}-k_{1}^{2})} \bigg(e^{i2\pi\big(q+n-(k_{1\text{s}_{f}+}^{1}-k_{1}^{2})\big)}-1\bigg) \nonumber \nonumber \\
\times & \int_{0}^{\infty} dr \ r^{k_{1\text{s}_{f}+}^{1}-k_{1}^{2}+1}e^{-\frac{|q_{f}B|r^{2}}{4}} L_{k_{1}^{1}}^{k_{1\text{s}_{f}+}^{1} -k_{1}^{2}} \bigg(\frac{|q_{f}B|r^{2}}{2}\bigg) \int_{0}^{\infty} dr' \ r'^{k_{2\text{s}_{f}-}^{1}-k_{1}^{2}+1} e^{-\frac{|q_{f}B| r'^{2}}{4}} L_{k_{2}^{1}}^{k_{2\text{s}_{f}-}^{1} -k_{2}^{2}} \bigg(\frac{|q_{f}B| r'^{2}}{2}\bigg) \nonumber \\
+ & \bigg(\frac{|q_{f}B|}{2}\bigg)^{(k_{1\text{s}_{f}-}^{1} - k_{1}^{2})/2} \bigg(\frac{|q_{f}B|}{2}\bigg)^{(k_{2\text{s}_{f}+}^{1} - k_{2}^{2})/2}  \frac{i^{\,m-1}e^{-im\varphi_{0}}}{m-n+(k_{2\text{s}_{f}+}^{1}-k_{2}^{2})} \Big(e^{i2\pi\big(m-n+(k_{2\text{s}_{f}+}^{1}-k_{2}^{2})\big)}-1\Big) \nonumber \\
\times & \frac{(-i)^{\,q+1}e^{-iq\varphi_{0}}}{q+n-(k_{1\text{s}_{f}-}^{1}-k_{1}^{2})} \bigg(e^{i2\pi\big(q+ n-(k_{1\text{s}_{f}-}^{1}-k_{1}^{2})\big)}-1\bigg) \int_{0}^{\infty} dr \ r^{k_{1\text{s}_{f}-}^{1}-k_{1}^{2}+1} e^{-\frac{|q_{f}B|r^{2}}{4}} L_{k_{1}^{1}}^{k_{1\text{s}_{f}-}^{1} -k_{1}^{2}} \bigg(\frac{|q_{f}B|r^{2}}{2}\bigg) \nonumber \\
\times & \int_{0}^{\infty} dr' \ r'^{k_{2\text{s}_{f}+}^{1}-k_{1}^{2}+1} e^{-\frac{|q_{f}B| r'^{2}}{4}} L_{k_{2}^{1}}^{k_{2\text{s}_{f}+}^{1} -k_{2}^{2}} \bigg(\frac{|q_{f}B| r'^{2}}{2}\bigg) +  \bigg(\frac{|q_{f}B|}{2}\bigg)^{(k_{1\text{s}_{f}-}^{1} - k_{1}^{2})/2} \bigg(\frac{|q_{f}B|}{2}\bigg)^{(k_{2\text{s}_{f}-}^{1} - k_{2}^{2})/2} \nonumber \\
\times & \frac{i^{\,m-1}e^{-im\varphi_{0}}}{m-n+(k_{2\text{s}_{f}-}^{1}-k_{2}^{2})} \Big(e^{i2\pi\big(m-n+(k_{2\text{s}_{f}-}^{1}-k_{2}^{2})\big)}-1\Big)\frac{(-i)^{\,q+1}e^{-iq\varphi_{0}}}{q+n-(k_{1\text{s}_{f}-}^{1}-k_{1}^{2})} \bigg(e^{i2\pi\big(q+n-(k_{1\text{s}_{f}-}^{1}-k_{1}^{2})\big)}-1\bigg) \nonumber \\
\times & \int_{0}^{\infty} dr \ r^{k_{1\text{s}_{f}-}^{1}-k_{1}^{2}+1} e^{-\frac{|q_{f}B|r^{2}}{4}} L_{k_{1}^{1}}^{k_{1\text{s}_{f}-}^{1} -k_{1}^{2}} \bigg(\frac{|q_{f}B|r^{2}}{2}\bigg)  \int_{0}^{\infty} dr' \ r'^{k_{2\text{s}_{f}-}^{1}-k_{1}^{2}+1} e^{-\frac{|q_{f}B| r'^{2}}{4}} L_{k_{2}^{1}}^{k_{2\text{s}_{f}-}^{1} -k_{2}^{2}} \bigg(\frac{|q_{f}B| r'^{2}}{2}\bigg) \Bigg] \nonumber \\
\times & \int\frac{d^{4}p_{2}}{(2\pi)^{4}} \int_{0}^{\infty} \frac{ds}{\cos(|q_{f}B|s)}e^{is\big(p_{2 \parallel}^{2}-p_{2 \perp}^{2} \frac{\tan(|q_{f}B| s)}{|q_{f}B|} -m_{f}^{2} +i\epsilon\big)} \int\frac{d^{4}p_{1}}{(2\pi)^{4}} \frac{i}{p_{1}^2-m_{\pi}^{2}+i\epsilon} \nonumber \\
\times & \bigg[\big(\cos(|q_{f}B|s)+ \text{s}_{f}\gamma^{1}\gamma^{2} \sin (|q_{f}B| s)\big)\big(m_{f} - \slashed{p}_{2 \parallel}\big) +\frac{\slashed{p}_{2 \perp}}{\cos(|q_{f}B| s)} \bigg]  \nonumber \\
\times & \bigg[ \frac{1}{2} \Big(1 + i \gamma^{1}\gamma^{2}\Big) N_{k_{2\text{s}_{f}+}^{1},k_{2}^{2}}\,+ \frac{1}{2} \Big(1 - i \gamma^{1}\gamma^{2}\Big) N_{k_{2\text{s}_{f}-}^{1},k_{2}^{2}}\, \bigg] \nonumber \\
\times & \frac{1}{\sqrt{2(E_{f}+m_{f})}} \bigg(\gamma^{0}k_{2}^{0} - \frac{\gamma^{1}\text{s}_{f}}{2}\sqrt{2\ell q_{f}B} + \frac{\gamma^{2}\text{s}_{f}}{2}\sqrt{2\ell q_{f}B} - \gamma^{3}k_{2}^{3} + m_{f} \mathbb{I}\bigg) 
\begin{pmatrix}
    1 \\ 0 \\ -1 \\ 0
\end{pmatrix} \nonumber \\
\times & (2\pi)^{4} \delta^{(2)}(p_{2\parallel} - k_{1\parallel} - p_{1\parallel}) \delta^{(2)}(k_{2\parallel} + p_{1\parallel} - p_{2\parallel}).
\label{D.6full}
\end{align}

Since, the expression in Eq.~(\ref{D.6full}) does not seem to admit further analytical simplification, we consider the inertial frame in which the external particle is at rest, namely $k_j^1=k_j^2=k_j^3=0$ and $k_j^2=(k_j^0)^2$, with $j=1,2$. Hence, the expression we need to compute becomes
\begin{align*}
-i\Pi_{f}^{(D.6)} = & \frac{g^{2}}{2\pi} \int d^{2}x_{\perp} \int d^{2}y_{\perp}  \bigg[ \frac{1}{2} \Big(1 - i \gamma^{1}\gamma^{2}\Big)  + \frac{1}{2} \Big(1 + i \gamma^{1}\gamma^{2}\Big)  \bigg] \sqrt{|q_{f}B|} e^{-\frac{\xi}{2}}  \frac{1}{\sqrt{2(E_{f}+m_{f})}} 
\begin{pmatrix}
    1, \ 0, \ 1, \ 0
\end{pmatrix} 
\Big(\gamma^{0}k_{1}^{0} + m_{f} \mathbb{I}\Big)\\
\times & e^{\frac{iq_{f}}{2}(x^{1}y^{2}-x^{2}y^{1})}  \int\frac{d^{4}p_{2}}{(2\pi)^{4}} e^{-i p_{2\perp} \cdot (x-y)_{\perp}} \int_{0}^{\infty} \frac{ds}{\cos(|q_{f}B|s)}e^{is\big(p_{2 \parallel}^{2}-p_{2 \perp}^{2} \frac{\tan(|q_{f}B| s)}{|q_{f}B|} -m_{f}^{2} +i\epsilon\big)} \\
\times & \bigg[\big(\cos(|q_{f}B|s)+ \text{s}_{f}\gamma^{1}\gamma^{2} \sin (|q_{f}B| s)\big)\big(m_{f} - \slashed{p}_{2 \parallel}\big) +\frac{\slashed{p}_{2 \perp}}{\cos(|q_{f}B| s)} \bigg] \int\frac{d^{4}p_{1}}{(2\pi)^{4}} \frac{ie^{-ip_{1\perp}\cdot(y-x)_{\perp}}}{p_{1}^2-m_{\pi}^{2}+i\epsilon} \\
\times & \bigg[ \frac{1}{2} \Big(1 + i \gamma^{1}\gamma^{2}\Big) +  \frac{1}{2} \Big(1 - i \gamma^{1}\gamma^{2}\Big)\bigg] \sqrt{|q_{f}B|} e^{-\frac{\xi'}{2}} \frac{1}{\sqrt{2(E_{f}+m_{f})}} \Big(\gamma^{0}k_{2}^{0} + m_{f} \mathbb{I}\Big)
\begin{pmatrix}
    1 \\ 0 \\ -1 \\ 0
\end{pmatrix}\\
\times & (2\pi)^{4} \delta^{(2)}(p_{2}^{0} - k_{1}^{0} - p_{1}^{0}) \delta(p_{2}^{3} - p_{1}^{3}) \delta(k_{2}^{0} + p_{1}^{0} - p_{2}^{0}) \delta(p_{1}^{3} - p_{2}^{3}).
\end{align*}
After some straightforward algebra, and using the following integrals,
\begin{align}
    \int dy^{2} e^{ i\Big( i\frac{|q_{f}B|}{4}(y^{2})^{2} + y^{2}\big[(p_{1}^{2} - p_{2}^{2})+\frac{q_{f}B}{2}x^{1}\big] \Big)} &= \sqrt{\frac{4\pi}{|q_{f}B|}} e^{-\frac{\big[(p_{2}^{2} - p_{1}^{2}) + \frac{q_{f}B}{2} x^{1}\big]^{2}}{|q_{f}B|}}, \nonumber \\
    \int dy^{1} e^{i\Big( i\frac{|q_{f}B|}{4}(y^{1})^{2} + y^{1}\big[(p_{1}^{1} - p_{2}^{1}) -\frac{q_{f}B}{2}x^{2}\big] \Big)} &= \sqrt{\frac{4\pi}{|q_{f}B|}} e^{-\frac{\big[(p_{1}^{1} - p_{2}^{1}) - \frac{q_{f}B}{2} x^{2}\big]^{2}}{|q_{f}B|}}, \nonumber \\
    e^{-\frac{(p_{1}^{1} - p_{2}^{1})^{2}}{|q_{f}B|}}\int dx^{2} e^{ - \frac{|q_{f}B|}{2}(x^{2})^{2} + x^{2} \big[ \frac{q_{f}B}{|q_{f}B|}(p_{1}^{1} - p_{2}^{1}) + i(p_{2}^{2} - p_{1}^{2})\big]} &= \sqrt{\frac{2\pi}{|q_{f}B|}} e^{-\frac{(p_{1}^{1} - p_{2}^{1})^{2}}{|q_{f}B|}} e^{\Big\{\frac{1}{2|q_{f}B|} \big[ \frac{q_{f}B}{|q_{f}B|}(p_{1}^{1} - p_{2}^{1}) + i(p_{2}^{2} - p_{1}^{2})\big]^{2} \Big\}}, \nonumber \\
    e^{-\frac{(p_{1}^{2} - p_{2}^{2})^{2}}{|q_{f}B|}} \int dx^{1} e^{-\frac{|q_{f}B|}{2}(x^{1})^{2} +x^{1} \big[ i(p_{2}^{1} - p_{1}^{1}) -\frac{q_{f}B}{|q_{f}B|}(p_{1}^{2}-p_{2}^{2})\big]} &= \sqrt{\frac{2\pi}{|q_{f}B|}} e^{-\frac{(p_{1}^{2} - p_{1}^{2})^{2}}{|q_{f}B|}} e^{\Big\{ \frac{1}{2|q_{f}B|} \big[ i(p_{2}^{1} - p_{1}^{1}) -\frac{q_{f}B}{|q_{f}B|}(p_{1}^{2}-p_{2}^{2})\big]^{2} \Big\}};
\end{align}
the expression for $D.6$ becomes
\begin{align}
-i\Pi_{f}^{(D.6)} = & \frac{g^{2}}{2\pi} \frac{|q_{f}B|}{4m_{f}} \frac{(2\pi)(4\pi)}{|q_{f}B|^{2}}
\begin{pmatrix}
    1, \ 0, \ 1, \ 0
\end{pmatrix} 
\Big(\gamma^{0}k_{1}^{0} + m_{f} \mathbb{I}\Big) \int\frac{d^{4}p_{2}}{(2\pi)^{4}} \int_{0}^{\infty} \frac{ds}{\cos(|q_{f}B|s)}e^{is\big(p_{2 \parallel}^{2}-p_{2 \perp}^{2} \frac{\tan(|q_{f}B| s)}{|q_{f}B|} -m_{f}^{2} +i\epsilon\big)} \nonumber \\
\times & \int\frac{d^{4}p_{1}}{(2\pi)^{4}} \frac{i}{p_{1}^2-m_{\pi}^{2}+i\epsilon} \bigg[\big(\cos(|q_{f}B|s) + \text{s}_{f}\gamma^{1}\gamma^{2} \sin (|q_{f}B| s)\big)\big(m_{f} - \slashed{p}_{2 \parallel}\big) + \frac{\slashed{p}_{2 \perp}}{\cos(|q_{f}B| s)} \bigg] \nonumber \\
\times &  e^{-\frac{(p_{1}^{1}-p_{2}^{1})^{2}}{|q_{f}B|}} e^{-\frac{(p_{1}^{2}-p_{2}^{2})^{2}}{|q_{f}B|}}\Big(\gamma^{0}k_{2}^{0} + m_{f} \mathbb{I}\Big) 
\begin{pmatrix}
    1 \\ 0 \\ -1 \\ 0
\end{pmatrix} (2\pi)^{4} \delta^{(2)}(p_{2}^{0} - k_{1}^{0} - p_{1}^{0}) \delta(p_{2}^{3} - p_{1}^{3}) \delta(k_{2}^{0} + p_{1}^{0} - p_{2}^{0}) \delta(p_{1}^{3} - p_{2}^{3}).
\end{align}
The integrations over the parallel components of one of the internal momenta can now be performed. Choosing $p_{2\parallel}$ and using the Dirac delta functions, we obtain the following expression in momentum space
\begin{align}
-i\Pi_{f}^{(D.6)} = &   g^{2} \frac{8\pi^2}{|q_{f}B|^{2}} \int\frac{d^{2}p_{2\perp}}{(2\pi)^{2}} \int_{0}^{\infty} \frac{ds}{\cos(|q_{f}B|s)}e^{is\big((p_{1}^{0}-k_{1}^{0})^{2} - (p_{1}^{3})^{2} -p_{2 \perp}^{2} \frac{\tan(|q_{f}B| s)}{|q_{f}B|} -m_{f}^{2} +i\epsilon\big)} \int\frac{d^{4}p_{1}}{(2\pi)^{4}} \frac{i}{p_{1}^2-m_{\pi}^{2}+i\epsilon} \nonumber \\
\times & e^{-\frac{(p_{1}^{1}-p_{2}^{1})^{2}}{|q_{f}B|}} e^{-\frac{(p_{1}^{2}-p_{2}^{2})^{2}}{|q_{f}B|}} \bigg[\big(\cos(|q_{f}B|s) + \text{s}_{f}\gamma^{1}\gamma^{2} \sin (|q_{f}B| s)\big) \big(m_{f} - \gamma^{0}(p_{1}^{0}+k_{1}^{0}) + \gamma^{3}p_{1}^{3}\big) + \frac{\slashed{p}_{2 \perp}}{\cos(|q_{f}B| s)} \bigg].
\end{align}
For the integrals over $p_{2\perp}$, we use the following results
\begin{align}
    \int d^{2}p_{2 \perp} e^{-\frac{(p_{2\perp})^{2}}{|q_{f}B|} \big[i \tan(|q_{f}B|s) + 1 \big]} e^{\frac{2 p_{2 \perp} \cdot p_{1 \perp}}{|q_{f}B|}} &= \frac{|q_{f}B| \pi}{i \tan  (|q_{f}B| s) + 1} e^{\frac{p_{1\perp}^{2}}{|q_{f}B| J}}, \nonumber \\
    \int d^{2}p_{2 \perp} \ p_{2 \perp}e^{-\frac{(p_{2\perp})^{2}}{|q_{f}B|} \left[i \tan(|q_{f}B|s) + 1 \right]} e^{\frac{2 p_{2 \perp} \cdot p_{1 \perp}}{|q_{f}B|}} &= \frac{p_{1 \perp}|q_{f}B| \pi}{(i \tan  (|q_{f}B| s) + 1)^2} e^{\frac{p_{1 \perp}^{2}}{|q_{f}B| J }},
\end{align}
and we obtain
\begin{align}
-i\Pi_{f}^{(D.6)} = &  g^{2} \frac{2}{|q_{f}B|} \int_{0}^{\infty} \frac{ds}{\cos(|q_{f}B| s)} \int\frac{d^{4}p_{1}}{(2\pi)^{4}} \frac{i}{p_{1}^2-m_{\pi}^{2}+i\epsilon} e^{\frac{p_{1\perp}^{2}}{|q_{f}B|} \big(\frac{1}{i \tan  (|q_{f}B| s) + 1} -1\big)} e^{is\big((p_{1}^{0}+k_{1}^{0})^{2} - (p_{1}^{3})^{2} - m_{f}^{2} +i\epsilon\big)}\nonumber \\
\times & \bigg(\frac{|q_{f}B| \pi}{i \tan  (|q_{f}B| s) + 1}\bigg) \bigg[\big(\cos(|q_{f}B|s) + \text{s}_{f} \gamma^{1}\gamma^{2} \sin(|q_{f}B| s)\big)  \big(m_{f} - \gamma^{0}(p_{1}^{0}+k_{1}^{0}) + \gamma^{3}p_{1}^{3}\big) \nonumber \\
+& \frac{1}{i \tan  (|q_{f}B| s) + 1} \frac{\gamma^{1}p_{1}^{1}+\gamma^{2}p_{1}^{2}}{\cos(|q_{f}B| s)} \bigg].
\end{align}
We now rewrite the neutral pion propagator in the Schwinger proper time representation and get
\begin{align}
-i\Pi_{f}^{(D.6)} = & g^{2} (2\pi) \int_{0}^{\infty} \frac{ds}{\cos(|q_{f}B| s)} \int\frac{d^{4}p_{1}}{(2\pi)^{4}} \int_{0}^{\infty}ds' e^{is'\big(p_{1}^{2}-m_{\pi}^{2}+i\epsilon\big)}  e^{\frac{(p_{1\perp})^{2}}{|q_{f}B|} \big(\frac{1}{i \tan  (|q_{f}B| s) + 1} -1\big)} e^{is\big((p_{1}^{0}+k_{1}^{0})^{2} - (p_{1}^{3})^{2} - m_{f}^{2} +i\epsilon\big)}\nonumber \\
\times & \bigg(\frac{1}{i \tan  (|q_{f}B| s) + 1}\bigg) \bigg[\big(\cos(|q_{f}B|s) + \text{s}_{f}\gamma^{1}\gamma^{2} \sin (|q_{f}B| s)\big) \big(m_{f} - \gamma^{0}(p_{1}^{0}+k_{1}^{0}) + \gamma^{3}p_{1}^{3}\big) \nonumber \\
+& \frac{1}{i \tan  (|q_{f}B| s) + 1} \frac{\gamma^{1}p_{1}^{1}+\gamma^{2}p_{1}^{2}}{\cos(|q_{f}B| s)} \bigg].
\end{align}
We now compute the integrals over $p_{1\perp}$ and $p_1^3$. For this purpose, we use
\begin{align}
    \int dp_{1}^{3} \ e^{-i(p_{1}^{3})^{2} (s + s')}&= \sqrt{\frac{\pi}{i(s+s')}}, \nonumber \\
    \int dp_{1 \perp} e^{-p_{1 \perp}^{2} \big[\frac{1}{|q_{f}B|}(1-\frac{1}{i \tan  (|q_{f}B| s) + 1}) + is' \big]} &= \frac{\pi}{\frac{1}{|q_{f}B|}(1-\frac{1}{i \tan  (|q_{f}B| s) + 1}) + is'}, \nonumber \\
    \int dp_{1 \perp}\ p_{1 \perp} e^{-p_{1 \perp}^{2}\big[\frac{1}{|q_{f}B|}(1-\frac{1}{i \tan  (|q_{f}B| s) + 1}) + is'\big]}&= \int dp_{1}^{3} \ p_{1}^{3} \ e^{-(p_{1}^{3})^{2}(s+s')}=0,
\end{align}
and the contribution $D.6$ becomes
\begin{align}
-i\Pi_{f}^{(D.6)} = & \frac{g^{2}}{4\pi} \int_{0}^{\infty} \frac{ds}{\cos(|q_{f}B|s)} \int_{0}^{\infty}ds' \int\frac{dp_{1}^{0}}{2\pi} \frac{e^{is'(p_{1}^{0})^{2}}   e^{is(p_{1}^{0}+k_{1}^{0})^{2}} e^{is'(-m_{\pi}^{2}+i\epsilon)} e^{is(- m_{f}^{2} +i\epsilon)}}{i\tan(|q_fB|s)\left(\frac{1}{|q_fB|}+is' \right)+is'}\nonumber \\
\times &\sqrt{\frac{\pi}{i(s+s')}} \bigg[\big(\cos(|q_{f}B|s) + \text{s}_{f}\gamma^{1}\gamma^{2} \sin(|q_{f}B| s)\big) \big(m_{f} - \gamma^{0}(p_{1}^{0}+k_{1}^{0}) \big)  \bigg].
\end{align}
We perform the analytic continuations $ s \rightarrow -i\tau$ and $s' \rightarrow -i\tau'$, and the temperature dependence is incorporated through the Matsubara formalism. This gives
\begin{align}
-i\Pi_{f}^{(D.6)} = & -\frac{g^{2}}{4\pi} \int_{0}^{\infty} \frac{d\tau}{\cosh(|q_{f}B|\tau)} \int_{0}^{\infty}d\tau' \, iT\sum_{n}  \frac{e^{\tau'(i\tilde{\omega}_{n})^{2}} e^{\tau[i(\tilde{\omega}_{n} - \omega)]^{2}}e^{-\tau'm_{\pi}^{2}} e^{- \tau m_{f}^{2}}}{\tanh(|q_fB|\tau)\left( \frac{1}{|q_fB|}+\tau' \right)+\tau'}\nonumber \\
\times & \sqrt{\frac{\pi}{\tau+\tau'}} \bigg[\big(\cosh(|q_{f}B|\tau) - i\text{s}_{f}\gamma^{1}\gamma^{2} \sinh (|q_{f}B|\tau)\big)  \big(m_{f} + \gamma^{0}(\tilde{\omega}_{n} + \omega) \big)  \bigg].
\label{D.6Matsubaraformalism}
\end{align}
In order to perform the sum over the Matsubara modes, we use the following two expressions
\begin{align}
    e^{-\tau\omega^{2}} \sum_{n=-\infty}^{\infty} e^{-(\tau+\tau')\tilde{\omega}_{n}^{2}-2\tilde{\omega}_{n}\omega\tau} &= \frac{1}{2T\sqrt{\pi (\tau+\tau')}} e^{-\frac{\tau \tau' \omega^{2}}{\tau + \tau'}}\bigg[1+2\sum_{n=1}^{\infty} (-1)^{n}e^{-\frac{n^{2}}{4T^{2}(\tau+\tau')}} \cos\Big( \frac{n \tau\omega}{T(\tau + \tau')}\Big) \bigg], \nonumber \\
    e^{-\tau\omega^{2}} \sum_{n=-\infty}^{\infty} \tilde{\omega}_{n} e^{-(\tau+\tau')\tilde{\omega}_{n}^{2}-2\tilde{\omega}_{n}\omega\tau}= & \ \frac{1}{2\sqrt{\pi (\tau+\tau')}} e^{-\frac{\tau \tau' \omega^{2}}{\tau + \tau'}} \bigg\{-\frac{\tau \omega}{\tau +\tau'}+ 2\sum_{n=1}^{\infty} (-1)^{n} e^{-\frac{n^{2}}{4T^{2}(\tau +\tau')}} \\
\times & \bigg[-\frac{\tau \omega}{\tau+\tau'} \cos\Big(\frac{n \tau \omega}{T(\tau+\tau')}\Big)+\frac{n}{2(\tau +\tau')T}\sin\Big(\frac{n \tau \omega}{T(\tau+\tau')} \Big)\bigg]\bigg\}.
\label{D.6sums}
\end{align}
Substituting Eq.~(\ref{D.6sums}) into Eq.~(\ref{D.6Matsubaraformalism}), we obtain
\begin{align}
-i\Pi_{f}^{(D.6)} = & -\frac{ig^{2}}{8\pi} \int_{0}^{\infty} \frac{d\tau}{\cosh(|q_{f}B|\tau)} \int_{0}^{\infty} \frac{d\tau'}{\tau+\tau'} \frac{e^{-\tau'm_{\pi}^{2}} e^{- \tau m_{f}^{2}}e^{-\frac{\tau \tau' \omega^{2}}{\tau + \tau'}}}{\tanh(|q_fB|\tau)\left( \frac{1}{|q_fB|}+\tau' \right)+\tau'}  \big(\cosh(|q_{f}B|\tau) - i\text{s}_{f}\gamma^{1}\gamma^{2} \sinh(|q_{f}B|\tau)\big) \nonumber \\
\times &\bigg[\big(m_{f} + \gamma^{0}\omega \big) \bigg( 1+2\sum_{n=1}^{\infty} (-1)^{n} e^{-\frac{n^{2}}{4T^{2}(\tau+\tau')}} \cos\Big( \frac{n \tau\omega}{T(\tau + \tau')}\Big) \bigg) +\gamma^{0} \bigg\{-\frac{\tau \omega}{\tau +\tau'}+ 2\sum_{n=1}^{\infty} (-1)^{n} e^{-\frac{n^{2}}{4T^{2}(\tau +\tau')}} \nonumber \\
\times & \bigg[-\frac{\tau \omega}{\tau+\tau'} \cos\Big(\frac{n \tau \omega}{T(\tau+\tau')}\Big)+\frac{n}{2(\tau +\tau')T}\sin\Big(\frac{n \tau \omega}{T(\tau+\tau')} \Big)\bigg]\bigg\}\bigg].
\end{align}
We perform the change of variables $\tau=u(1-v)$ and $\tau'=uv$, and obtain the final expression
\begin{align}
-i\Pi_{f}^{(D.6)} = & -\frac{ig^{2}}{8\pi} \int_{0}^{\infty}du \int_{0}^{1} dv  \frac{e^{-uvm_{\pi}^{2}} e^{-u(1-v)m_{f}^{2}} e^{-uv(1-v)(i\omega)^{2}}}{\cosh(|q_{f}B|u(1-v)) \Big(\tanh(|q_{f}B| u(1-v))\left( \frac{1}{|q_fB|}+uv \right)+uv\Big)}  \nonumber \\
\times & \big(\cosh(|q_{f}B|u(1-v)) - i\text{s}_{f}\gamma^{1}\gamma^{2} \sinh(|q_{f}B| u(1-v))\big) \bigg[\big(m_{f} - i\gamma^{0}(i\omega) \big) \nonumber \\
\times &  \bigg( 1+2\sum_{n=1}^{\infty} (-1)^{n}e^{-\frac{n^{2}}{4T^{2}u}} \cosh\Big( \frac{n(i\omega)(1-v)}{T}\Big) \bigg) -i\gamma^{0} \bigg\{(1-v)(i\omega) + 2\sum_{n=1}^{\infty}(-1)^{n} e^{-\frac{n^{2}}{4T^{2}u}} \nonumber \\
\times & \bigg[(1-v)(i\omega) \cosh\Big(\frac{n(i\omega)(1-v)}{T}\Big)+\frac{n}{2uT} \sinh\Big(\frac{n(i\omega)(1-v)}{T}\Big)\bigg]\bigg\}\bigg].
\label{D.6final}
\end{align}
In Eq.~(\ref{D.6final}), the temperature-independent terms contain ultraviolet divergences, since they include the vacuum contribution. The corresponding vacuum piece is obtained by taking the limit $|q_fB|\rightarrow 0$, yielding
\begin{equation}
-i\Pi_{f,B=0}^{(D.6)} =-\frac{ig^{2}}{16\pi^{2}} \int_{0}^{\infty}du \int_{0}^{1} dv  e^{-uvm_{\pi}^{2}} e^{-u(1-v)m_{f}^{2}} e^{-uv(1-v)(i\omega)^{2}} \Big[m_{f} - i\gamma^{0}(2-v)(i\omega)\Big].
\end{equation}

The next contribution corresponds to the loop formed by a sigma meson and a quark. This term is analogous to the previous case, with the neutral pion propagator replaced by the sigma meson propagator and with the corresponding change in the vertices. In contrast to $D.6$, no $\gamma^5$ matrices appear in this contribution, and therefore no anticommutation through the Dirac matrices of the quark propagator is required. Hence, the initial expression can be written as
\begin{align}
-i\Pi_{f}^{(D.7)} = & - g^{2} \int d^{4}x \int d^{4}y  \bigg[ \frac{1}{2} \Big(1 - i \gamma^{1}\gamma^{2}\Big) \frac{1}{\sqrt{2\pi}} e^{i(k_{1}^{0}x^{0} - k_{1}^{3}x^{3})}e^{i\text{s}_{f}(k_{1\text{s}_{f}+}^{1} -k_{1}^{2})\theta} N_{k_{1\text{s}_{f}+}^{1},k_{1}^{2}}\, \xi^{(k_{1\text{s}_{f}+}^{1} -k_{1}^{2})/2} e^{-\frac{\xi}{2}}L_{k_{1}^{2}}^{k_{1\text{s}_{f}+}^{1} -k_{1}^{2}} (\xi) \nonumber \\
+ & \frac{1}{2} \Big(1 + i \gamma^{1}\gamma^{2}\Big) \frac{1}{\sqrt{2\pi}} e^{i(k_{1}^{0}x^{0} - k_{1}^{3}x^{3})}e^{i\text{s}_{f}(k_{1\text{s}_{f}-}^{1} -k_{1}^{2})\theta} N_{k_{1\text{s}_{f}-}^{1},k_{1}^{2}}\, \xi^{(k_{1\text{s}_{f}-}^{1} -k_{1}^{2})/2} e^{-\frac{\xi}{2}}L_{k_{1}^{2}}^{k_{1\text{s}_{f}-}^{1} -k_{1}^{2}} (\xi) \bigg] \nonumber \\
\times & \frac{1}{\sqrt{2(E_{f}+m_{f})}} 
\begin{pmatrix}
    1, \ 0, \ 1, \ 0
\end{pmatrix} 
\bigg( \gamma^{0}k_{1}^{0} + \frac{\gamma^{1}\text{s}_{f}}{2}\sqrt{2\ell q_{f}B} - \frac{\gamma^{2}\text{s}_{f}}{2}\sqrt{2\ell q_{f}B} + \gamma^{3}k_{1}^{3} + m_{f} \mathbb{I}\bigg) \nonumber \\
\times & \, e^{iq_{f}\int_{x}^{y} d\xi'_{\mu} \big[A^{\mu}(\xi')+\frac{1}{2}F^{\mu\,\nu}(\xi'-y)_{\nu}\big]}  \int\frac{d^{4}p_{2}}{(2\pi)^{4}} e^{-i p_{2} \cdot (x-y)} \int_{0}^{\infty} \frac{ds}{\cos(|q_{f}B|s)}e^{is\big(p_{2 \parallel}^{2}-p_{2 \perp}^{2} \frac{\tan(|q_{f}B| s)}{|q_{f}B|} -m_{f}^{2} +i\epsilon\big)} \nonumber \\
\times & \bigg[\big(\cos(|q_{f}B|s)+ \text{s}_{f}\gamma^{1}\gamma^{2} \sin (|q_{f}B| s)\big)\big(m_{f} + \slashed{p}_{2 \parallel}\big)-\frac{\slashed{p}_{2 \perp}}{\cos(|q_{f}B| s)} \bigg] \int\frac{d^{4}p_{1}}{(2\pi)^{4}} \frac{ie^{-ip_{1}\cdot(y-x)}}{p_{1}^2-m_{\sigma}^{2}+i\epsilon} \nonumber \\
\times & \bigg[ \frac{1}{2} \Big(1 + i \gamma^{1}\gamma^{2}\Big) \frac{1}{\sqrt{2\pi}} e^{-i(k_{2}^{0}y^{0} - k_{2}^{3}y^{3})}e^{-i\text{s}_{f}(k_{2\text{s}_{f}+}^{1} -k_{2}^{2})\theta'} N_{k_{2\text{s}_{f}+}^{1},k_{2}^{2}}\, \xi'^{(k_{2\text{s}_{f}+}^{1} -k_{2}^{2})/2} e^{-\frac{\xi'}{2}}L_{k_{1}^{2}}^{k_{1\text{s}_{f}+}^{1} -k_{1}^{2}} (\xi') \nonumber \\
+ & \frac{1}{2} \Big(1 - i \gamma^{1}\gamma^{2}\Big) \frac{1}{\sqrt{2\pi}} e^{-i(k_{2}^{0}y^{0} - k_{2}^{3}y^{3})}e^{-i\text{s}_{f}(k_{2\text{s}_{f}-}^{1} -k_{2}^{2})\theta'} N_{k_{2\text{s}_{f}-}^{1},k_{2}^{2}}\, \xi'^{(k_{2\text{s}_{f}-}^{1} -k_{2}^{2})/2} e^{-\frac{\xi'}{2}}L_{k_{2}^{2}}^{k_{2\text{s}_{f}-}^{1} -k_{2}^{2}}(\xi') \bigg] \nonumber \\
\times & \frac{1}{\sqrt{2(E_{f}+m_{f})}} \bigg( \gamma^{0}k_{2}^{0} - \frac{\gamma^{1}\text{s}_{f}}{2}\sqrt{2\ell q_{f}B} + \frac{\gamma^{2}\text{s}_{f}}{2}\sqrt{2\ell q_{f}B} - \gamma^{3}k_{2}^{3} + m_{f} \mathbb{I}\bigg)
\begin{pmatrix}
    1 \\ 0 \\ -1 \\ 0
\end{pmatrix}.
\label{D.7initial}
\end{align}
The notation in Eq.~(\ref{D.7initial}) is the same as that used for contribution $D.6$. Since the only structural difference with respect to the previous contribution is the absence of the $\gamma^5$ matrices, the subsequent steps follow in complete analogy. We can therefore write the corresponding general expression as
\begin{align}
-i\Pi_{f}^{(D.7)} = & \ \frac{g^{2}}{2\pi} \int d^{2}x_{\perp} \int d^{2}y_{\perp} \bigg[ \frac{1}{2} \Big(1 - i \gamma^{1}\gamma^{2}\Big) N_{k_{1\text{s}_{f}+}^{1},k_{1}^{2}}\, + \frac{1}{2} N_{k_{1\text{s}_{f}-}^{1},k_{1}^{2}}\, \bigg] \nonumber \\
\times & \frac{1}{\sqrt{2(E_{f}+m_{f})}} 
\begin{pmatrix}
    1, \ 0, \ 1, \ 0
\end{pmatrix} 
\bigg( \gamma^{0}k_{1}^{0} + \frac{\gamma^{1}\text{s}_{f}}{2}\sqrt{2\ell q_{f}B} - \frac{\gamma^{2}\text{s}_{f}}{2}\sqrt{2\ell q_{f}B} + \gamma^{3}k_{1}^{3} + m_{f} \mathbb{I}\bigg) \nonumber \\
\times & \sum_{n}\sum_{m}\sum_{q}  J_{q}\bigg(r\sqrt{(p_{2}^{1}-p_{1}^{1})^{2} + (p_{2}^{2}-p_{1}^{2})^{2}}\bigg) J_{m}\bigg(r'\sqrt{(p_{2}^{1}-p_{1}^{1})^{2} + (p_{2}^{2}-p_{1}^{2})^{2}}\bigg) J_{n}\bigg(\frac{|q_{f}B|}{2} r r'\bigg) \nonumber \\
\times & \Bigg[ \bigg(\frac{|q_{f}B|}{2}\bigg)^{(k_{1\text{s}_{f}+}^{1} - k_{1}^{2})/2} \bigg(\frac{|q_{f}B|}{2}\bigg)^{(k_{2\text{s}_{f}+}^{1} - k_{2}^{2})/2} \frac{i^{\,m-1}e^{-im\varphi_{0}}}{m-n+(k_{2\text{s}_{f}+}^{1}-k_{2}^{2})} \Big(e^{i2\pi\big(m-n+(k_{2\text{s}_{f}+}^{1}-k_{2}^{2})\big)}-1\Big) \nonumber \\
\times & \frac{(-i)^{\,q+1}e^{-iq\varphi_{0}}}{q+n-(k_{1\text{s}_{f}+}^{1}-k_{1}^{2})} \bigg(e^{i2\pi\big(q+n-( k_{1\text{s}_{f}+}^{1}-k_{1}^{2})\big)}-1\bigg) \int_{0}^{\infty} dr \ r^{k_{1\text{s}_{f}+}^{1}-k_{1}^{2}+1} e^{-\frac{|q_{f}B|r^{2}}{4}} L_{k_{1}^{1}}^{k_{1\text{s}_{f}+}^{1} -k_{1}^{2}}\bigg(\frac{|q_{f}B|r^{2}}{2}\bigg) \nonumber \\
\times & \int_{0}^{\infty} dr' \ r'^{k_{2\text{s}_{f}+}^{1}-k_{1}^{2}+1} e^{-\frac{|q_{f}B| r'^{2}}{4}} L_{k_{2}^{1}}^{k_{2\text{s}_{f}+}^{1} -k_{2}^{2}} \bigg(\frac{|q_{f}B| r'^{2}}{2}\bigg) + \bigg(\frac{|q_{f}B|}{2}\bigg)^{(k_{1\text{s}_{f}+}^{1} - k_{1}^{2})/2} \bigg(\frac{|q_{f}B|}{2}\bigg)^{(k_{2\text{s}_{f}-}^{1} - k_{2}^{2})/2}  \nonumber \\
\times & \frac{i^{\,m-1}e^{-im\varphi_{0}}}{m-n+(k_{2\text{s}_{f}-}^{1}-k_{2}^{2})} \Big(e^{i2\pi\big(m-n+(k_{2\text{s}_{f}-}^{1}-k_{2}^{2})\big)}-1\Big)\frac{(-i)^{\,q+1}e^{-iq\varphi_{0}}}{q+n-(k_{1\text{s}_{f}+}^{1}-k_{1}^{2})} \bigg(e^{i2\pi\big(q+n-(k_{1\text{s}_{f}+}^{1}-k_{1}^{2})\big)}-1\bigg) \nonumber \\
\times & \int_{0}^{\infty} dr \ r^{k_{1\text{s}_{f}+}^{1}-k_{1}^{2}+1}e^{-\frac{|q_{f}B|r^{2}}{4}} L_{k_{1}^{1}}^{k_{1\text{s}_{f}+}^{1} -k_{1}^{2}} \bigg(\frac{|q_{f}B|r^{2}}{2}\bigg) \int_{0}^{\infty} dr' \ r'^{k_{2\text{s}_{f}-}^{1}-k_{1}^{2}+1} e^{-\frac{|q_{f}B| r'^{2}}{4}} L_{k_{2}^{1}}^{k_{2\text{s}_{f}-}^{1} -k_{2}^{2}} \bigg(\frac{|q_{f}B| r'^{2}}{2}\bigg) \nonumber \\
+ & \bigg(\frac{|q_{f}B|}{2}\bigg)^{(k_{1\text{s}_{f}-}^{1} - k_{1}^{2})/2} \bigg(\frac{|q_{f}B|}{2}\bigg)^{(k_{2\text{s}_{f}+}^{1} - k_{2}^{2})/2}  \frac{i^{\,m-1}e^{-im\varphi_{0}}}{m-n+(k_{2\text{s}_{f}+}^{1}-k_{2}^{2})} \Big(e^{i2\pi\big(m-n+(k_{2\text{s}_{f}+}^{1}-k_{2}^{2})\big)}-1\Big) \nonumber \\
\times & \frac{(-i)^{\,q+1}e^{-iq\varphi_{0}}}{q+n-(k_{1\text{s}_{f}-}^{1}-k_{1}^{2})} \bigg(e^{i2\pi\big(q- n-(k_{1\text{s}_{f}-}^{1}-k_{1}^{2})\big)}-1\bigg) \int_{0}^{\infty} dr \ r^{k_{1\text{s}_{f}-}^{1}-k_{1}^{2}+1} e^{-\frac{|q_{f}B|r^{2}}{4}} L_{k_{1}^{1}}^{k_{1\text{s}_{f}-}^{1} -k_{1}^{2}} \bigg(\frac{|q_{f}B|r^{2}}{2}\bigg) \nonumber \\
\times & \int_{0}^{\infty} dr' \ r'^{k_{2\text{s}_{f}+}^{1}-k_{1}^{2}+1} e^{-\frac{|q_{f}B| r'^{2}}{4}} L_{k_{2}^{1}}^{k_{2\text{s}_{f}+}^{1} -k_{2}^{2}} \bigg(\frac{|q_{f}B| r'^{2}}{2}\bigg) +  \bigg(\frac{|q_{f}B|}{2}\bigg)^{(k_{1\text{s}_{f}-}^{1} - k_{1}^{2})/2} \bigg(\frac{|q_{f}B|}{2}\bigg)^{(k_{2\text{s}_{f}-}^{1} - k_{2}^{2})/2} \nonumber \\
\times & \frac{i^{\,m-1}e^{-im\varphi_{0}}}{m-n+(k_{2\text{s}_{f}-}^{1}-k_{2}^{2})} \Big(e^{i2\pi\big(m-n+(k_{2\text{s}_{f}-}^{1}-k_{2}^{2})\big)}-1\Big)\frac{(-i)^{\,q+1}e^{-iq\varphi_{0}}}{q+n-(k_{1\text{s}_{f}-}^{1}-k_{1}^{2})} \bigg(e^{i2\pi\big(q+n-(k_{1\text{s}_{f}-}^{1}-k_{1}^{2})\big)}-1\bigg) \nonumber \\
\times & \int_{0}^{\infty} dr \ r^{k_{1\text{s}_{f}-}^{1}-k_{1}^{2}+1} e^{-\frac{|q_{f}B|r^{2}}{4}} L_{k_{1}^{1}}^{k_{1\text{s}_{f}-}^{1} -k_{1}^{2}} \bigg(\frac{|q_{f}B|r^{2}}{2}\bigg)  \int_{0}^{\infty} dr' \ r'^{k_{2\text{s}_{f}-}^{1}-k_{1}^{2}+1} e^{-\frac{|q_{f}B| r'^{2}}{4}} L_{k_{2}^{1}}^{k_{2\text{s}_{f}-}^{1} -k_{2}^{2}} \bigg(\frac{|q_{f}B| r'^{2}}{2}\bigg) \Bigg] \nonumber \\
\times & \int\frac{d^{4}p_{2}}{(2\pi)^{4}} \int_{0}^{\infty} \frac{ds}{\cos(|q_{f}B|s)}e^{is\big(p_{2 \parallel}^{2}-p_{2 \perp}^{2} \frac{\tan(|q_{f}B| s)}{|q_{f}B|} -m_{f}^{2} +i\epsilon\big)} \int\frac{d^{4}p_{1}}{(2\pi)^{4}} \frac{i}{p_{1}^2-m_{\sigma}^{2}+i\epsilon} \nonumber \\
\times & \bigg[\big(\cos(|q_{f}B|s)+ \text{s}_{f}\gamma^{1}\gamma^{2} \sin (|q_{f}B| s)\big)\big(m_{f} + \slashed{p}_{2 \parallel}\big) -\frac{\slashed{p}_{2 \perp}}{\cos(|q_{f}B| s)} \bigg]  \nonumber \\
\times & \bigg[ \frac{1}{2} \Big(1 + i \gamma^{1}\gamma^{2}\Big) N_{k_{2\text{s}_{f}+}^{1},k_{2}^{2}}\,+ \frac{1}{2} \Big(1 - i \gamma^{1}\gamma^{2}\Big) N_{k_{2\text{s}_{f}-}^{1},k_{2}^{2}}\, \bigg] \nonumber \\
\times & \frac{1}{\sqrt{2(E_{f}+m_{f})}} \bigg(\gamma^{0}k_{2}^{0} - \frac{\gamma^{1}\text{s}_{f}}{2}\sqrt{2\ell q_{f}B} + \frac{\gamma^{2}\text{s}_{f}}{2}\sqrt{2\ell q_{f}B} - \gamma^{3}k_{2}^{3} + m_{f} \mathbb{I}\bigg) 
\begin{pmatrix}
    1 \\ 0 \\ -1 \\ 0
\end{pmatrix} \nonumber \\
\times & (2\pi)^{4} \delta^{(2)}(p_{2\parallel} - k_{1\parallel} - p_{1\parallel}) \delta^{(2)}(k_{2\parallel} + p_{1\parallel} - p_{2\parallel}),
\label{D.7full}
\end{align}
and in the inertial frame in which the external particle is at rest, namely $k_j^1=k_j^2=k_j^3=0$ and $k_j^2=(k_j^0)^2$, with $j=1,2$, the final expression is
\begin{align}
-i\Pi_{f}^{(D.7)} = & \frac{ig^{2}}{8\pi} \int_{0}^{\infty}du \int_{0}^{1} dv  \frac{e^{-uvm_{\sigma}^{2}} e^{-u(1-v)m_{f}^{2}} e^{-uv(1-v)(i\omega)^{2}}}{\cosh(|q_fB|u(1-v))\left( \tanh(|q_fB|u(1-v))\left( \frac{1}{|q_fB|}+uv \right)+uv \right)}  \nonumber \\
\times & \big(\cosh(|q_{f}B|u(1-v)) - i\text{s}_{f}\gamma^{1}\gamma^{2} \sinh(|q_{f}B| u(1-v))\big) \bigg[\big(m_{f} + i \gamma^{0}(i\omega)\big) \nonumber \\
\times &  \bigg( 1+2\sum_{n=1}^{\infty} (-1)^{n}e^{-\frac{n^{2}}{4T^{2}u}} \cosh\Big(\frac{n(i\omega)(1-v)}{T}\Big) \bigg) + i\gamma^{0} \bigg\{(1-v)(i\omega) + 2\sum_{n=1}^{\infty}(-1)^{n} e^{-\frac{n^{2}}{4T^{2}u}} \nonumber \\
\times & \bigg[(1-v)(i\omega) \cosh\Big(\frac{n(i\omega)(1-v)}{T}\Big) + \frac{n}{2uT} \sinh\Big(\frac{n(i\omega)(1-v)}{T}\Big)\bigg]\bigg\}\bigg],
\label{D.7final}
\end{align}
whose vacuum piece is
\begin{equation}
-i\Pi_{f,B=0}^{(D.7)} = \frac{ig^{2}}{16\pi^{2}} \int_{0}^{\infty}du \int_{0}^{1} dv e^{-uvm_{\sigma}^{2}} e^{-u(1-v)m_{f}^{2}} e^{-uv(1-v)(i\omega)^{2}} \Big(m_{f} +i\gamma^{0} (2-v)(i\omega)\Big).
\label{D.7vacuum}
\end{equation}
The last contribution corresponds to the loop formed by a charged pion and a quark. Since both particles are charged, both propagators must be written in the Schwinger proper-time representation. As a result, two Schwinger phases appear, each associated with a different electric charge, and they do not cancel. This is the main difference with respect to contribution $D.6$, where only one charged propagator is present. The masses are the same as in that case, while the vertices must be replaced according to the Feynman rules shown in Fig.~\ref{fig2}. We nevertheless present the relevant steps explicitly in order to make the derivation of the final result transparent. The initial expression is
\begin{equation}
-i\Pi_{f}^{(D.8)} = (\sqrt{2}g)^{2} \int d^{4}x \int d^{4}y  \; \overline{U}(x,k_{1}) \gamma^{5} S^{B}(x-y) \gamma^{5} D^{B}(y-x) U(y,k_{2}),
\label{D.8initial}
\end{equation}
where the propagators are
\begin{align}
S^{B}(x-y) = & e^{iq_{f} \int_{x}^{y} d\xi_{\mu} \Bigl[ A^{\mu}(\xi) + \frac{1}{2} F^{\mu\nu}(\xi - y)_{\nu} \Bigr]} \int \frac{d^{4}p_{2}}{(2\pi)^{4}} e^{-ip_{2}\cdot(x-y)} \int_{0}^{\infty} \frac{ds'}{\cos(|q_{f}B| s')} e^{is'\bigl(p_{2\parallel}^{2} -p_{2\perp}^{2}\frac{\tan(|q_{f}B| s')}{|q_{f}B| s'} -m_{f}^{2} + i\epsilon \bigr)} \nonumber \\
\times & \biggl[ \Bigl(\cos(|q_{f}B| s') + \text{s}_{f}\gamma^{1}\gamma^{2}\sin(|q_{f}B| s')\Bigr)\Bigl(m_{f}+\slashed{p}_{2\parallel}\Bigr) - \frac{\slashed{p}_{2\perp}}{\cos(|q_{f}B| s')} \biggr],
\label{SqfBD.8}
\end{align}
\begin{equation}
D^{B}(y-x) = e^{ie \int_{y}^{x} d\xi'_{\mu} \Bigl[ A^{\mu}(\xi') + \frac{1}{2} F^{\mu\nu}(\xi' - x)_{\nu} \Bigr]} \int \frac{d^{4}p_{1}}{(2\pi)^{4}} e^{-ip_{1}\cdot(y-x)} \int_{0}^{\infty} \frac{ds}{\cos(|eB| s)} e^{is\bigl(p_{1\parallel}^{2} -p_{1\perp}^{2}\frac{\tan(|eB| s)}{|eB| s} -m_{\pi}^{2} + i\epsilon \bigr)},
\label{DeBD.8}
\end{equation}
and the Ritus spinor functions are
\begin{align}
\bar{U}(x,k_{1}) = & \biggl[ \frac{1}{2} \big(1 - i \gamma^{1}\gamma^{2}\big) \frac{1}{\sqrt{2\pi}} e^{i(k_{1}^{0}x^{0} - k_{1}^{3}x^{3})}e^{i\text{s}_{f'}(k_{1\text{s}_{f'}+}^{1} -k_{1}^{2})\theta} N_{k_{1\text{s}_{f'}+}^{1},k_{1}^{2}}\, \xi^{(k_{1\text{s}_{f'}+}^{1} -k_{1}^{2})/2} e^{-\frac{\xi}{2}}L_{k_{1}^{2}}^{k_{1\text{s}_{f'}+}^{1} -k_{1}^{2}} (\xi) \nonumber \\
+& \,\frac{1}{2} \big(1 + i \gamma^{1}\gamma^{2}\big) \frac{1}{\sqrt{2\pi}} e^{i(k_{1}^{0}x^{0} - k_{1}^{3}x^{3})}e^{i\text{s}_{f'}(k_{1\text{s}_{f'}-}^{1} -k_{1}^{2})\theta} N_{k_{1\text{s}_{f'}-}^{1},k_{1}^{2}}\, \xi^{(k_{1\text{s}_{f'}-}^{1} -k_{1}^{2})/2} e^{-\frac{\xi}{2}}L_{k_{1}^{2}}^{k_{1\text{s}_{f'}-}^{1} -k_{1}^{2}}(\xi) \biggr] \nonumber \\
\times & \frac{1}{\sqrt{2(E_{f'}+m_{f'})}} 
\begin{pmatrix}
    1, \ 0, \ 1, \ 0
\end{pmatrix} 
\Bigl( \gamma^{0}k_{1}^{0} + \frac{\gamma^{1}\text{s}_{f'}}{2}\sqrt{2\ell q_{f'}B} - \frac{\gamma^{2}\text{s}_{f'}}{2}\sqrt{2\ell q_{f'}B} + \gamma^{3}k_{1}^{3} + m_{f'} \mathbb{I}\Bigr),
\label{Ritusout}
\end{align}
\begin{align}
U(y,k_{2}) = & \biggl[\frac{1}{2} \big(1 + i \gamma^{1}\gamma^{2}\big) \frac{1}{\sqrt{2\pi}} e^{-i(k_{2}^{0}y^{0} - k_{2}^{3}y^{3})}e^{-i\text{s}_{f'}(k_{2\text{s}_{f'}+}^{1} -k_{2}^{2})\theta'} N_{k_{2\text{s}_{f'}+}^{1},k_{2}^{2}}\, \xi'^{(k_{2\text{s}_{f'}+}^{1} -k_{2}^{2})/2} e^{-\frac{\xi'}{2}}L_{k_{1}^{2}}^{k_{1\text{s}_{f'}+}^{1} -k_{1}^{2}} (\xi') \nonumber \\
+ & \,\frac{1}{2} \big(1 - i \gamma^{1}\gamma^{2}\big) \frac{1}{\sqrt{2\pi}} e^{-i(k_{2}^{0}y^{0} - k_{2}^{3}y^{3})}e^{-i\text{s}_{f'}(k_{2\text{s}_{f'}-}^{1} -k_{2}^{2})\theta'} N_{k_{2\text{s}_{f'}-}^{1},k_{2}^{2}}\, \xi'^{(k_{2\text{s}_{f'}-}^{1} -k_{2}^{2})/2} e^{-\frac{\xi'}{2}}L_{k_{2}^{2}}^{k_{2\text{s}_{f'}-}^{1} -k_{2}^{2}} (\xi') \bigg] \nonumber \\
\times & \frac{1}{\sqrt{2(E_{f'}+m_{f'})}} \Big( \gamma^{0}k_{2}^{0} - \frac{\gamma^{1}\text{s}_{f'}}{2}\sqrt{2\ell q_{f'}B} + \frac{\gamma^{2}\text{s}_{f'}}{2}\sqrt{2\ell q_{f'}B} - \gamma^{3}k_{2}^{3} + m_{f'} \mathbb{I}\Big) 
\begin{pmatrix}
    1 \\ 0 \\ -1 \\ 0
\end{pmatrix}.
\label{Ritusin}
\end{align}
We now clarify the notation used in this contribution. the external quark which appears in the Ritus function has charge $q_{f'} = q_{\{d,u\}}$, the internal quark which appears in the propagator has charge $q_{f} = q_{\{u,d\}}$, and the corresponding charged pion has charge $e = e_{\{-,+\}}$, respectively. Thus, if the external quark is an up quark, the internal quark is a down quark and the pion carries positive charge. Conversely, if the external quark is a down quark, the internal quark is an up quark and the pion carries negative charge. This assignment satisfies charge conservation at each vertex.

Substituting Eqs.~(\ref{SqfBD.8})-(\ref{Ritusin}) into Eq.~(\ref{D.8initial}), the contribution $D.8$ becomes
\begin{align}
-i\Pi_{f}^{(D.8)} = & \,2g^{2} \int d^{4}x \int d^{4}y \biggl[ \frac{1}{2} \big(1 - i \gamma^{1}\gamma^{2}\big) \frac{1}{\sqrt{2\pi}} e^{i(k_{1}^{0}x^{0} - k_{1}^{3}x^{3})}e^{i\text{s}_{f'}(k_{1\text{s}_{f'}+}^{1} -k_{1}^{2})\theta} N_{k_{1\text{s}_{f'}+}^{1},k_{1}^{2}}\, \xi^{(k_{1\text{s}_{f'}+}^{1} -k_{1}^{2})/2} e^{-\frac{\xi}{2}}L_{k_{1}^{2}}^{k_{1\text{s}_{f'}+}^{1} -k_{1}^{2}} (\xi) \nonumber \\
+& \,\frac{1}{2} \big(1 + i \gamma^{1}\gamma^{2}\big) \frac{1}{\sqrt{2\pi}} e^{i(k_{1}^{0}x^{0} - k_{1}^{3}x^{3})}e^{i\text{s}_{f'}(k_{1\text{s}_{f'}-}^{1} -k_{1}^{2})\theta} N_{k_{1\text{s}_{f'}-}^{1},k_{1}^{2}}\, \xi^{(k_{1\text{s}_{f'}-}^{1} -k_{1}^{2})/2} e^{-\frac{\xi}{2}}L_{k_{1}^{2}}^{k_{1\text{s}_{f'}-}^{1} -k_{1}^{2}}(\xi) \biggr] \nonumber \\
\times & \frac{1}{\sqrt{2(E_{f'}+m_{f'})}} 
\begin{pmatrix}
    1, \ 0, \ 1, \ 0
\end{pmatrix} 
\Bigl( \gamma^{0}k_{1}^{0} + \frac{\gamma^{1}\text{s}_{f'}}{2}\sqrt{2\ell q_{f'}B} - \frac{\gamma^{2}\text{s}_{f'}}{2}\sqrt{2\ell q_{f'}B} + \gamma^{3}k_{1}^{3} + m_{f'} \mathbb{I}\Bigr) \nonumber \\
\times & \, \gamma^{5} e^{iq_{f} \int_{x}^{y} d\xi_{\mu} \Bigl[ A^{\mu}(\xi) + \frac{1}{2} F^{\mu\nu}(\xi - y)_{\nu} \Bigr]} \int \frac{d^{4}p_{2}}{(2\pi)^{4}} e^{-ip_{2}\cdot(x-y)} \int_{0}^{\infty} \frac{ds'}{\cos(|q_{f}B| s')} e^{is\bigl(p_{2\parallel}^{2} -p_{2\perp}^{2}\frac{\tan(|q_{f}B| s')}{|q_{f}B| s'} -m_{f}^{2} + i\epsilon \bigr)} \nonumber \\
\times & \biggl[ \Bigl(\cos(|q_{f}B| s') + \text{s}_{f}\gamma^{1}\gamma^{2}\sin(|q_{f}B|s')\Bigr)\Bigl(m_{f}+\slashed{p}_{2\parallel}\Bigr) - \frac{\slashed{p}_{2\perp}}{\cos(|q_{f}B| s')} \biggr] \gamma^{5} \nonumber \\ 
\times & e^{ie \int_{y}^{x} d\xi'_{\mu} \Bigl[ A^{\mu}(\xi') + \frac{1}{2} F^{\mu\nu}(\xi' - x)_{\nu} \Bigr]} \int \frac{d^{4}p_{1}}{(2\pi)^{4}} e^{-ip_{1}\cdot(y-x)} \int_{0}^{\infty} \frac{ds}{\cos(|eB| s)} e^{is\bigl(p_{1\parallel}^{2} -p_{1\perp}^{2}\frac{\tan(|eB| s)}{|eB| s} -m_{\pi}^{2} + i\epsilon \bigr)} \nonumber \\
\times & \biggl[\frac{1}{2} \big(1 + i \gamma^{1}\gamma^{2}\big) \frac{1}{\sqrt{2\pi}} e^{-i(k_{2}^{0}y^{0} - k_{2}^{3}y^{3})}e^{-i\text{s}_{f'}(k_{2\text{s}_{f'}+}^{1} -k_{2}^{2})\theta'} N_{k_{2\text{s}_{f'}+}^{1},k_{2}^{2}}\, \xi'^{(k_{2\text{s}_{f'}+}^{1} -k_{2}^{2})/2} e^{-\frac{\xi'}{2}}L_{k_{1}^{2}}^{k_{1\text{s}_{f'}+}^{1} -k_{1}^{2}} (\xi') \nonumber \\
+ & \,\frac{1}{2} \big(1 - i \gamma^{1}\gamma^{2}\big) \frac{1}{\sqrt{2\pi}} e^{-i(k_{2}^{0}y^{0} - k_{2}^{3}y^{3})}e^{-i\text{s}_{f'}(k_{2\text{s}_{f'}-}^{1} -k_{2}^{2})\theta'} N_{k_{2\text{s}_{f'}-}^{1},k_{2}^{2}}\, \xi'^{(k_{2\text{s}_{f'}-}^{1} -k_{2}^{2})/2} e^{-\frac{\xi'}{2}}L_{k_{2}^{2}}^{k_{2\text{s}_{f'}-}^{1} -k_{2}^{2}} (\xi') \bigg] \nonumber \\
\times & \frac{1}{\sqrt{2(E_{f'}+m_{f'})}} \Big( \gamma^{0}k_{2}^{0} - \frac{\gamma^{1}\text{s}_{f'}}{2}\sqrt{2\ell q_{f'}B} + \frac{\gamma^{2}\text{s}_{f'}}{2}\sqrt{2\ell q_{f'}B} - \gamma^{3}k_{2}^{3} + m_{f'} \mathbb{I}\Big) 
\begin{pmatrix}
    1 \\ 0 \\ -1 \\ 0
\end{pmatrix}.
\end{align}

We now simplify this expression. First, we commute one of the $\gamma^5$ matrices through the Dirac matrices appearing in the quark propagator and then use $(\gamma^5)^2=\mathbb{I}$. The Schwinger phases are then computed as
\begin{align}
    e^{ie\int_{y}^{x} d\xi'_{\mu} \Bigl[ A^{\mu}(\xi') + \frac{1}{2} F^{\mu\nu}(\xi' - x)_{\nu} \Bigr]} &= e^{-i\frac{eB}{2}(x^{1}y^{2} - x^{2}y^{1})}, \nonumber \\
    e^{iq_{f} \int_{x}^{y} d\xi_{\mu} \Bigl[ A^{\mu}(\xi) + \frac{1}{2} F^{\mu\nu}(\xi - y)_{\nu} \Bigr]} &= e^{i\frac{q_{f}B}{2}(x^{1}y^{2} - x^{2}y^{1})}.
\end{align}
After performing these steps, the expression for $D.8$ can be written
\begin{align}
-i\Pi_{f}^{(D.8)} = & \,2g^{2} \int d^{4}x \int d^{4}y \biggl[ \frac{1}{2} \big(1 - i \gamma^{1}\gamma^{2}\big) \frac{1}{\sqrt{2\pi}} e^{i(k_{1}^{0}x^{0} - k_{1}^{3}x^{3})}e^{i\text{s}_{f'}(k_{1\text{s}_{f'}+}^{1} -k_{1}^{2})\theta} N_{k_{1\text{s}_{f'}+}^{1},k_{1}^{2}}\, \xi^{(k_{1\text{s}_{f'}+}^{1} -k_{1}^{2})/2} e^{-\frac{\xi}{2}}L_{k_{1}^{2}}^{k_{1\text{s}_{f'}+}^{1} -k_{1}^{2}} (\xi) \nonumber \\
+& \,\frac{1}{2} \big(1 + i \gamma^{1}\gamma^{2}\big) \frac{1}{\sqrt{2\pi}} e^{i(k_{1}^{0}x^{0} - k_{1}^{3}x^{3})}e^{i\text{s}_{f'}(k_{1\text{s}_{f'}-}^{1} -k_{1}^{2})\theta} N_{k_{1\text{s}_{f'}-}^{1},k_{1}^{2}}\, \xi^{(k_{1\text{s}_{f'}-}^{1} -k_{1}^{2})/2} e^{-\frac{\xi}{2}}L_{k_{1}^{2}}^{k_{1\text{s}_{f'}-}^{1} -k_{1}^{2}}(\xi) \biggr] \nonumber \\
\times & \frac{1}{\sqrt{2(E_{f'}+m_{f'})}} 
\begin{pmatrix}
    1, \ 0, \ 1, \ 0
\end{pmatrix} 
\Bigl( \gamma^{0}k_{1}^{0} + \frac{\gamma^{1}\text{s}_{f'}}{2}\sqrt{2\ell q_{f'}B} - \frac{\gamma^{2}\text{s}_{f'}}{2}\sqrt{2\ell q_{f'}B} + \gamma^{3}k_{1}^{3} + m_{f'} \mathbb{I}\Bigr) \nonumber \\
\times & \,e^{i\frac{q_{f}B}{2}\bigl(x^{1}y^{2} - x^{2}y^{1}\bigr)} \int \frac{d^{4}p_{2}}{(2\pi)^{4}} e^{-ip_{2}\cdot(x-y)} \int_{0}^{\infty} \frac{ds'}{\cos(|q_{f}B| s')} e^{is'\bigl(p_{2\parallel}^{2} -p_{2\perp}^{2}\frac{\tan(|q_{f}B| s')}{|q_{f}B| s'} -m_{f}^{2} + i\epsilon \bigr)} \nonumber \\
\times & \biggl[ \Bigl(\cos(|q_{f}B| s') + \text{s}_{f}\gamma^{1}\gamma^{2}\sin(|q_{f}B| s')\Bigr)\Bigl(m_{f} -\slashed{p}_{2\parallel}\Bigr) + \frac{\slashed{p}_{2\perp}}{\cos(|q_{f}B| s')} \biggr] \nonumber \\
\times & \,e^{-i\frac{eB}{2}\bigl(x^{1}y^{2} - x^{2}y^{1}\bigr)} \int \frac{d^{4}p_{1}}{(2\pi)^{4}} e^{-ip_{1}\cdot(y-x)} \int_{0}^{\infty} \frac{ds}{\cos(|eB| s)} e^{is\bigl(p_{1\parallel}^{2} - p_{1\perp}^{2}\frac{\tan(|eB|s)}{|eB|s} -m_{\pi}^{2} + i\epsilon \bigr)} \nonumber \\
\times & \biggl[\frac{1}{2} \big(1 + i \gamma^{1}\gamma^{2}\big) \frac{1}{\sqrt{2\pi}} e^{-i(k_{2}^{0}y^{0} - k_{2}^{3}y^{3})}e^{-i\text{s}_{f'}(k_{2\text{s}_{f'}+}^{1} -k_{2}^{2})\theta'} N_{k_{2\text{s}_{f'}+}^{1},k_{2}^{2}}\, \xi'^{(k_{2\text{s}_{f'}+}^{1} -k_{2}^{2})/2} e^{-\frac{\xi'}{2}}L_{k_{1}^{2}}^{k_{1\text{s}_{f'}+}^{1} -k_{1}^{2}} (\xi') \nonumber \\
+ & \,\frac{1}{2} \big(1 - i \gamma^{1}\gamma^{2}\big) \frac{1}{\sqrt{2\pi}} e^{-i(k_{2}^{0}y^{0} - k_{2}^{3}y^{3})}e^{-i\text{s}_{f'}(k_{2\text{s}_{f'}-}^{1} -k_{2}^{2})\theta'} N_{k_{2\text{s}_{f'}-}^{1},k_{2}^{2}}\, \xi'^{(k_{2\text{s}_{f'}-}^{1} -k_{2}^{2})/2} e^{-\frac{\xi'}{2}}L_{k_{2}^{2}}^{k_{2\text{s}_{f'}-}^{1} -k_{2}^{2}} (\xi') \bigg] \nonumber \\
\times & \frac{1}{\sqrt{2(E_{f'}+m_{f'})}} \Big( \gamma^{0}k_{2}^{0} - \frac{\gamma^{1}\text{s}_{f'}}{2}\sqrt{2\ell q_{f'}B} + \frac{\gamma^{2}\text{s}_{f'}}{2}\sqrt{2\ell q_{f'}B} - \gamma^{3}k_{2}^{3} + m_{f'} \mathbb{I}\Big) 
\begin{pmatrix}
    1 \\ 0 \\ -1 \\ 0
\end{pmatrix}.
\end{align}
The integrations over the parallel coordinates, $x_\parallel$ and $y_\parallel$, are performed as in contribution $D.6$, yielding
\begin{align}
-i\Pi_{f}^{(D.8)} = & \frac{2g^{2}}{2\pi} \int d^{2}x_{\perp} \int d^{2}y_{\perp} \biggl[ \frac{1}{2} \big(1 - i \gamma^{1}\gamma^{2}\big)  e^{i\text{s}_{f'}(k_{1\text{s}_{f'}+}^{1} -k_{1}^{2})\theta} N_{k_{1\text{s}_{f'}+}^{1},k_{1}^{2}}\, \xi^{(k_{1\text{s}_{f'}+}^{1} -k_{1}^{2})/2} e^{-\frac{\xi}{2}}L_{k_{1}^{2}}^{k_{1\text{s}_{f'}+}^{1} -k_{1}^{2}} (\xi) \nonumber \\
+& \,\frac{1}{2} \big(1 + i \gamma^{1}\gamma^{2}\big) e^{i\text{s}_{f'}(k_{1\text{s}_{f'}-}^{1} -k_{1}^{2})\theta} N_{k_{1\text{s}_{f'}-}^{1},k_{1}^{2}}\, \xi^{(k_{1\text{s}_{f'}-}^{1} -k_{1}^{2})/2} e^{-\frac{\xi}{2}}L_{k_{1}^{2}}^{k_{1\text{s}_{f'}-}^{1} -k_{1}^{2}}(\xi) \biggr] \nonumber \\
\times & \frac{1}{\sqrt{2(E_{f'}+m_{f'})}} 
\begin{pmatrix}
    1, \ 0, \ 1, \ 0
\end{pmatrix} 
\Bigl( \gamma^{0}k_{1}^{0} + \frac{\gamma^{1}\text{s}_{f'}}{2}\sqrt{2\ell q_{f'}B} - \frac{\gamma^{2}\text{s}_{f'}}{2}\sqrt{2\ell q_{f'}B} + \gamma^{3}k_{1}^{3} + m_{f'} \mathbb{I}\Bigr) \nonumber \\
\times & \,e^{i\frac{q_{f}B}{2}\bigl(x^{1}y^{2} - x^{2}y^{1}\bigr)} \int \frac{d^{4}p_{2}}{(2\pi)^{4}} e^{-ip_{2\perp}\cdot(x-y)_{\perp}} \int_{0}^{\infty} \frac{ds'}{\cos(|q_{f}B| s')} e^{is'\bigl(p_{2\parallel}^{2} -p_{2\perp}^{2}\frac{\tan(|q_{f}B| s')}{|q_{f}B| s'} -m_{f}^{2} + i\epsilon \bigr)} \nonumber \\
\times & \biggl[ \Bigl(\cos(|q_{f}B| s') + \text{s}_{f}\gamma^{1}\gamma^{2}\sin(|q_{f}B| s')\Bigr)\Bigl(m_{f} -\slashed{p}_{2\parallel}\Bigr) + \frac{\slashed{p}_{2\perp}}{\cos(|q_{f}B| s')} \biggr] \nonumber \\
\times & \,e^{-i\frac{eB}{2}\bigl(x^{1}y^{2} - x^{2}y^{1}\bigr)} \int \frac{d^{4}p_{1}}{(2\pi)^{4}} e^{-ip_{1\perp}\cdot(y-x)_{\perp}} \int_{0}^{\infty} \frac{ds}{\cos(|eB| s)} e^{is\bigl(p_{1\parallel}^{2} - p_{1\perp}^{2}\frac{\tan(|eB|s)}{|eB|s} -m_{\pi}^{2} + i\epsilon \bigr)} \nonumber \\
\times &\biggl[\frac{1}{2} \big(1 + i \gamma^{1}\gamma^{2}\big) e^{-i\text{s}_{f'}(k_{2\text{s}_{f'}+}^{1} -k_{2}^{2})\theta'} N_{k_{2\text{s}_{f'}+}^{1},k_{2}^{2}}\, \xi'^{(k_{2\text{s}_{f'}+}^{1} -k_{2}^{2})/2} e^{-\frac{\xi'}{2}}L_{k_{1}^{2}}^{k_{1\text{s}_{f'}+}^{1} -k_{1}^{2}} (\xi') \nonumber \\
+ & \,\frac{1}{2} \big(1 - i \gamma^{1}\gamma^{2}\big)e^{-i\text{s}_{f'}(k_{2\text{s}_{f'}-}^{1} -k_{2}^{2})\theta'} N_{k_{2\text{s}_{f'}-}^{1},k_{2}^{2}}\, \xi'^{(k_{2\text{s}_{f'}-}^{1} -k_{2}^{2})/2} e^{-\frac{\xi'}{2}}L_{k_{2}^{2}}^{k_{2\text{s}_{f'}-}^{1} -k_{2}^{2}} (\xi') \bigg] \nonumber \\
\times & \frac{1}{\sqrt{2(E_{f'}+m_{f'})}} \Big( \gamma^{0}k_{2}^{0} - \frac{\gamma^{1}\text{s}_{f'}}{2}\sqrt{2\ell q_{f'}B} + \frac{\gamma^{2}\text{s}_{f'}}{2}\sqrt{2\ell q_{f'}B} - \gamma^{3}k_{2}^{3} + m_{f'} \mathbb{I}\Big) 
\begin{pmatrix}
    1 \\ 0 \\ -1 \\ 0
\end{pmatrix}\nonumber \\
\times & (2\pi)^{4} \delta^{(2)}(p_{2\parallel} - k_{1\parallel} - p_{1\parallel}) \delta^{(2)}(k_{2\parallel} + p_{1\parallel} - p_{2\parallel}).
\end{align}
For the perpendicular coordinates, we introduce polar variables, following the same strategy used in the previous contributions. The angular integrations can then be performed, giving
\begin{align}
-i\Pi_{f}^{(D.8)} = & \frac{2g^{2}}{2\pi} \int d^{2}x_{\perp} \int d^{2}y_{\perp} \biggl[ \frac{1}{2} \big(1 - i \gamma^{1}\gamma^{2}\big) N_{k_{1\text{s}_{f'}+}^{1},k_{1}^{2}}\,+ \frac{1}{2} \big(1 + i \gamma^{1}\gamma^{2}\big) N_{k_{1\text{s}_{f'}-}^{1},k_{1}^{2}}\biggr] \nonumber \\
\times & \frac{1}{\sqrt{2(E_{f'}+m_{f'})}} 
\begin{pmatrix}
    1, \ 0, \ 1, \ 0
\end{pmatrix} 
\Bigl( \gamma^{0}k_{1}^{0} + \frac{\gamma^{1}\text{s}_{f'}}{2}\sqrt{2\ell q_{f'}B} - \frac{\gamma^{2}\text{s}_{f'}}{2}\sqrt{2\ell q_{f'}B} + \gamma^{3}k_{1}^{3} + m_{f'} \mathbb{I}\Bigr) \nonumber \\
\times & \sum_{n,m,q} J_{q}\left(r\sqrt{(p_{2}^{1}-p_{1}^{1})^{2} + (p_{2}^{2}-p_{1}^{2})^{2}}\right) J_{m}\left(r'\sqrt{(p_{2}^{1}-p_{1}^{1})^{2} + (p_{2}^{2}-p_{1}^{2})^{2}}\right) J_{n}\left(\frac{|q_{f'}B|}{2}r r'\right) \nonumber \\
\times & \Bigg[ \bigg(\frac{|q_{f'}B|}{2}\bigg)^{(k_{1\text{s}_{f'}+}^{1} -k_{1}^{2})/2}\bigg(\frac{|q_{f'}B|}{2}\bigg)^{(k_{2\text{s}_{f'}+}^{1} -k_{2}^{2})/2} \frac{i^{\,m-1}e^{-im\varphi_{0}}}{m-n+(k_{2\text{s}_{f'}+}^{1}-k_{2}^{2})} \Big(e^{i2\pi\big(m-n+(k_{2\text{s}_{f'}+}^{1}-k_{2}^{2})\big)}-1\Big) \nonumber \\
\times & \frac{(-i)^{\,q+1}e^{-iq\varphi_{0}}}{q+n-(k_{1\text{s}_{f'}+}^{1}-k_{1}^{2})} \bigg(e^{i2\pi\big(q+n-(k_{1\text{s}_{f'}+}^{1}-k_{1}^{2})\big)}-1\bigg) \int_{0}^{\infty} dr \ r^{(k_{1\text{s}_{f'}+}^{1}-k_{1}^{2})-1} e^{-\frac{|q_{f'}B|r^{2}}{4}} L_{k_{1}^{2}}^{k_{1\text{s}_{f'}+}^{1} -k_{1}^{2}} \bigg(\frac{|q_{f'}B|r^{2}}{2}\bigg) \nonumber \\
\times & \int_{0}^{\infty} dr' \ r'^{(k_{1\text{s}_{f'}+}^{1}-k_{2}^{2})-1} e^{-\frac{|q_{f'}B| r'^{2}}{4}} L_{k_{1}^{2}}^{k_{2\text{s}_{f'}+}^{1} -k_{2}^{2}} \left(\frac{|q_{f'}B| r'^{2}}{2}\right) + \bigg(\frac{|q_{f'}B|}{2}\bigg)^{(k_{1\text{s}_{f'}+}^{1} -k_{1}^{2})/2}\bigg(\frac{|q_{f'}B|}{2}\bigg)^{(k_{2\text{s}_{f'}-}^{1} -k_{2}^{2})/2} \nonumber \\
\times & \frac{i^{\,m-1}e^{-im\varphi_{0}}}{m-n+(k_{2\text{s}_{f'}-}^{1}-k_{2}^{2})} \Big(e^{i2\pi\big(m-n+(k_{2\text{s}_{f'}-}^{1}-k_{2}^{2})\big)}-1\Big)\frac{(-i)^{\,q+1}e^{-iq\varphi_{0}}}{q+n-k_{1\text{s}_{f'}+}^{1}-k_{1}^{2}} \bigg(e^{i2\pi\big(q+n-(k_{1\text{s}_{f'}+}^{1}-k_{1}^{2})\big)}-1\bigg) \nonumber \\
\times & \int_{0}^{\infty} dr \ r^{(k_{1\text{s}_{f'}+}^{1}-k_{1}^{2})-1} e^{-\frac{|q_{f'}B|r^{2}}{4}} L_{k_{1}^{2}}^{k_{1\text{s}_{f'}+}^{1} -k_{1}^{2}} \bigg(\frac{|q_{f'}B|r^{2}}{2}\bigg) \int_{0}^{\infty} dr' \ r'^{(k_{1\text{s}_{f'}-}^{1}-k_{2}^{2})-1} e^{-\frac{|q_{f'}B| r'^{2}}{4}} L_{k_{1}^{2}}^{k_{2\text{s}_{f'}-}^{1} -k_{2}^{2}} \left(\frac{|q_{f'}B| r'^{2}}{2}\right) \nonumber \\
\times & \bigg(\frac{|q_{f'}B|}{2}\bigg)^{(k_{1\text{s}_{f'}-}^{1} -k_{1}^{2})/2}\bigg(\frac{|q_{f'}B|}{2}\bigg)^{(k_{2\text{s}_{f'}+}^{1} -k_{2}^{2})/2} \frac{i^{\,m-1}e^{-im\varphi_{0}}}{m-n+(k_{2\text{s}_{f'}-}^{1}-k_{2}^{2})} \Big(e^{i2\pi\big(m-n+(k_{2\text{s}_{f'}-}^{1}-k_{2}^{2})\big)}-1\Big) \nonumber \\
\times & \frac{(-i)^{\,q+1}e^{-iq\varphi_{0}}}{q+n-(k_{1\text{s}_{f'}-}^{1}-k_{1}^{2})} \bigg(e^{i2\pi\big(q+n-(k_{1\text{s}_{f'}-}^{1}-k_{1}^{2})\big)}-1\bigg) \int_{0}^{\infty} dr \ r^{(k_{1\text{s}_{f'}-}^{1}-k_{1}^{2})-1} e^{-\frac{|q_{f'}B|r^{2}}{4}} L_{k_{1}^{2}}^{k_{1\text{s}_{f'}-}^{1} -k_{1}^{2}} \bigg(\frac{|q_{f'}B|r^{2}}{2}\bigg) \nonumber \\
\times & \int_{0}^{\infty} dr' \ r'^{(k_{2\text{s}_{f'}-}^{1}-k_{2}^{2})-1} e^{-\frac{|q_{f'}B| r'^{2}}{4}} L_{k_{1}^{2}}^{k_{2\text{s}_{f'}-}^{1} -k_{2}^{2}} \left(\frac{|q_{f'}B| r'^{2}}{2}\right) + \bigg(\frac{|q_{f'}B|}{2}\bigg)^{(k_{1\text{s}_{f'}-}^{1} -k_{1}^{2})/2}\bigg(\frac{|q_{f'}B|}{2}\bigg)^{(k_{2\text{s}_{f'}-}^{1} -k_{2}^{2})/2} \nonumber \\
\times & \frac{i^{\,m-1}e^{-im\varphi_{0}}}{m-n+(k_{2\text{s}_{f'}-}^{1}-k_{2}^{2})} \Big(e^{i2\pi\big(m-n+(k_{2\text{s}_{f'}-}^{1}-k_{2}^{2})\big)}-1\Big) \frac{(-i)^{\,q+1}e^{-iq\varphi_{0}}}{q+n-(k_{1\text{s}_{f'}-}^{1}-k_{1}^{2})} \bigg(e^{i2\pi\big(q+n-(k_{1\text{s}_{f'}-}^{1}-k_{1}^{2})\big)}-1\bigg) \nonumber \\
\times & \int_{0}^{\infty} dr \ r^{(k_{1\text{s}_{f'}+}^{1}-k_{1}^{2})-1} e^{-\frac{|q_{f'}B|r^{2}}{4}} L_{k_{1}^{2}}^{k_{1\text{s}_{f'}+}^{1} -k_{1}^{2}} \bigg(\frac{|q_{f'}B|r^{2}}{2}\bigg) \int_{0}^{\infty} dr' \ r'^{(k_{2\text{s}_{f'}-}^{1}-k_{2}^{2})-1} e^{-\frac{|q_{f'}B| r'^{2}}{4}} L_{k_{1}^{2}}^{k_{2\text{s}_{f'}-}^{1} -k_{2}^{2}} \left(\frac{|q_{f'}B| r'^{2}}{2}\right) \nonumber \\
\times & \int \frac{d^{4}p_{2}}{(2\pi)^{4}} \int_{0}^{\infty} \frac{ds'}{\cos(|q_{f}B| s')} e^{is'\bigl(p_{2\parallel}^{2} -p_{2\perp}^{2}\frac{\tan(|q_{f}B| s')}{|q_{f}B| s'} -m_{f}^{2} + i\epsilon \bigr)}\biggl[ \Bigl(\cos(|q_{f}B| s') + \text{s}_{f}\gamma^{1}\gamma^{2}\sin(|q_{f}B| s')\Bigr) \nonumber \\
\times & \Bigl(m_{f} -\slashed{p}_{2\parallel}\Bigr) + \frac{\slashed{p}_{2\perp}}{\cos(|q_{f}B| s')} \biggr] \int \frac{d^{4}p_{1}}{(2\pi)^{4}} \int_{0}^{\infty} \frac{ds}{\cos(|eB| s)} e^{is\bigl(p_{1\parallel}^{2} - p_{1\perp}^{2}\frac{\tan(|eB|s)}{|eB|s} -m_{\pi}^{2} + i\epsilon \bigr)} \nonumber \\
\times & \biggl[\frac{1}{2} \big(1 + i \gamma^{1}\gamma^{2}\big) N_{k_{2\text{s}_{f'}+}^{1},k_{2}^{2}} +\frac{1}{2} \big(1 - i \gamma^{1}\gamma^{2}\big) N_{k_{2\text{s}_{f'}-}^{1},k_{2}^{2}}\bigg] \nonumber \\
\times & \frac{1}{\sqrt{2(E_{f'}+m_{f'})}} \Big( \gamma^{0}k_{2}^{0} - \frac{\gamma^{1}\text{s}_{f'}}{2}\sqrt{2\ell q_{f'}B} + \frac{\gamma^{2}\text{s}_{f'}}{2}\sqrt{2\ell q_{f'}B} - \gamma^{3}k_{2}^{3} + m_{f'} \mathbb{I}\Big) 
\begin{pmatrix}
    1 \\ 0 \\ -1 \\ 0
\end{pmatrix}\nonumber \\
\times & (2\pi)^{4} \delta^{(2)}(p_{2\parallel} - k_{1\parallel} - p_{1\parallel}) \delta^{(2)}(k_{2\parallel} + p_{1\parallel} - p_{2\parallel}).
\label{D.8afterangularintegration}
\end{align}
where $J_l(X)$ are the Bessel functions of first kind; these functions arise from the angular integrations, where the Jacobi-Anger identity is used to expand plane waves in terms of cylindrical harmonics. Since the expression in Eq.~(\ref{D.8afterangularintegration}) does not seem to admit further analytical simplification, we consider the inertial frame in which the external particle is at rest, namely $k_j^1=k_j^2=k_j^3=0$ and $k_j^2=(k_j^0)^2$, with $j=1,2$. Thus, the expression we need to compute becomes
\begin{align}
-i\Pi_{f}^{(D.8)} = & \frac{2g^{2}}{2\pi} \int d^{2}x_{\perp} \int d^{2}y_{\perp} \frac{\sqrt{|q_{f'}B|} e^{-\frac{\xi}{2}}}{\sqrt{4m_{f'}}} 
\begin{pmatrix}
    1, \ 0, \ 1, \ 0
\end{pmatrix} \Bigl( \gamma^{0}k_{1}^{0} + m_{f'} \mathbb{I}\Bigr)  \,e^{i\frac{q_{f}B}{2}\bigl(x^{1}y^{2} - x^{2}y^{1}\bigr)} \nonumber \\
\times & \int \frac{d^{4}p_{2}}{(2\pi)^{4}} e^{-ip_{2\perp}\cdot(x-y)_{\perp}} \int_{0}^{\infty} \frac{ds'}{\cos(|q_{f}B| s')} e^{is'\bigl(p_{2\parallel}^{2} -p_{2\perp}^{2}\frac{\tan(|q_{f}B| s')}{|q_{f}B| s'} -m_{f}^{2} + i\epsilon \bigr)} \nonumber \\
\times & \biggl[ \Bigl(\cos(|q_{f}B| s') + \text{s}_{f}\gamma^{1}\gamma^{2}\sin(|q_{f}B| s')\Bigr)\Bigl(m_{f} -\slashed{p}_{2\parallel}\Bigr) + \frac{\slashed{p}_{2\perp}}{\cos(|q_{f}B| s')} \biggr] e^{-i\frac{eB}{2}\bigl(x^{1}y^{2} - x^{2}y^{1}\bigr)} \nonumber \\
\times &\int \frac{d^{4}p_{1}}{(2\pi)^{4}} e^{-ip_{1\perp}\cdot(y-x)_{\perp}} \int_{0}^{\infty} \frac{ds}{\cos(|eB| s)} e^{is\bigl(p_{1\parallel}^{2} - p_{1\perp}^{2}\frac{\tan(|eB|s)}{|eB|s} -m_{\pi}^{2} + i\epsilon \bigr)} \nonumber \\
\times & \frac{\sqrt{|q_{f'}B|} e^{-\frac{\xi'}{2}}}{\sqrt{4m_{f'}}} \Big( \gamma^{0}k_{2}^{0} + m_{f'} \mathbb{I}\Big) 
\begin{pmatrix}
    1 \\ 0 \\ -1 \\ 0
\end{pmatrix} (2\pi)^{4} \delta(p_{2}^{0} - k_{1}^{0} - p_{1}^{0})\delta(p_{2}^{3} - p_{1}^{3}) \delta(k_{2}^{0} + p_{1}^{0} - p_{2}^{0})\delta(p_{1}^{3} - p_{2}^{3}).
\end{align}
We proceed to integrate over the perpendicular spatial components, using the following results
\begin{align}
    \int dy^{2} e^{ i\Big( i\frac{|q_{f'}B|}{4}(y^{2})^{2} + y^{2}\big[(p_{1}^{2} - p_{2}^{2})+\frac{e'B}{2}x^{1}\big] \Big)} &= \sqrt{\frac{4\pi}{|q_{f'}B|}} \exp{\Bigg\{ -\frac{\big[(p_{1}^{2} - p_{2}^{2})+\frac{e'B}{2}x^{1}\big]^{2}}{|q_{f'}B|}\Bigg\}}, \nonumber \\
    \int dy^{1} e^{ i\Big( i\frac{|q_{f'}B|}{4}(y^{1})^{2} + y^{1}\big[(p_{2}^{1} - p_{1}^{1})-\frac{e'B}{2}x^{2}\big] \Big)} &= \sqrt{\frac{4\pi}{|q_{f'}B|}} \exp{\Bigg\{-\frac{\big[(p_{1}^{1} - p_{2}^{1})-\frac{e'B}{2}x^{2}\big]^{2}}{|q_{f'}B|}\Bigg\}},\nonumber \\
    e^{-\frac{(p_{1}^{1}-p_{2}^{1})^{2}}{|q_{f'}B|}}\int dx^{2} e^{-\frac{(x^{2})^{2}}{4} \big(|q_{f'}B|+ \frac{(e'B)^{2}}{|q_{f'}B|}\big)} e^{x^{2}\Big[ \frac{e'B}{|q_{f'}B|}(p_{1}^{1} - p_{2}^{1}) + i(p_{2}^{2} - p_{1}^{2})\Big]} &= \sqrt{\frac{4\pi}{M}} e^{-\frac{(p_{1}^{1}-p_{2}^{1})^{2}}{|q_{f'}B|}} e^{\frac{1}{M} \Big[ \frac{e'B}{|q_{f'}B|}(p_{1}^{1} - p_{2}^{1}) + i(p_{2}^{2} - p_{1}^{2})\Big]^{2}}, \nonumber \\
    e^{-\frac{(p_{1}^{2}-p_{2}^{2})^{2}}{|q_{f'}B|}}\int dx^{1} e^{-\frac{(x^{1})^{2}}{4} \big(|q_{f'}B|+ \frac{(e'B)^{2}}{|q_{f'}B|}\big)} e^{x^{2}\Big[i(p_{2}^{1} - p_{1}^{1}) - \frac{e'B}{|q_{f'}B|}(p_{1}^{2} - p_{2}^{2})\Big]} &= \sqrt{\frac{4\pi}{M}} e^{-\frac{(p_{1}^{2}-p_{2}^{2})^{2}}{|q_{f'}B|}} e^{\frac{1}{M} \Big[i(p_{2}^{1} - p_{1}^{1}) - \frac{e'B}{|q_{f'}B|}(p_{1}^{2} - p_{2}^{2})\Big]^{2}},
\end{align}
with $e' = q_{\{u,d\}} + e_{\{-,+\}}$, $q_{f'} = q_{\{d,u\}}$ and $M = |q_{f'}B|+ \frac{(e'B)^{2}}{|q_{f'}B|}$. Therefore, after a straightforward algebra, the expression for $D.8$ becomes
\begin{align}
-i\Pi_{f}^{(D.8)} = & \frac{2g^{2}}{2\pi} \frac{\sqrt{|q_{f'}B||q_{f'}B|}}{4m_{f'}} \frac{(4\pi)(4\pi)}{ |q_{f'}B|M} 
\begin{pmatrix}
    1, \ 0, \ 1, \ 0
\end{pmatrix} 
\Bigl( \gamma^{0}k_{1}^{0} + m_{f'} \mathbb{I}\Bigr) \nonumber \\
\times & \int \frac{d^{4}p_{2}}{(2\pi)^{4}} \int_{0}^{\infty} \frac{ds'}{\cos(|q_{f}B| s')} e^{is\bigl(p_{2\parallel}^{2} - p_{2\perp}^{2}\frac{\tan(|q_{f}B| s')}{|q_{f}B| s'} - m_{f}^{2} + i\epsilon \bigr)} \nonumber \\
\times & \biggl[ \Bigl(\cos(|q_{f}B| s') + \text{s}_{f}\gamma^{1}\gamma^{2}\sin(|q_{f}B| s')\Bigr)\Bigl(m_{f} -\slashed{p}_{2\parallel}\Bigr) + \frac{\slashed{p}_{2\perp}}{\cos(|q_{f}B| s')} \biggr] \nonumber \\
\times & \,\int \frac{d^{4}p_{1}}{(2\pi)^{4}} \int_{0}^{\infty} \frac{ds}{\cos(|eB|s)} e^{is\bigl(p_{1\parallel}^{2} -p_{1\perp}^{2}\frac{\tan(|eB|s)}{|eB|s} -m_{\pi}^{2} + i\epsilon \bigr)} e^{-\frac{(p_{1}^{1}-p_{2}^{1})^{2}}{|q_{f'}B|}} e^{-\frac{(p_{1}^{2}-p_{2}^{2})^{2}}{|q_{f'}B|}} \nonumber \\
\times & \exp{\Bigg\{ -\frac{1}{M} \Big[(p_{2}^{1}-p_{1}^{1})^{2}\Big(1+\frac{(e'B)^{2}}{|q_{f'}B|^{2}}\Big)+(p_{2}^{2}-p_{1}^{2})^{2}\Big(1+\frac{(e'B)^{2}}{|q_{f'}B|^{2}}\Big)\Big]\Bigg\}} \Big(\gamma^{0}k_{2}^{0} + m_{f'} \mathbb{I}\Big) 
\begin{pmatrix}
    1 \\ 0 \\ -1 \\ 0
\end{pmatrix} \nonumber \\
\times & (2\pi)^{4} \delta(p_{2}^{0} - k_{1}^{0} - p_{1}^{0})\delta(p_{2}^{3} - p_{1}^{3}) \delta(k_{2}^{0} + p_{1}^{0} - p_{2}^{0})\delta(p_{1}^{3} - p_{2}^{3}).
\end{align}
We now integrate over one of the internal momenta. Choosing $p_2$, we first perform the integrations over its parallel components, obtaining the following momentum-space expression
\begin{align}
-i\Pi_{f}^{(D.8)} = &  2g^{2} \frac{(4\pi)(4\pi)}{ |q_{f'}B|M} \int \frac{d^{2}p_{2\perp}}{(2\pi)^{2}}\int \frac{d^{4}p_{1}}{(2\pi)^{4}} \int_{0}^{\infty} \frac{ds'}{\cos(|q_{f}B| s')} e^{is'\bigl((p_{1}^{0} + k_{1}^{0})^{2} - (p_{1}^{3})^{2} -p_{2\perp}^{2}\frac{\tan(|q_{f}B| s')}{|q_{f}B| s'} - m_{f}^{2} + i\epsilon \bigr)} \nonumber\\
\times & \biggl[ \Bigl(\cos(|q_{f}B| s') + \text{s}_{f}\gamma^{1}\gamma^{2}\sin(|q_{f}B| s')\Bigr)\Bigl(m_{f}-\gamma^{0}(p_{1}^{0} + k_{1}^{0}) + \gamma^{3}p_{1}^{3} \Bigr) + \frac{\slashed{p}_{2\perp}}{\cos(|q_{f}B| s')} \biggr] \nonumber \\
\times &  \int_{0}^{\infty} \frac{ds}{\cos(|eB|s)} e^{is\bigl(p_{1\parallel}^{2} -p_{1\perp}^{2}\frac{\tan(|eB|s)}{|eB|s} -m_{\pi}^{2} + i\epsilon \bigr)} e^{-\frac{(p_{1}^{1}-p_{2}^{1})^{2}}{|q_{f'}B|}} e^{-\frac{(p_{1}^{2}-p_{2}^{2})^{2}}{|q_{f'}B|}}  e^{-\frac{\Big(1+\frac{(e'B)^{2}}{|q_{f'}B|^{2}}\Big)}{M} \Big[(p_{2}^{1}-p_{1}^{1})^{2}+(p_{2}^{2}-p_{1}^{2})^{2}\Big]}.
\label{D.8afterparallelmomentumintegrals}
\end{align}
We then integrate over the perpendicular components, using
\begin{align}
    \int d^{2}p_{2\perp} e^{-i p_{2\perp}^{2} \frac{\tan(|q_{f}B| s)}{|q_{f}B| }} e^{-Q(p_{2\perp} - p_{1\perp})^{2}}& = \frac{\pi|q_{f}B||q_{f'}B|}{2|q_{f}B| + i|q_{f'}B|\tan(|q_{f}B| s')} e^{-\frac{2ip_{1\perp}^{2}\tan(|q_{f}B| s')}{2|q_{f}B| + i|q_{f'}B|\tan(|q_{f}B| s')}}, \nonumber \\
    \int d^{2}p_{2\perp} \ \slashed{p}_{2\perp}e^{-(R p_{2\perp}^{2}-2Qp_{1\perp}\cdot p_{2\perp})} &=\frac{\pi|q_{f}B||q_{f'}B|}{2|q_{f}B| + i|q_{f'}B|\tan(|q_{f}B| s')} e^{-\frac{2ip_{1\perp}^{2}\tan(|q_{f}B| s')}{2|q_{f}B| + i|q_{f'}B|\tan(|q_{f}B| s')}} \nonumber \\
\times & \frac{2|q_{f}B|}{2|q_{f}B| + i|q_{f'}B|\tan(|q_{f}B| s')} (\gamma^{1}p_{1}^{1} + \gamma^{2}p_{1}^{2}),
\label{D.8momentumperpintegrals}
\end{align}
with 
\begin{equation}
    Q = \frac{2}{|q_{f'}B|}, \qquad R = \frac{2|q_{f}B| + i|q_{f'}B|\tan(|q_{f}B| s')}{|q_{f}B||q_{f'}B|}.
\end{equation}
Substituting Eq.~(\ref{D.8momentumperpintegrals}) into Eq.~(\ref{D.8afterparallelmomentumintegrals}), we get
\begin{align}
-i\Pi_{f}^{(D.8)} = &   \frac{8g^2}{|q_{f}B|M} \int \frac{d^{4}p_{1}}{(2\pi)^{4}} \int_{0}^{\infty} \frac{ds'}{\cos(|q_{f}B| s')} e^{is'\bigl((p_{1}^{0} + k_{1}^{0})^{2} - (p_{1}^{3})^{2} - m_{f}^{2} + i\epsilon \bigr)} \int_{0}^{\infty} \frac{ds}{\cos(|eB|s)} e^{is\bigl(p_{1\parallel}^{2} -p_{1\perp}^{2}\frac{\tan(|eB|s)}{|eB|s} -m_{\pi}^{2} + i\epsilon \bigr)}\nonumber \\
\times & \frac{\pi|q_{f}B||q_{f'}B|}{2|q_{f}B| + i|q_{f'}B|\tan(|q_{f}B| s')} e^{-\frac{2ip_{1\perp}^{2}\tan(|q_{f}B| s')}{2|q_{f}B| + i|q_{f'}B|\tan(|q_{f}B| s')}}\biggl[ \Bigl(\cos(|q_{f}B| s') + \text{s}_{f}\gamma^{1}\gamma^{2}\sin(|q_{f}B| s')\Bigr)\nonumber \\
\times &\Bigl(m_{f}-\gamma^{0}(p_{1}^{0} + k_{1}^{0}) +\gamma^{3}p_{1}^{3} \Bigr)+\frac{2|q_{f}B|}{2|q_{f}B| + i|q_{f'}B|\tan(|q_{f}B| s')} (\gamma^{1}p_{1}^{1} + \gamma^{2}p_{1}^{2}) \frac{\gamma^{1}{p}_{1}^{1}+\gamma^{2}p_{1}^{2}}{\cos(|q_{f}B| s')} \biggr] .
\end{align}

We proceed to compute the integrals over $p_{1\perp}$ and $p_1^3$, for which we use
\begin{align}
    \int d^{2}p_{1\perp} e^{-p_{2\perp}^{2} \left[\frac{2i\tan(|q_{f}B| s')}{2|q_{f}B| + i|q_{f'}B|\tan(|q_{f}B| s')} +i \frac{\tan(|eB| s')}{|eB|}\right]} &= \frac{\pi}{Z}, \nonumber \\
    \int dp_{1}^{3} e^{-i(p_{1}^{3})^{2}(s+s')} &= \sqrt{\frac{\pi}{i(s+s')}}, \nonumber \\
    \int d^{2}p_{1\perp} \ \slashed{p}_{1\perp} e^{-p_{1\perp}^{2} \left[Q -\frac{Q^{2}}{R} +i \frac{\tan(|eB| s')}{|eB|}\right]} &= 0, \nonumber \\
    \int dp_{1}^{3} \ p_{1}^{3} e^{-i(p_{1}^{3})^{2}(s+s')} &= 0,
\end{align}
where
\begin{equation}
    Z = \frac{2i|eB|\tan(|q_{f}B|s')+2i|q_{f}B|\tan(|eB|s)-|q_{f'}B|\tan(|eB|s)\tan(|q_{f}B|s')}{|eB|(2|q_{f}B| + i|q_{f'}B|\tan(|q_{f}B| s'))},
\end{equation}
and the expression for $D.8$ becomes
\begin{align}
-i\Pi_{f}^{(D.8)} = & \frac{g^{2}}{\pi^3M}  \int \frac{dp_{1}^{0}}{2\pi} \int_{0}^{\infty} \frac{ds'}{\cos(|q_{f}B| s')} e^{is'(p_{1}^{0} + k_{1}^{0})^{2}}e^{is'(-m_{f}^{2} + i\epsilon)} \int_{0}^{\infty} \frac{ds}{\cos(|eB|s)} e^{is(p_{1}^{0})^{2}}e^{is(-m_{\pi}^{2} + i\epsilon)} \sqrt{\frac{\pi}{i(s+s')}}  \nonumber \\
\times &\frac{\pi|q_{f}B||q_{f'}B|}{2|q_{f}B| + i|q_{f'}B|\tan(|q_{f}B| s')} \Bigl(\cos(|q_{f}B| s') + \text{s}_{f}\gamma^{1}\gamma^{2}\sin(|q_{f}B| s')\Bigr) \Bigl(m_{f}-\gamma^{0}(p_{1}^{0} + k_{1}^{0}) \Bigr) \nonumber \\
\times & \frac{\pi|eB|(2|q_{f}B| + i|q_{f'}B|\tan(|q_{f}B| s'))}{2i|eB|\tan(|q_{f}B|s')+2i|q_{f}B|\tan(|eB|s)-|q_{f'}B|\tan(|eB|s)\tan(|q_{f}B|s')}.
\label{D.8withoutinternal3momentum}
\end{align}

The next step is to implement the change of variables $s\rightarrow -i\tau$ and $s'\rightarrow -i\tau'$ and then to perform a Wick rotation and rewrite the Eq.~(\ref{D.8withoutinternal3momentum}) within the Matsubara formalism as follows
\begin{align}
-i\Pi_{f}^{(D.8)} = & -\frac{ig^2}{\pi M} T\sum_{n}  \int_{0}^{\infty} \frac{d\tau'}{\cosh(|q_{f}B| \tau')} e^{\tau'[i(\omega_{n} + \omega)]^{2}}e^{-\tau'm_{f}^{2}} \int_{0}^{\infty} \frac{d\tau}{\cosh(|eB|\tau)} e^{\tau(i\omega_{n})^{2}}e^{-\tau m_{\pi}^{2}} \sqrt{\frac{\pi}{\tau+\tau'}}\nonumber \\
\times & \frac{|eB||q_{f}B||q_{f'}B|}{2|eB|\tanh(|q_{f}B|\tau')+2|q_{f}B|\tanh(|eB|\tau) +|q_{f'}B|\tanh(|eB|\tau)\tanh(|q_{f}B|\tau')}\nonumber \\
\times & \Bigl(\cosh(|q_{f}B| \tau') - i\text{s}_{f}\gamma^{1}\gamma^{2}\sinh(|q_{f}B|\tau')\Bigr) \Bigl(m_{f} + \gamma^{0}(\omega_{n} + \omega) \Bigr). 
\end{align}
We perform the sums
\begin{align}
    e^{-\tau'\omega^{2}} \sum_{n=-\infty}^{\infty} e^{-(\tau+\tau')\omega_{n}^{2}-2\omega_{n}\omega\tau'} = & \frac{1}{2T\sqrt{\pi(\tau+\tau')}} e^{-\frac{\tau\tau' \omega^{2}}{\tau + \tau'}} \Big[ 1 + 2\sum_{n=1}^{\infty} e^{-\frac{n^{2}}{4(\tau + \tau')T^{2}}} \cos\Big(\frac{n\omega\tau'}{(\tau+\tau')T}\Big)\Big], \nonumber \\
    e^{-\tau'\omega^{2}} \sum_{n=-\infty}^{\infty} \omega_{n}  e^{-(\tau+\tau')\omega_{n}^{2}-2\omega_{n}\omega\tau'}= & \ \frac{1}{2T\sqrt{\pi (\tau+\tau')}} e^{-\frac{\tau \tau' \omega^{2}}{\tau + \tau'}} \bigg\{-\frac{\tau \omega}{\tau +\tau'}+ 2\sum_{n=1}^{\infty} e^{-\frac{n^{2}}{4T^{2}(\tau +\tau')}} \\
\times & \bigg[-\frac{\tau' \omega}{\tau+\tau'} \cos\Big(\frac{n \tau' \omega}{T(\tau+\tau')}\Big)+\frac{n}{2(\tau +\tau')T}\sin\Big(\frac{n \tau' \omega}{T(\tau+\tau')} \Big)\bigg]\bigg\},
\end{align}
and the change of variables $\tau=u(1-v)$ and $\tau'=uv$, and obtain the final expression
\begin{align}
-i\Pi_{f}^{(D.8)} = &  -\frac{ig^{2}}{4\pi^{2}} \frac{1}{M} \int_{0}^{\infty} du \int_{0}^{1} dv \frac{1}{\cosh(|q_{f}B| uv)} \frac{1}{\cosh(|eB|u(1-v))} e^{-u(1-v)m_{\pi}^{2}}  e^{-uvm_{f}^{2}} e^{-uv(1-v)(i\omega)^{2}} \nonumber \\
\times & \frac{|eB||q_{f}B||q_{f'}B|}{2|eB|\tanh(|q_{f}B|uv)+2|q_{f}B|\tanh(|eB|u(1-v)) +|q_{f'}B|\tanh(|eB|u(1-v))\tanh(|q_{f}B|uv)}\nonumber \\
\times & \Bigl(\cosh(|q_{f}B| uv) - i \text{s}_{f}\gamma^{1}\gamma^{2}\sinh(|q_{f}B| uv)\Bigr)\nonumber \\
\times & \bigg[\Big(m_{f} - i\gamma^{0}(i\omega)\Big) \Big[ 1 + 2\sum_{n=1}^{\infty} e^{-\frac{n^{2}}{4uT^{2}}} \cosh\Big(\frac{v(i\omega)n}{T}\Big)\Big] + \gamma^{0} \Big\{ iv(i\omega) \nonumber \\
+ & \, 2\sum_{n=1}^{\infty} (-1)^{n}e^{-\frac{n^{2}}{4uT^{2}}} \Big[iv(i\omega) \cosh\Big(\frac{v(i\omega)n}{T}\Big) - \frac{in}{2uT}\sinh\Big(\frac{v(i\omega)n}{T}\Big)\Big]\Big\} \biggr].
\label{D.8final}
\end{align}
The purely magnetic terms in Eq.~(\ref{D.8final}) contain the ultraviolet-divergent vacuum contribution, which can be identified by taking the limit when the magnetic field goes to zero
\begin{equation}
-i\Pi_{f,B=0}^{(D.8)} = -\frac{ig^{2}}{8\pi^{2}} \int_{0}^{\infty} du  \int_{0}^{1} dv\,e^{-u(1-v) m_{\pi}^{2}}   e^{-uvm_{f}^{2}}  e^{-uv(1-v)(i\omega)^{2}} \frac{1}{u}  \Big(m_{f} - i\gamma^{0}(1-v)(i\omega) \Big).
\end{equation}
This completes the one-loop calculation of the quark self-energy.
\end{widetext}

Before concluding this subsection, it is useful to discuss the general Dirac structure of the quark self-energy in Euclidean space. In the most general case, the one-loop quark self-energy can be written as
\begin{equation}
-i\Pi_{f}(i\omega,\vec{k}) = A(i\omega,\vec{k})\, i\gamma^{0} - B(i\omega,\vec{k})\,\vec{\gamma}\cdot\vec{k} + C(i\omega,\vec{k}),
\label{generalquarkselfenergy}
\end{equation}
where the scalar functions $A$, $B$, and $C$ encode the thermal and magnetic corrections to the fermion propagator.

In the present work, all contributions have been evaluated in the inertial frame where the external particle is at rest, namely $\vec{k}=0$. Therefore, the term proportional to $\vec{\gamma}\cdot\vec{k}$ vanishes identically, implying
\begin{equation}
B(i\omega,\vec{k}=0)=0.
\end{equation}
Consequently, the complete one-loop self-energy is fully determined by the coefficients multiplying the structures $i\gamma^{0}$ and the identity matrix. From the explicit results obtained in contributions $D.1$--$D.8$, the coefficient $A$ is identified from all terms proportional to $i\gamma^{0}$, whereas the coefficient $C$ is obtained from the remaining scalar contributions proportional to the identity matrix in Dirac space. Hence, the quark self-energy in the rest frame can be written in the simplified form
\begin{equation}
-i\Pi_{f}(i\omega,\vec{k}=0)=A(i\omega)\, i\gamma^{0}+C(i\omega).
\label{quarkselfenergyrestframe}
\end{equation}
The explicit expressions for $A(i\omega)$ and $C(i\omega)$ follow directly from the decomposition of the final expressions for the contributions $D.1-D.8$ into their corresponding Dirac structures. 

\section{Conclusions \label{sec4}}

We have presented a complete calculation of the one-loop self-energies for all fields in the linear sigma model coupled to quarks at finite temperature and in the presence of a uniform magnetic field. The analysis consistently incorporates thermal and magnetic effects for neutral and charged degrees of freedom, providing a unified treatment valid for arbitrary values of the temperature and the field strength.

The calculation is carried out using the Matsubara formalism to account for thermal effects and the Schwinger proper-time representation for charged propagators in a magnetic background. Particular attention has been devoted to contributions involving particles with different electric charges in the loop. In these cases, the associated Schwinger phases do not cancel, preventing a direct evaluation in momentum space. We have shown that a proper treatment of such contributions requires describing charged external states in the Ritus basis, rather than as plane waves, so that the calculation can be systematically performed in coordinate space and consistently mapped into momentum space. To the best of our knowledge, this constitutes the first explicit derivation of these terms within the linear sigma model including both thermal and magnetic effects.

The structure of the resulting self-energies reflects the nontrivial interplay between thermal fluctuations and the magnetic background. In particular, the separation between vacuum and matter contributions allows for a clear identification of ultraviolet divergences and their origin, providing a solid basis for subsequent renormalization procedures.

The results obtained in this work provide the necessary building blocks for a fully self-consistent and non-perturbative treatment of thermodynamic properties in effective models of QCD under extreme conditions.

Finally, this work establishes a consistent and systematic framework for the computation of thermomagnetic one-loop corrections in effective models of QCD, capturing the full interplay between thermal and magnetic effects for all dynamical degrees of freedom.

\begin{acknowledgments}
Support for this work was received in part by the Secretaria de Ciencia, Humanidades, Tecnología e Innovación Grant No. CBF-2025-G-1718. RZ acknowledges support from ANID/CONICYT FONDECYT Regular (Chile) under Grant No. 1241436, and AA-F and JCM acknowledge the financial support of a fellowship granted by Secretaria de Ciencia, Humanidades, Tecnología e Innovación as part of the Sistema Nacional de Posgrados.
\end{acknowledgments}

\bibliography{mybibliography}

\end{document}